\documentclass[pdflatex,sn-mathphys-num]{sn-jnl}% Math and Physical Sciences Numbered Reference Style
\usepackage{graphicx}%
\usepackage{multirow}%
\usepackage{amsmath,amssymb,amsfonts}%
\usepackage{amsthm}%
\usepackage{mathrsfs}%
\usepackage[title]{appendix}%
\usepackage{xcolor}%
\usepackage{textcomp}%
\usepackage{manyfoot}%
\usepackage{booktabs}%
\usepackage{physics}%
\usepackage{algorithm}%
\usepackage{algorithmicx}%
\usepackage{algpseudocode}%
\usepackage{listings}%

\usepackage{ulem}

\theoremstyle{thmstyleone}%
\theoremstyle{thmstyletwo}%

\theoremstyle{thmstylethree}%

\begin{document}

\title[Article Title]{Bennett Vorticity: A family of nonlinear Shear-Flow Stabilized Z-Pinch Equilibria}


\author*[1]{\fnm{Matt} \sur{Russell}}\email{russm66@uw.edu}

% \author[2]{\fnm{Second} \sur{Author}}\email{iiauthor@gmail.com}


% \affil*[1]{}

% \affil*[1]{\orgdiv{Computational Plasma Dynamics Lab}, \orgname{University of Washington}, \orgaddress{\city{Seattle}, \postcode{98105}, \state{WA}, \country{USA}}}

% \affil[2]{\orgdiv{Department}, \orgname{Organization}, \orgaddress{\street{Street}, \city{City}, \postcode{10587}, \state{State}, \country{Country}}}

% \affil[3]{\orgdiv{Department}, \orgname{Organization}, \orgaddress{\street{Street}, \city{City}, \postcode{610101}, \state{State}, \country{Country}}}


\abstract{Plasma equilibria are typically treated as arising from distinct mechanisms across different regimes. Here we demonstrate that a single analytic axial flow profile, obtained by exchanging the Bennett nonlinearity from density to flow, generates a family of shear-flow stabilized Z-pinch equilibria in which the properties are determined directly by the flow. This analytic profile reconstructs the axial velocity, and magnetic structure of shear-flow stabilized fusion plasma experiments, reproduces the spatial structure of emission intensity in the front, wake, and needletip structures of an air plasma streamer head, and a toroidal pre-ELM edge pedestal. Explorations of nanoscale observables illustrate both the reach and limitations of the ideal model, while the emergence of sawtooth structures when multiple of these profiles are chained together further supports its internal consistency. These results suggest a common shear-organized component across disparate regimes, with potential implications for both laboratory and natural plasmas.}

\keywords{Ideal MHD, Bennett Pinch, Shear-Flow Stabilized Z-Pinch, Nonlinear Science, Analytic}



\maketitle

\section*{\centering Introduction}\label{intro}
\qquad Plasma equilibria across widely separated regimes are typically described using distinct physical mechanisms. Here we show that a single analytic axial flow profile generates a family of shear-flow-stabilized Z-pinch equilibria\cite{uri_1995} from which density, current, magnetic field, and pressure follow directly. This same profile reconstructs experimental shear-flow stabilized Z-pinch structure\cite{uri_2001}\cite{uri_2009}, and captures streamer dynamics\cite{Li_2021}\cite{fridman2008plasma}, suggesting a common flow-generated origin of plasma structure across regimes.

\qquad The magnetohydrodynamic Z-pinch equilibrium\cite{Haines_2011} is one of the chief reasons why fusion scientists, buoyed by the success of nuclear fission, predicted they would have fusion harnessed by the end of the decade when the international community came together and declassified the field in the 50s. Unfortunately, the axisymmetric Z-pinch suffers from $m=0$ kink, and $m=1$ sausage instabilities\cite{Shafranov_1958}\cite{kruskal_1958} which limit the device lifetime unless stabilized. A means of stabilizing the Z-pinch in a manner viable for a fusion reactor\cite{uri1998}\cite{uri_2003}\cite{uri_2005}\cite{uri_2020} was discovered about a decade or so after researchers abandoned it in a heap of bitter frustration for toroidal approaches, instead. Namely, by the means of a sheared axial flow which satisfies the criterion,
\begin{equation}
    \label{eqn:shumlak_criterion}
    \dv{u_{z}}{r} > 0.1kV_{A} = \frac{\pi}{5L}\frac{B_{\theta}}{\sqrt{\rho\mu_{0}}}
\end{equation}
Otherwise the axial transport of magnetic energy will be sufficient to destabilize the pinch via virulent kink and sausage MHD modes.

\section*{\centering Theory}
\qquad A Z-pinch is the simplest kind of magnetohydrodynamic equilibrium\cite{FriedbergIMHD2014}\cite{Goedbloed_Keppens_Poedts_2019}, featuring only an axial current density which creates an azimuthal magnetic field that confines the flow. With an axisymmetric system where the flow properties only depend on the radial coordinate the equations are simple enough to integrate, leaving solutions which correspond to plasma flows that in reality will collapse on nanosecond timescales unless they are shear-flow stabilized.  

\qquad In the case of a high-$R_{m}$ flow\cite{DavidsonMHD2017}\cite{DavidsonTurbulence}, where the magnetic Reynolds number is of course,
\begin{equation}
    R_{m} = \mu u\sigma L    
\end{equation}
representing the relative impact of the magnetic forces against the inertial ones in a fluid of charged particles, then the governing equations are given by,
\begin{align}
    J_{z} &= -en(r)u_{z}(r) \\
    B_{\theta} &= \frac{\mu_{0}}{r}\int_{0}^{r}r'J_{z}(r')dr' \\
    \dv{p}{r} &= -J_{z}B_{\theta} 
\end{align}
The Bennett pinch is one interesting analytic solution to this equilibrium which considers a two temperature system that carries a uniform flow with a density of the form\cite{Bennett1934}\cite{Allen2018},
\begin{align}
    n(r) &= \frac{n_{0}}{(1 + \xi^{2}r^{2})^{2}} \label{eqn:bennett_density} \\
    \xi^{2} &= bn_{0} \\
    b &= \frac{\mu_{0}e^{2}u_{z,0}^{2}}{8k_{B}(T_{e} + T_{i})}
\end{align}
giving the plasma current density,
\begin{equation}
    J_{z}(r) = \frac{-en_{0}u_{z,0}}{(1 + \xi^{2}r^{2})^{2}}
\end{equation}
Let us choose to swap the profile in Equation (\ref{eqn:bennett_density}) from density to flow. The flow then will entirely generate the structure of the equilibrium. In order for the momentum balance to remain consistent with the physics of the equilibrium, the temperature now must become non-uniform to compensate for the uniform density. 

\qquad From the conditional in Equation (\ref{eqn:shumlak_criterion}) we can ask what is required of this non-uniform temperature in order for the flow profile,
\begin{equation}
    \label{eqn:pureflow_uz}
    u_{z}(r) = \frac{u_{z,0}}{(1 + \xi^{2}r^{2})^{2}}
\end{equation}
to be a shear-flow stabilized Z-pinch. Derivations of the weak and strong forms that answer this question are presented in the supplemental appendix, and merely quoted here for brevity. 

\qquad The weak form of the answer is,
\begin{equation}
    \label{eqn:weak_criterion}
    T(r) > r^{2}
\end{equation}
Assuming the temperature profile to be a power-law as suggested by the above,
\begin{equation}
    T(r) = C_{T}r^{n}
\end{equation}
then the strong form is merely that the power of the exponent satisfies,
\begin{equation}
    \label{eqn:strong_criterion}
    n > 2
\end{equation}
Inserting a cubic temperature into Equation (\ref{eqn:pureflow_uz}),
\begin{equation}
    \label{eqn:cubic_temperature}
    T(r) = \frac{T_{p}}{r_{p}^{3}}r^{3}
\end{equation}
we have after some algebra,
\begin{align}
    u_{z}(r) &= u_{z,0}\frac{r^{2}}{(r + C^{(3)}_{B,T})^{2}} \\
    C_{B,T}^{(3)} &= (1.45967*10^{-22})\frac{n_{0}u_{z,0}^{2}r_{p}^{3}}{T_{p}} \ [m]
\end{align}
This gives a current density, 
\begin{equation}
    J_{z} = -en_{0}u_{z,0}\frac{r^{2}}{(r + C_{B,T})^{2}}
\end{equation}
a magnetic field,
\begin{equation}
    B_{\theta}(r) = -\frac{\mu_{0}en_{0}u_{z,0}}{2r(r + C_{B,T})}f(r,C_{B,T})\hat{\theta} \label{eqn:btheta_Trcubed}
\end{equation}
and a plasma pressure whose exact form is left for an appendix due to the sake of brevity. The function in the magnetic field expression illustrates the logic behind this decision as it introduces natural logarithms into the kernel for the pressure,
\begin{equation}
    \label{eqn:f_Btheta}
    f(r,C_{B,T}) = f_{1} + f_{2} + f_{3} + f_{4} \\
\end{equation}
with,
\begin{align}
    f_{1}(r) &= r^{3} \\
    f_{2}(r, C_{B,T}) &= -3r^{2}C_{B,T} \\
    f_{3}(r, C_{B,T}) &= -6rC_{B,T}^{2}(1 + \ln(\frac{C_{B,T}}{r + C_{B,T}})) \\
    f_{4}(r, C_{B,T}) &= -6C_{B,T}^{3}\ln(\frac{C_{B,T}}{r + C_{B,T}})
\end{align}
\qquad Besides this "pure-flow" profile we can also add a uniform background flow to this equilibrium, as well as a sign to the shear, 
\begin{equation}
    \label{eqn:bulk_pureflow}
    u_{z}^{(2,\pm)} = u_{0}\pm u_{z,0}\frac{r^{2}}{(r + C_{B,T})^{2}}
\end{equation}
Because the current densities constructed by the above are analytic, then any combination of density and flow profile which produces one of these current densities creates exactly the same magnetic field and plasma pressure as is produced in the specific case under consideration, giving a self-similar characteristic to the system, as well as the existence of a mixed one to the fundamental character behind this result. We identify this aforementioned pureflow flow pattern with the situation when the nonlinearity is entirely distributed to the flow. Since, 
\begin{equation}
    \chi + \nu = 2
\end{equation}
is required if we identify these as the exponents which are distributed to the flow, and density, respectively, then $\chi = 2$ corresponds to the pureflow case. 

When the bulk flow is introduced, the magnetic field picks up a term proportional to the radius, but the pressure becomes complex, and involves a Junquiere function of second order. This is noteworthy because this function typically arises in the context of quantum statistics, and this is a classical equilibrium. 

\qquad The boundary conditions for the equilibrium are simple,
\begin{align}
    p(0) &= p_{0} \\
    p(r_{p}) &= 0 \\
    p_{0} &= \int_{0}^{r_{p}}J_{z}B_{\theta}dr \\
    u_{z}(r_{p}) &= u_{edge} = u_{z,0}\frac{r_{p}^{2}}{(r_{p} + C_{B,T})^{2}}
\end{align}
but the flow boundary condition expresses a quartic system whose roots give the value of the flow speed. If the parameter, $C_{B,T}$, is small compared to the pinch radius then these values become degenerate, and only the edge speed solution remains. However, in general there will be four roots to this system, and they may be complex-valued. For brevity, the investigation of this system is found in the supplemental appendix. 

\qquad The minimum pinch length that these vortices need to be shear-flow stabilized can be investigated, not just for the pureflow vortices discussed, but also for bulk vortices, and even higher-order vortices than cubic. However, the cubic vortex is the minimum energy state among this shear-flow stabilized family as the parabolic vortex has no shear, and the thermal power of an n-vortex is given by,
\begin{equation}
    S_{n} = -2\pi Ln\kappa T_{p}
\end{equation}
with the derivation left for the appendix.

\qquad The aforementioned lengths can be found straightforwardly from re-arranging the shear-flow criterion,
\begin{equation}
    L > \frac{\pi}{5}\bigg(\dv{u_{z}}{r}\bigg)^{-1}V_{A}
\end{equation}
Taking the limit of this expression as we approach the pinch axis, we find for the respective systems,
\begin{align}
    L^{(2)}_{3} &> 0 \\
    \tilde{L}^{(2,\pm)}_{3} &> -\frac{en_{0}\mu_{0}C_{B,T}^{2}u_{0}}{4u_{z,0}} \\
    \tilde{L}^{(2)}_{n} &> 0 \\
    \tilde{L}^{(2,\pm)}_{n} &> \lim_{r\rightarrow0} \frac{(r^{n} + C_{B,T}^{(n)}r^{2})^{3}}{r^{2n}}
\end{align}
The lengths for the vacuum forms are naturally zero, and the lengths for the bulk forms go to zero when the pinch radius does. The tildes represent normalizations which are described in the supplement alongside the derivation of these expressions. Evidently, within the idealized framework presented here these vortices are all shear-flow stabilized for an arbitrarily small space. We now test whether this analytically-generated equilibrium reproduces experimentally observed plasma structure.

\section*{\centering ZaP Reconstructions}
\qquad This analytic flow profile reconstructs the axial velocity profiles observed in the shear-flow stabilized Zap Z-pinch experiments from 2001\cite{uri_2001} and 2009\cite{uri_2009}, cited here again for reference. Either a single form of vortex can be solved for across an entire half-chord, or the solution can be localized in each cell of the experimental data using the associated speeds as boundary conditions. In doing this, the edge temperature of the system is chosen to coincide with the core temperature of the plasma. If this approach fails due to the natural limitations of the MHD model here, e.g., neglecting viscosity, kinetic effects, relativistic physics, etc. then the parameter $C_{B,T}$, which represents the placement of the shear layer, can be tuned so that the flow rises to the edge state when it needs to. This reflects the inherent self-similarity of the model based on a physically-bound parameter, rather than an unconstrained curve-fitting exercise using a free value. Unless noted otherwise, the magnitude of the flow root is taken rather than just the real part as complex values influence the location of the shear layer. Consequently, the accuracy of these solutions will suffer greatly as the shear layer will be located beyond the pinch radius due to the largeness of the flow magnitude from these imaginary velocities. They arise due to the structure of the quartic equations which must be solved when obtaining the flow roots.

\qquad To evaluate the closeness of fit, the relative rate of mean squared error is calculated between the analytic solution, and linear interpolant, for the roots of each segment before these values are averaged respectively,
\begin{equation}
    \overline{RRMSE} = \frac{1}{N_{V}}\sum_{i=0}^{N_{V}-1}\frac{\sqrt{MSE_{i}}}{mean(u_{lin,i})}
\end{equation}
A cubic, pure-flow vortex will have four roots per problem unless $u_{0} = u_{edge}$, in which case there will be no roots because the problem is made trivial. Tables (\ref{tab:rrmse_zap2001}) - (\ref{tab:rrmse_swtc}) list the experimental reconstructions made, and the closeness of fit between the analytic solution expressed by a chain of given roots, and the experimental sawtooth data. 
\begin{table}[ht!]
    \centering
    \begin{tabular}{c|c|c|c|c|c|c}
         Experiment & HC & $T_{p}$ (keV) & Root 1 (\%) & Root 2 (\%) & Root 3 (\%) & Root 4 (\%) \\
         Zap 2001 (FS) & + & 0.2 & 21.086 & 21.086 & 21.483 & 21.483 \\
         Zap 2001 (FS) & + & 2 & 23.211 & 23.211 & 33.221 & 33.221 \\
         Zap 2001 (FS) & + & 10 & 26.606 & 26.606 & 44.046 & 50.305  \\
         Zap 2001 & + & 0.2 & 8.766 & 8.766 & 5.349 & 5.349 \\
         Zap 2001 & + & 2 & 4.753 & 4.753 & 13.057 & 13.057 \\
         Zap 2001 & + & 10 & 1.565 & 1.565 & 18.249 & 37.433 \\
         Zap 2001 (ELV) & + & 0.2 & 13.761 & 13.761 & 6.789 & 22.871 \\
         Zap 2001 (FS) & - & 0.2 & 28.092 & 28.092 & 24.822 & 24.822 \\
         Zap 2001 (FS) & - & 2 & 22.729 & 22.729 & 19.412 & 19.412 \\
         Zap 2001 (FS) & - & 10 & 17.901 & 17.901 & 20.548 & 44.886 \\
         Zap 2001 & - & 0.2 & 22.807 & 22.807 & 19.931 & 19.931  \\
         Zap 2001 & - & 2 & 18.629 & 18.629 & 20.992 & 20.992  \\
         Zap 2001 & - & 10 & 15.284 & 15.284 & 23.051 & 50.271  \\
         Zap 2001 (ELV) & - & 0.2 & 17.812 & 17.812 & 8.463 & 14.599 \\
    \end{tabular}
    \caption{Closeness of fit between the experimental shear-flow stabilized Z-pinch velocity data, and analytic single-vortex reconstructions for the positive (+) and negative (-) half-chords (HC) based on positive and negative bulk, pureflow vortices, respectively. The closeness is compared both on the full span, and when data points near the edge of the pinch are neglected with this latter category indicated by the absence of (FS) in the 'Experiment' field. Reconstructions that use an edge-localized vortex (ELV) to improve the fit at the shear layer are marked accordingly.}
    \label{tab:rrmse_zap2001}
\end{table}

\begin{figure}[ht!]
    \centering
    \includegraphics[width=\linewidth]{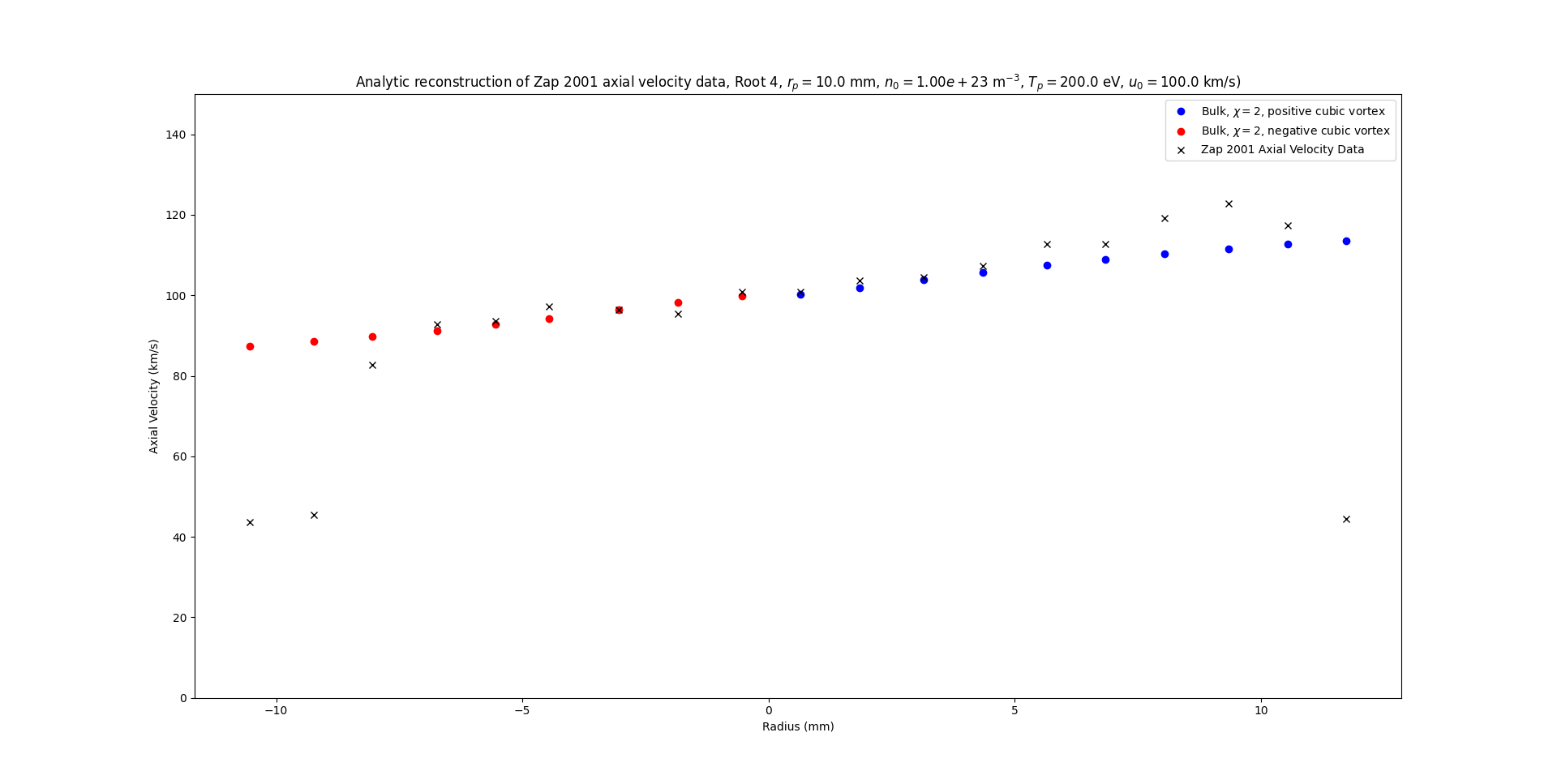}
    \caption{Analytic, single-vortex reconstruction of the Zap 2001 experimental velocity profile based on the most accurate solution. Evidently, the shear of this solution naturally captures subtleties in the shear of the experimental profile, however, the location of the shear layer is too large in the analytic solution which is what leads to the large discrepancies seen in Table \ref{tab:rrmse_zap2001}.}
    \label{fig:zap2001_Tp200eV_root4}
\end{figure}

\begin{figure}[ht!]
    \centering
    \includegraphics[width=\linewidth]{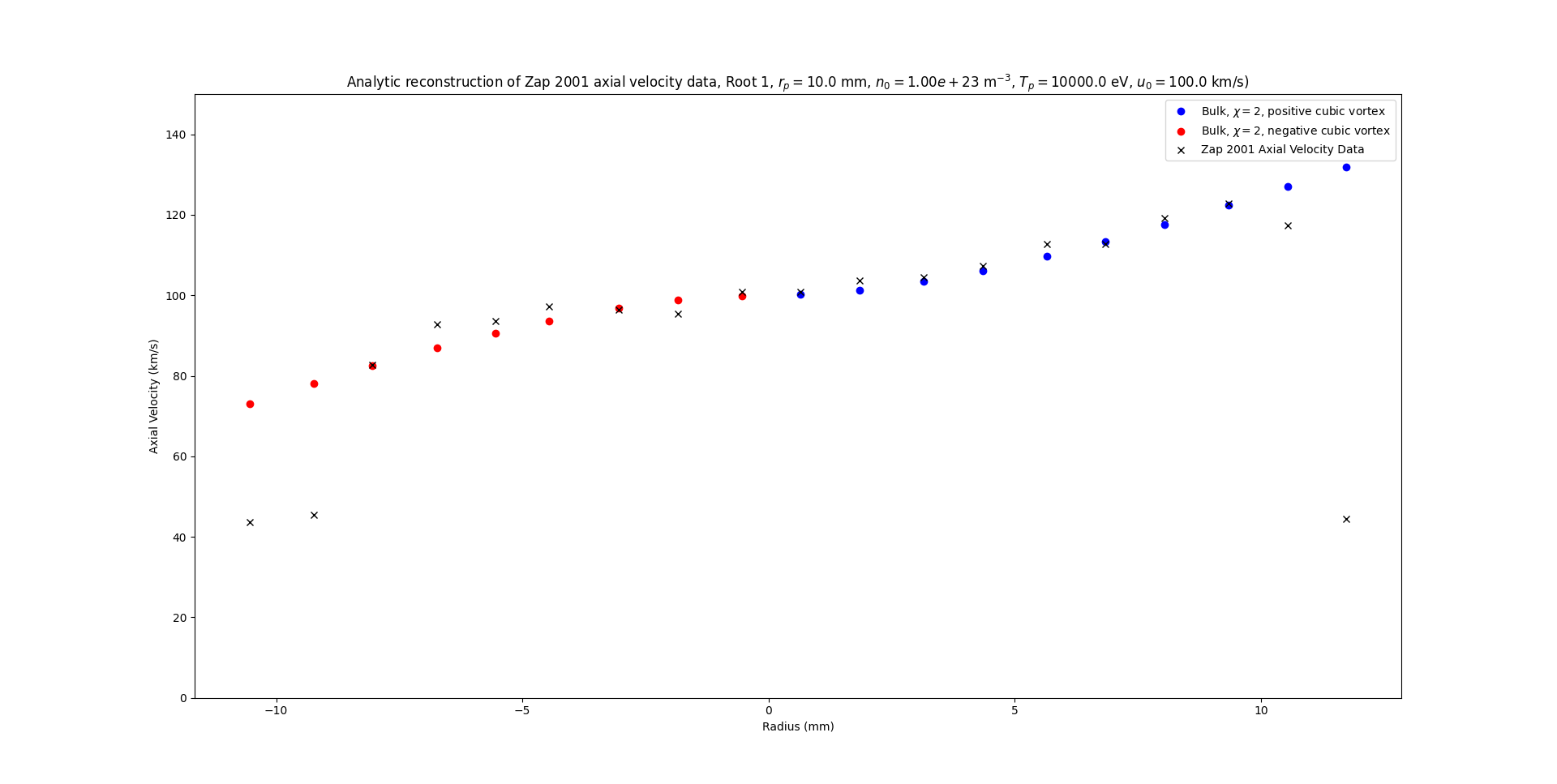}
    \caption{Analytic, single-vortex reconstruction of the Zap 2001 experimental velocity profile based on altering the plasma properties in order to tune the location of the shear layer. This process improves the accuracy in the negative half-chord, but degrades it in the positive half-chord because the positive vortex being fit there does not model the negative shear seen in the experiment. }
    \label{fig:zap2001_Tp10keV_root1}
\end{figure}

\begin{figure}[ht!]
    \centering
    \includegraphics[width=\linewidth]{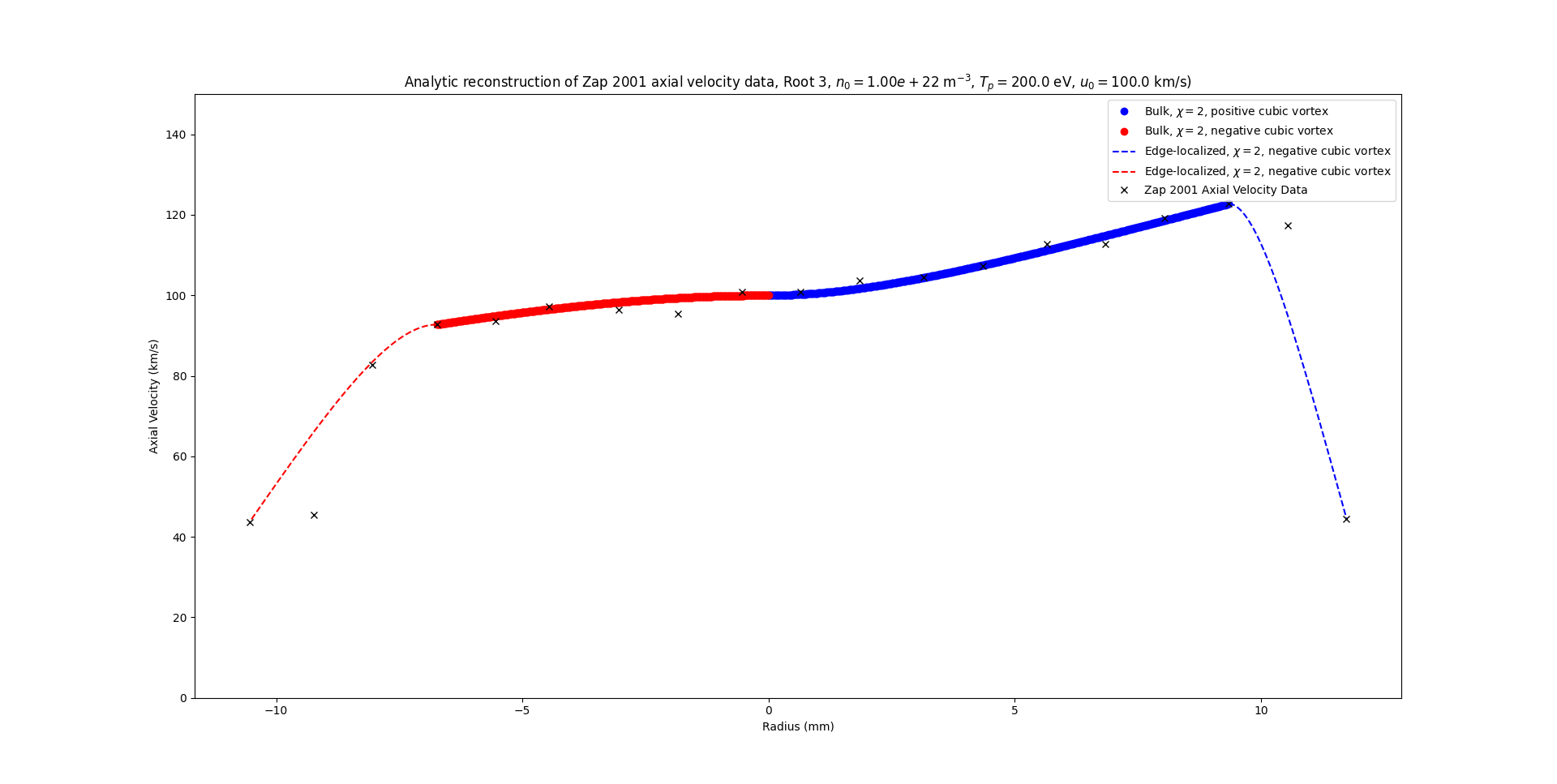}
    \caption{Edge-localized vortex (ELV) reconstruction of the Zap 2001 experimental velocity profile based on grids with $N_{r} = 500$ points for each vortex. Just like with the sawteeth reconstructions of the 2009 data the third branch of solutions is the most accurate, suggesting nature has a preferred path. }
    \label{fig:placeholder}
\end{figure}

\begin{table}[ht!]
    \centering
    \begin{tabular}{c|c|c|c|c|c}
         Experiment & Shot & Root 1 (\%) & Root 2 (\%) & Root 3 (\%) & Root 4 (\%) \\
         Zap 2009 & $\tau$ = 29.6 $\mu s$ & 2.463 & 2.463 & 1.414 & 4.926 \\
         Zap 2009 & $\tau$ = 38.4 $\mu s$ & 4.682 & 4.683 & 3.278 & 11.104 \\
         Zap 2009 & $\tau$ = 41.4 $\mu s$ & 7.336 & 7.336 & 4.577 & 15.814 \\
         Zap 2009 & $\tau$ = 48.96 $\mu s$ & 4.655 & 4.655 & 3.427 & 11.464 \\
         Zap 2009 & $\tau$ = 58.64 $\mu s$ & 10.369 & 10.369 & 5.919 & 20.613 \\       
    \end{tabular}
    \caption{Closeness of fit between the experimental shear-flow stabilized Z-pinch velocity data, and the analytic reconstructions. The average RRMSE of each reconstruction is calculated from the RRMSEs of each individual segment for each root to give these numbers.}
    \label{tab:rrmse_swtc}
\end{table}

\begin{figure}[ht!]
    \centering
    \includegraphics[width=\linewidth]{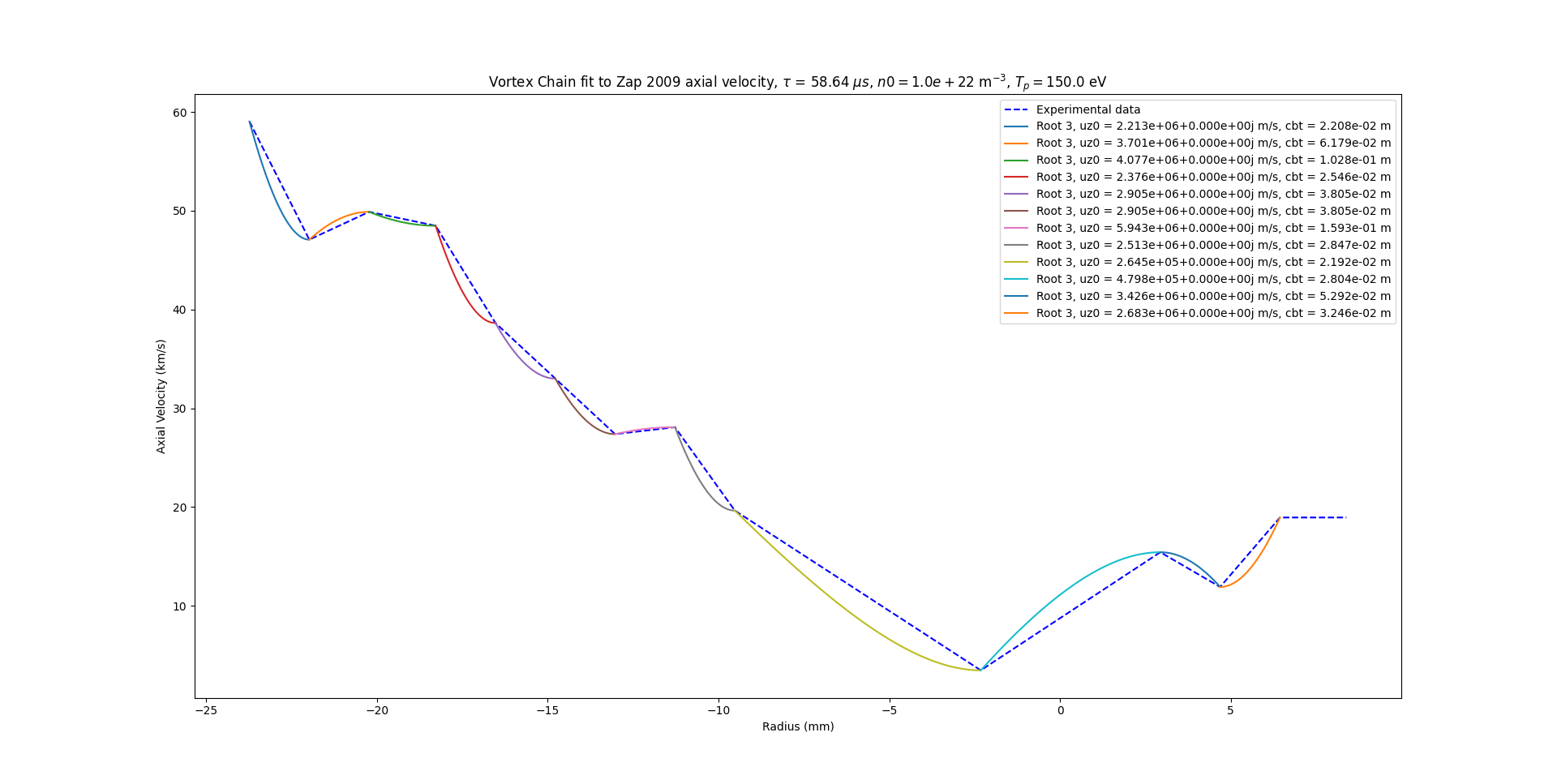}
    \caption{Analytic reconstruction using the third solution branch for representative Zap 2009 data. The model captures the global structure measured by the experiment, and suggests a sub-grid structure to the flow.}
    \label{fig:zap2009_fits_tau0pt56}
\end{figure}

\begin{figure}[ht!]
    \centering
    \includegraphics[width=\linewidth]{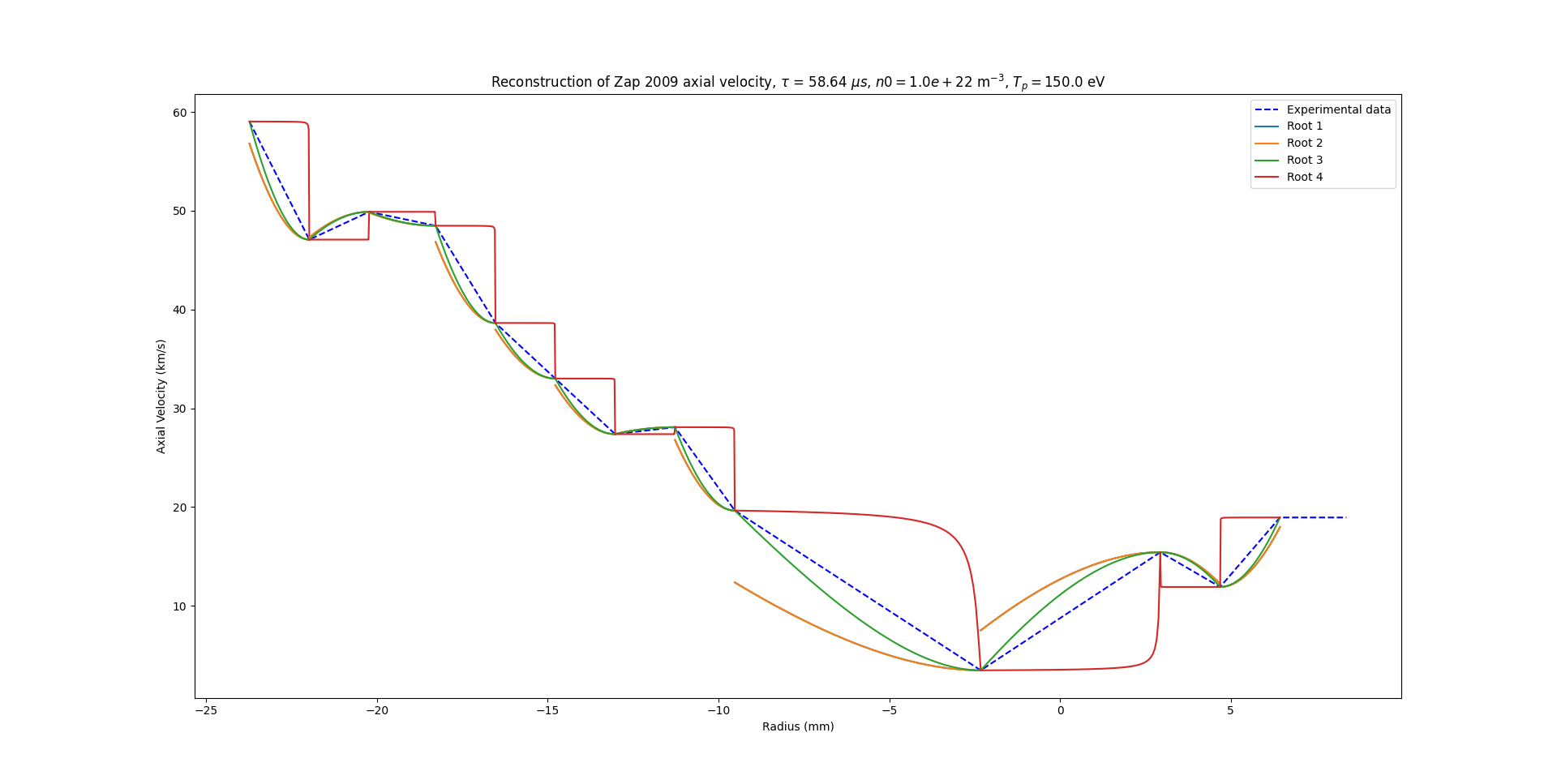}
    \caption{Analytic reconstructions with all solution branches for the representative Zap 2009 data as in Figure (\ref{fig:zap2009_fits_tau0pt56}). The first and second solution branches are identical, and possess roots which are complex conjugates of each other. The third branch naturally gives the best fit, indicating that nature has a preferred selection. This idea is supported by the data in Table (\ref{tab:rrmse_swtc}), and these features are repeated throughout the sawtooth reconstructions.}
    \label{fig:zap2009_fits_tau0pt56_ALL}
\end{figure}
% \qquad By solving for this analytic profile, the axial velocity profiles of the various shear-flow stabilized Zap Z-pinch experiments can be fit to the model. This fitting can be done in one of two ways, either a single form of vortex can be solved for, or pairs of the axial velocity data can be used as boundary conditions for a vortex which is localized inside that particular cell. When the fit is not perfect for a single vortex, $C_{B,T}$ can be treated as a control parameter in order to produce a more accurate single-form fit either by elevating the temperature to decrease the parameter, or elevating the density to increase it. 

\newpage

\section*{\centering Streamer Validation}
\qquad This model also reconstructs the intensity of emitted light in the head of an air plasma streamer\cite{Li_2021}. A linear model connects the plasma current density with the emission of photons from the plasma, $I$,
\begin{equation}
    I = \alpha J_{z}
\end{equation}
which is justified by assuming that the primary source of photonic emission in a plasma discharge are electron-impact collisions\cite{Raizer1991GasDP}, so that the local structure of the emission is directly related to the local electron plasma current. 

\begin{figure}
    \centering
    \includegraphics[width=\linewidth]{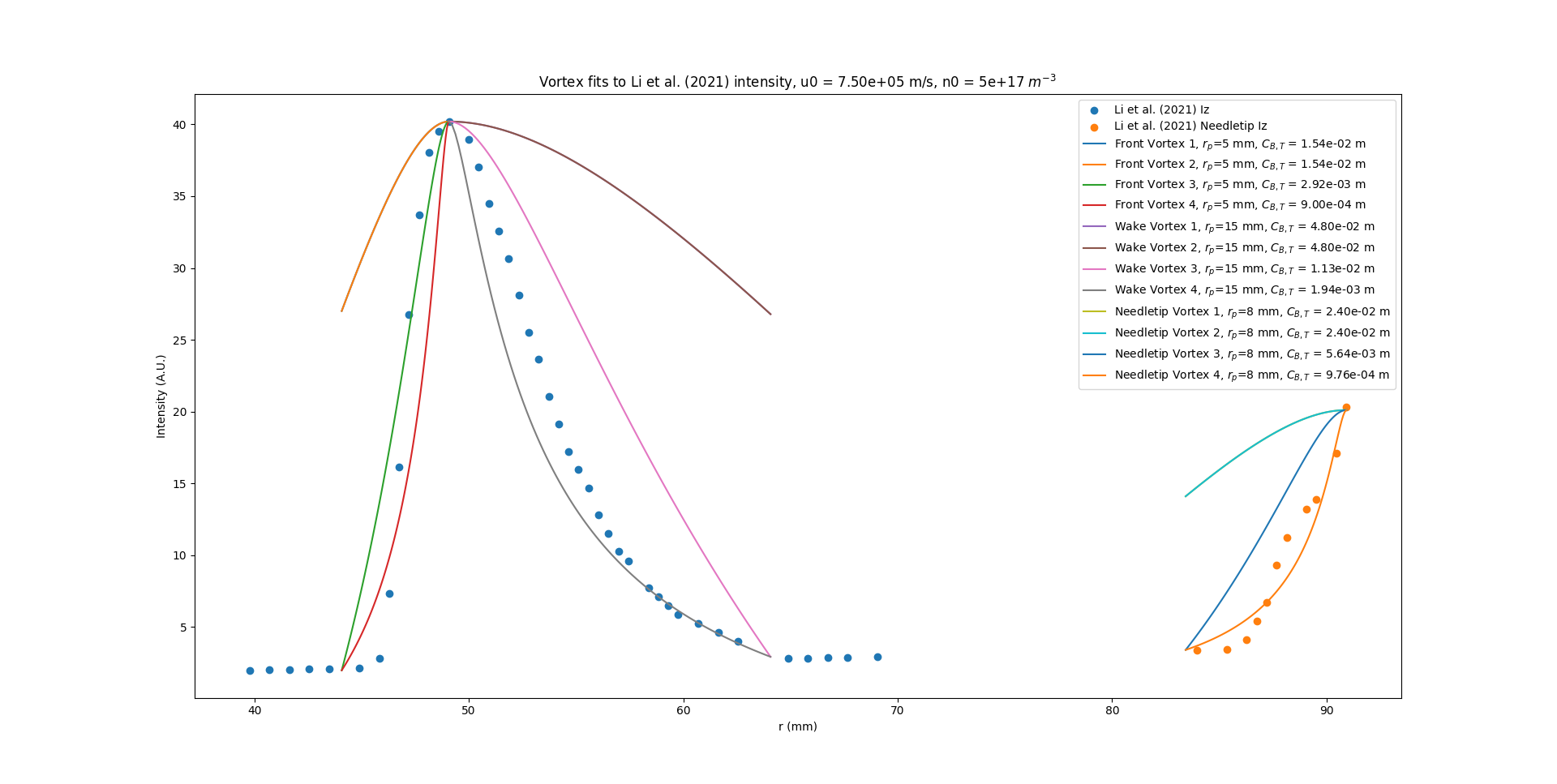}
    \caption{Analytic single-vortex solutions to the axial intensity of an air plasma streamer viewed in the $yz$-plane with $z$ interpreted as a plasma radius. The streamer front shows the same pattern as observed in many of the Zap reconstructions, namely that the third branch gives the most accurate reconstruction. However, in the wake and needletip it is the fourth branch which is the most accurate. The first and second remain complex conjugates across all three structures. The needletip can be modelled as a single vortex, but it actually appears to be comprised of two as there is a distinct form caused by a second shear layer following the rise to the edge of the first.}
    \label{fig:li2021_Izfits_all}
\end{figure}

\qquad The intensity falls off very rapidly in the streamer head so a range normalization is used, instead of the prior mean normalization, in the calculation of the relative mean squared error to avoid introducing an extraneous penalty. Visually, solutions to the wake, front, and needletip are presented in Figure (\ref{fig:li2021_Izfits_all}). Large $C_{B,T}$ values associated with complex conjugate roots cause large deviations in some of the solutions, but what is noteworthy about this reconstruction is that it represents the axial current density as a function of axial coordinate which is interpreted as a pinch radius for the purposes of fitting this ideal MHD model to the plasma. Table (\ref{tab:li2021_table}) presents the data.

\qquad Why this is exceptionally noteworthy is because the assumptions behind this equilibrium include that of an axial symmetry. However, here we see no axial symmetry, and the flow pattern still makes an accurate fit when the shear layer is suitably located to match the experimental shear. This suggests that the requirement of axial symmetry can be relaxed, and supports the surprising conclusion that axial symmetry is not required for this equilibrium even though that is a fundamental mathematical assumption leading up to it. 

\qquad The choice of edge plasma temperature in this is taken arbitrarily, and for a given $C_{B,T}$ that represents an accurate solution, the temperature can be tuned by varying, e.g., $n_{0}$. For example, in the wake and needletip if the density was reduced by a factor of 5, then the temperature could be as well, and the same $C_{B,T}$ would remain. With the flow constant, and pinch radius being untouched the same solution would appear, albeit with the more reasonable $T_{p} = 20000 \ ^{o}K$ instead.  

\begin{table}[]
    \centering
    \begin{tabular}{c|c|c|c|c|c}
         Structure & $T_{p}$ ($^{o}$K) & Root 1 (\%) & Root 2 (\%) & Root 3 (\%) & Root 4 (\%) \\
         Front & 10000 & 48.38 & 48.38 & 14.94 & 19.56 \\
         Wake & 100000 & 54.87 & 54.87 & 21.34 & 10.22  \\
         Needletip & 100000 & 55.81 & 55.81 & 23.17 & 8.46  
    \end{tabular}
    \caption{Accuracies for the analytic solutions to the air plasma streamer head structures. The mean normalization used in the Zap reconstructions is exchanged for a range normalization as the relative change in intensity is very large over a very small span so this high variance will cause the mean normalization to introduce an extraneous penalty into the accuracy. The large deviation in some of the roots results from the shear layer being located at too large a value so that the analytic vortex does not reach the edge state. This coincides with the complex conjugate roots. The large temperatures attained at the wake and needletip can be lowered by lowering $n_{0}$ while keeping the same $C_{B,T}$, and $r_{p}$ so that the solutions are the same.}
    \label{tab:li2021_table}
\end{table}

\newpage

\section*{MAST Edge Pedestal pre-ELM}
\qquad Tokamaks operating in H-mode develop an edge confinement barrier at sufficiently high heating power, and this phenomenon is associated with the development of enhanced thermodynamic gradients\cite{Wagner1982,JWConnor_2000,Groebner1998} and shear layers\cite{Leonard2014}. These barriers periodically relax through edge-localized modes (ELMs)\cite{HZohm_1996}, which are associated with the large edge pressure and current gradients that develop in the pedestal before reforming under continued plasma heating. This cyclic behavior is qualitatively suggestive of a localized shear structure which experiences collapse after exhaustion of the sustaining edge gradient before reforming once the source of free energy is restored.

\begin{figure}
    \centering
    \includegraphics[width=\linewidth]{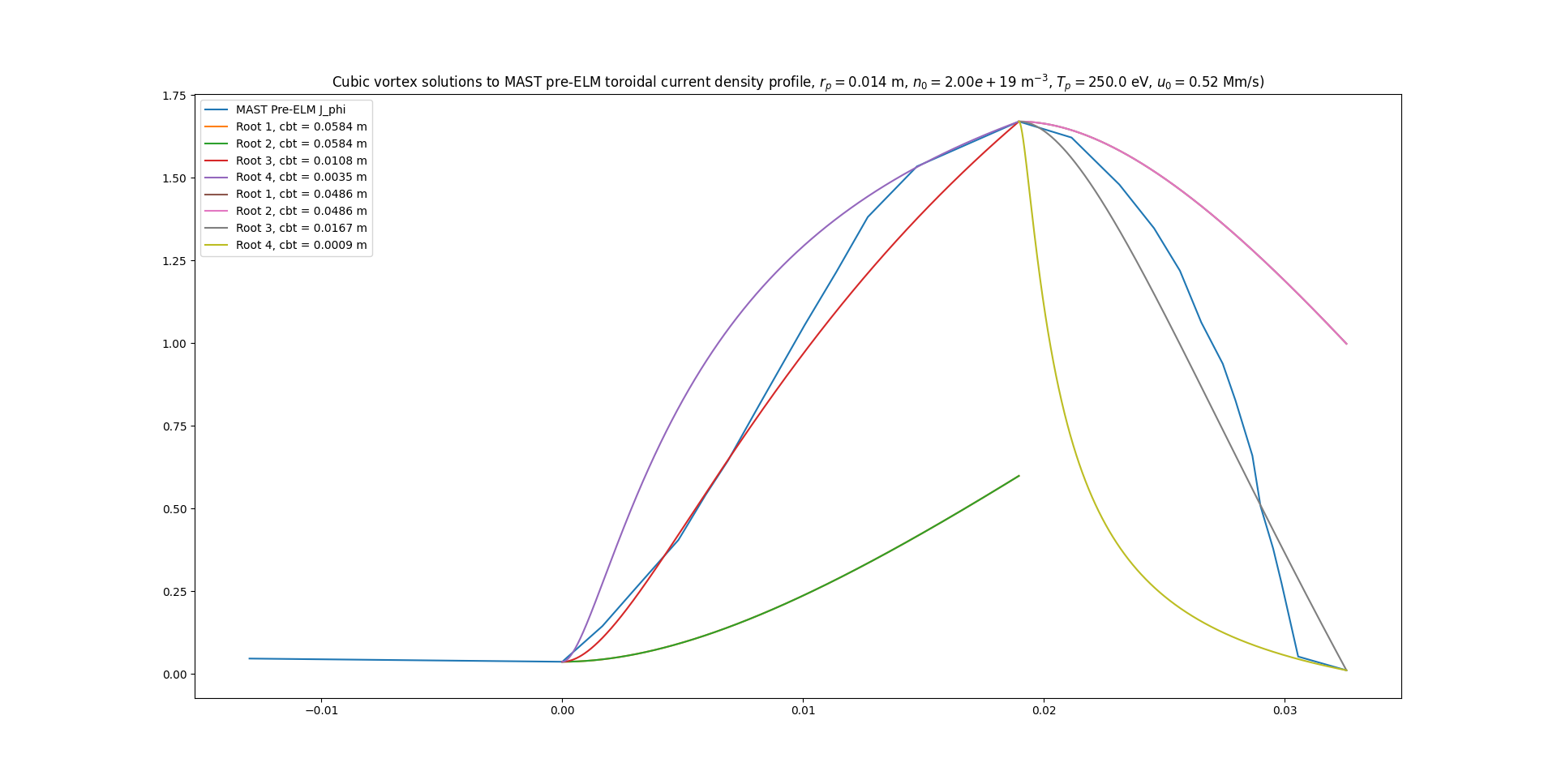}
    \caption{Analytic cubic vortex solutions to the MAST pedestal pre-ELM for experimental plasma conditions.}
    \label{fig:mast2023_preELM_Tp250eV_n02e19}
\end{figure}

\begin{figure}
    \centering
    \includegraphics[width=\linewidth]{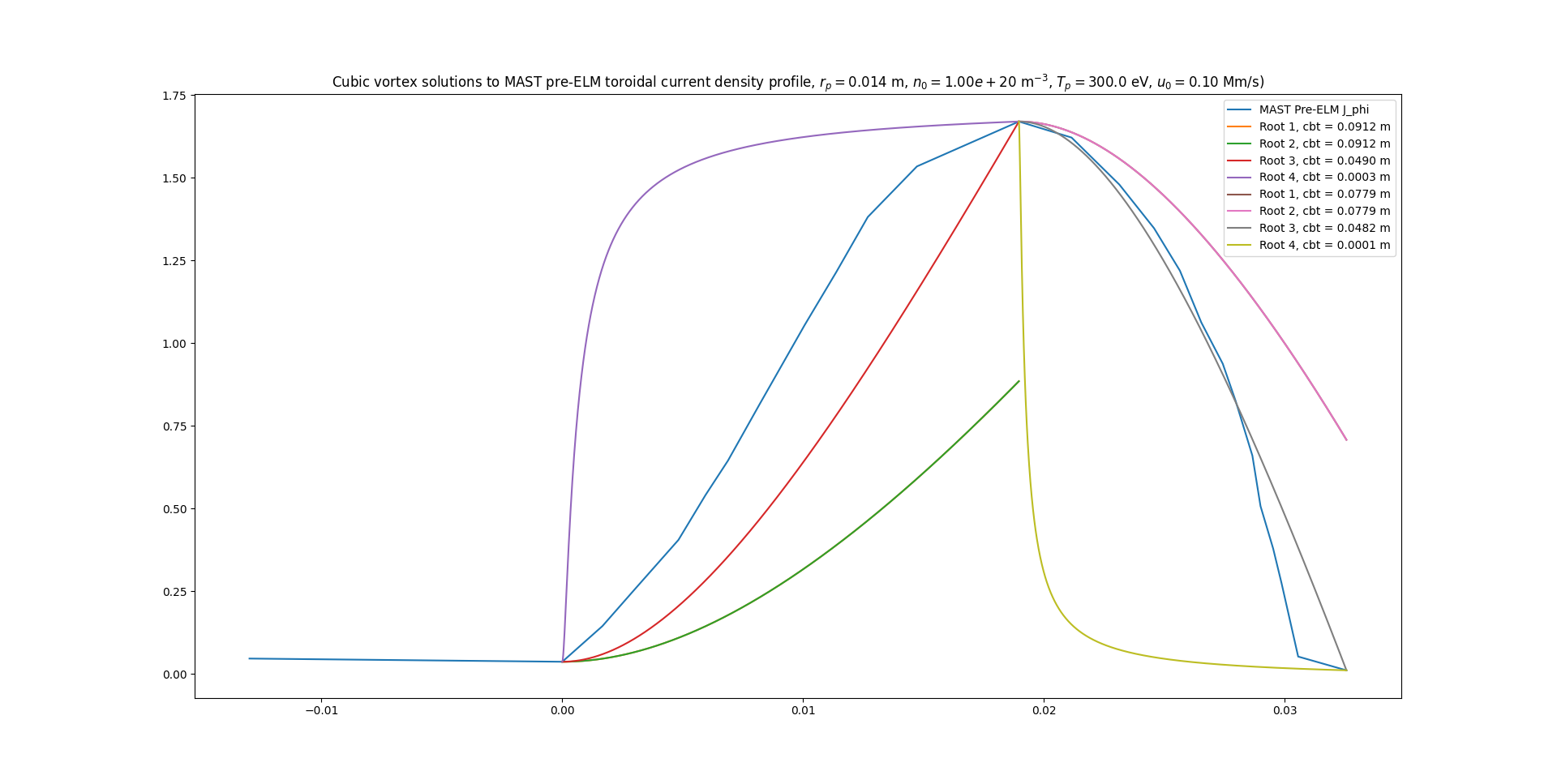}
    \caption{Cubic vortex solution to MAST pedestal pre-ELM for a second set experimental plasma conditions.}
    \label{fig:mast2023_preELM_Tp300eV_n01e20}
    
\end{figure}

\qquad Figures (\ref{fig:mast2023_preELM_Tp250eV_n02e19}) and (\ref{fig:mast2023_preELM_Tp300eV_n01e20}) show analytic solutions made from experimental data of the MAST edge pedestal\cite{Groebner_2023}. Tables (\ref{tab:mast2023_preELM_Tp250eV_n02e19}) and (\ref{tab:mast2023_preELM_Tp300eV_n01e20}) show the range-normalized relative root mean square error for the solutions. The front is taken to be the portion occurring after the peak plasma current density, and the wake is taken to be the portion occurring before. A grid of $Nr=1000$ points was taken as the basis for the solution. The third root consistently produces the lowest reconstruction error across the cases considered.

\qquad The asymmetry in reconstruction quality between the front and wake suggests that a single global equilibrium may be insufficient to fully capture the pedestal structure, supporting a multi-scale or matched-locality interpretation of the edge dynamics.

\begin{table}[]
    \centering
    \begin{tabular}{c|c|c|c|c}
        Region & Root & $u_{z,0}$ (m/s) & $C_{B,T}$ (m) & RRMSE-RN (\%) \\
        Front & 1 & -1.577e6 + 4.086e6j & 0.0486 & 31.4 \\
        Front & 2 & -1.577e6 - 4.086e6j & 0.0486 & 31.4 \\
        Front & 3 & 2.567e6 + 0j & 0.0167 & 22.02 \\
        Front & 4 & 5.86e5 + 0j & 0.00087 & 115.74 \\
        Wake & 1 & -9.848e5+2.746e6j & 0.0584 & 42.604 \\
        Wake & 2 & -9.848e5-2.746e6j & 0.0584 & 42.604 \\
        Wake & 3 & 1.253e6+0.j & 0.0108 & 5.146  \\         
        Wake & 4 & 7.165e6+0.j & 0.00352 & 19.7 
    \end{tabular}
    \caption{Accuracies for the analytic solutions to the Bennett-Shumlak cubic vortex model fit to the front and wake of the toroidal plasma current density of a pre-ELM edge pedestal for the experimental conditions in Figure (\ref{fig:mast2023_preELM_Tp250eV_n02e19}). Very small $C_{B,T}$ vortices lead to large reconstruction error by shear localization failure.}
    \label{tab:mast2023_preELM_Tp250eV_n02e19}
\end{table}

\begin{table}[]
    \centering
    \begin{tabular}{c|c|c|c|c}
         Region & Root & $u_{z,0}$ (m/s) & $C_{B,T}$ (m) & RRMSE-RN (\%) \\
         Front & 1 & -1.121e6 + 2.475e6j & 0.0779 & 23.7 \\
         Front & 2 & -1.121e6 - 2.475e6j & 0.0779 & 23.7 \\
         Front & 3 & 2.137e6 + 0j & 0.0482 & 7.905 \\
         Front & 4 & 1.053e5 + 0j &  0.000117 & 145.8 \\
         Wake & 1 & -7.073e5 +1.64e6j & 0.0912 & 38.3 \\
         Wake & 2 & -7.073e5 -1.64e6j &  0.0912 & 38.3 \\
        Wake & 3 & 1.309e6 +0.j & 0.0489 & 21.3 \\          
        Wake & 4 & 1.054e5 +0.j & 0.0003 & 58.8        
    \end{tabular}
    \caption{Accuracies for the analytic solutions to the Bennett-Shumlak cubic vortex model fit to the front and wake of the toroidal plasma current density of a pre-ELM edge pedestal for the experimental conditions in Figure (\ref{fig:mast2023_preELM_Tp300eV_n01e20}). Very small $C_{B,T}$ vortices lead to large reconstruction error by shear localization failure. }
    \label{tab:mast2023_preELM_Tp300eV_n01e20}
\end{table}

\section*{\centering Applications}
\qquad Because the equilibrium is entirely generated by the axial flow profile, and remains valid at arbitrarily small scales under shear-flow stabilization, this suggests that its applicability is not restricted to a specific device or regime. The scale-free nature of this equilibrium suggests broader relevance across plasma systems, from laboratory discharges to astrophysical filaments, and its inherent structure provides several novel avenues where it can naturally be employed in a critical technological capacity beyond that of the context of magnetic fusion energy science for baseband power generation.

\qquad The list presented here is by no means claimed to be exhaustive, as this phenomena is tied to the mathematical theory of non-relativistic high-$R_{m}$ MHD flows, which fusion plasmas behave similar to because they are fully-ionized, and other systems also could behave similar to, to varying degrees. Fusion plasmas also require an understanding of the relativistic theory of the equilibrium as the experimental data suggest there are significant relativistic populations\cite{uri_2017}. The dynamics of the fast compression process at play is dominated by the growth of strong electric fields, and the presence of these is a common feature across the fusion, air plasma streamer head, and edge pedestal experiments although the structure differs exactly. 

\subsection*{Plasma Filament Formation}
\qquad For example, we can posit that this equilibrium provides a candidate mechanism for filament formation at all scales, including galactic ones. Ideal magnetohydrodynamic systems are known to possess similar evolutions if they are geometrically similar in structure despite describing different scales\cite{ryutov2000}. Astrophysical jets are known to emerge from a wide variety of systems, but stable and narrow collimation of the flow can result from the presence of an azimuthal magnetic field\cite{albertazzi2014laboratory} after emergence from a standing shock.

\qquad Treating a test plasma mass as fixed, we can calculate the relative impact of electromagnetic forces against gravitational forces in the pinch,
\begin{equation}
    \eta_{GL} = \frac{\int_{V}|\vec{F}_{G}(r)|dV}{\int_{V}|\vec{F}_{L,lab}(r)|dV} = \frac{I_{G}}{I_{EM}} \label{eqn:eta_GL}
\end{equation}
In order to lift the singularity in the denominator that comes from Ohm's Law in this idealized treatment of the plasma,
\begin{equation}
    \vec{E} = -\vec{u}\times\vec{B}
\end{equation}
we must boost to a frame that is travelling at a velocity, $\vec{v}_{boost} = -2\vec{u}$, relative to the plasma. This calculation is involved, but can be carried out. The result is we find,
\begin{equation}
    \eta_{GL} \sim \frac{L}{r_{p}}
\end{equation}
For a relativistic flow in a flat spacetime the pinch length contracts\cite{jackson1998classical} in the flow frame to a new length, $L'$, 
\begin{equation}
    L' = \frac{1}{\gamma(v)}L = L\sqrt{1 - \frac{v^{2}}{c^{2}}}
\end{equation}
In the ultra-relativistic limit, $v \approx c$, this length contracts strongly, and particles may effectively traverse the pinch on timescales short compared to the MHD timescale. This suggests a possible mechanism contributing to the filamentary structure formation, namely, high-energy electrons propagating on the fastest timescale in a plasma. 

\qquad The proper time of the plasma particles also dilates\cite{goldstein2002classical} relative to the time kept in the lab frame,
\begin{equation}
    t = \gamma(v)\tau = \frac{\tau}{\sqrt{1 - \frac{v^{2}}{c^{2}}}}
\end{equation}
where $t$ is the lab time, and $\tau$ is the time interval measured in a frame at rest with respect to the plasma current. If a fast electron population were to establish such an equilibrium, then this may drive the collective motion of the surrounding plasma for the duration of the thermal lifetime. This lifetime will also be increased in the lab frame relative to the plasma frame for relativistic plasmas. 

\subsection*{High-Thrust Fusion Propulsion}
\qquad If a fusing vortex plasma were to split into $N$ distinct vortices, analogous to the branching structures observed in red sprite and other mesospheric discharge phenomena\cite{pasko2007red}\cite{pasko2010recent}, then conservation of energy suggests a possible mechanism for a high-thrust fusion jet,
\begin{equation}
    T = \dot{m}u_{e}
\end{equation}
When the core plasma splits into a number of $N$ distinct heads, this causes a redistribution of energy amongst the exhaust vortices which define the exhaust plane so that an adiabatic treatment of the system at the plasma edge yields,
\begin{equation}
    \frac{1}{2}mu_{z,0}^{2} + \frac{3}{2}k_{B}T_{p} = \frac{N}{2}mu_{e}^{2} + \frac{3}{2}k_{B}T_{e}
\end{equation}
If each exhaust vortex remains at the same edge temperature as the core vortex, then $T_{e} = NT_{p}$, and
\begin{equation}
    u_{e} =   \sqrt{u_{z,0}^{2} - \frac{3k_{B}T_{p}(N-1)}{m}}
\end{equation}
If the total thermal energy in the plasma remains fixed instead, then,
\begin{equation}
    u_{e} = \frac{u_{z,0}}{\sqrt{N}}
\end{equation}
In both cases, a lowered exhaust velocity is obtained. Then we can see from the fusion jet thrust,
\begin{equation}
    T = \frac{2\eta_{J}P_{fusion}}{u_{e}}
\end{equation}
that for a power-limited thruster this will result in an elevated thrust at the expense of the need for a large mass flow rate. 

\qquad A central challenge is how to get this action to produce the maximum number of exhaust vortices. The exhaust structure can be imagined as a spatially periodic plasma lattice, and these exhaust vortices are conceptually analogous to the streamer heads propagating forward in streamer discharges, but driven here by fusion energy rather than atmospheric discharge physics. This morphology further connects the equilibrium theory of the core pinch to the dynamic branching regime in which individual exhaust pinches evolve independently. 

% \qquad This is also of interest to study in the context of contemporary plasma rifle inventions, e.g., the American inventor Carlos Gaines, who uses a triaxial electrode design to generate a kind of branching plasma structure based on the tazing of a core chemical jet. In observations of these tests the onset of recoil, i.e., thrust, is coincident with the expulsion of a large amount of mass from the core that blooms into a large plume.

\qquad Similar branching plume morphologies have also been observed in pulsed plasma discharge devices employing multi-electrode geometries, e.g., the exploratory engineering work of the American inventor Carlos Gaines\cite{carlosgaines}, where the onset of recoil coincides with the rapid expulsion of ionized mass into an expanding plume structure. While highly preliminary, such observations qualitatively suggest that these branching plasma exhaust structures may be engineered to arise from a single core jet.

\qquad The scaling is also particularly poor here in the ideal case, which is a definite motivation to study the intrinsic physics of the pinch process further in order to continue to better understand the exact dynamics of the sprite transition process. Lastly, stagnation of the exhaust vortices is undesirable as it will result in the collapse of the exhaust structure.  

\qquad The adiabatic structure of this system suggests application to both fusion propulsion and magnetic fusion energy systems more broadly. The MW-scale nature of this specific concept challenges existing research that argues fusion propulsion is unfeasible for a manned mission to Mars\cite{Petkow2009} as well as extends existing discussions on the subject of fusion-powered shear-flow stabilized spaceflight\cite{uri2000}\cite{uri2003_rocket}. These earlier proposals note the potential that this approach has to bring both high specific impulse and high thrust levels to a rocket technology. 

\qquad The proposed concept also improves on the power budget of this earlier research which requires TW-scale fusion reactors, and very high power efficiencies, in order to provide high thrust alongside fast exhaust velocity. Instead, it focuses on redistribution of the exhaust energy through structured branching so that high thrust can be achieved with MW-scale reactors at the expense of a large mass flow rate.   

\subsection*{GHG Thermal Plasma Chemistry - Waterfall Reactors}
\qquad By aligning the pinch axis with a uniform local gravitational field, such as found on the surface of the Earth, strong radial plasma drifts will result because this gravitationally induced radial "waterfall drift" scales inversely with cyclotron frequency\cite{chen},
\begin{equation}
    \vec{v}_{wf} =  \frac{g_{0}}{\omega_{cj}}\hat{r}
\end{equation}
These drifts may be used in a thermal plasma reactor based on a pureflow Bennett vortex to deposit ionized greenhouse gases outside of the pinch volume as the interior of these structures possesses a naturally small magnetic field amplitude\cite{Crews_2025_torbit}
\begin{equation}
    \omega_{cj} = \frac{|q_{j}|}{m_{j}}B
\end{equation}
Collisions interrupt the picture of this waterfall drift transporting ionized pollutants out of a shear-flow stabilized Z-pinch plasma in order to deposit them on a material surface for sequestration. If the collisionality in the plasma is low, then the transport is ballistic, and assuming perfect sticking, i.e., no reflections, the rate of mass being deposited out of a layer of the plasma is taken to be,
\begin{align}
    \dot{M}_{s}(r) &= 2\pi rLm_{s}\Gamma_{wf}(r) \\
    &\Gamma_{wf}(r) = n(r)v_{wf}(r)
\end{align}
% if the plasma current density is that of a pureflow Bennett vortex.

\qquad When the collisionality in the plasma is high, then the transport becomes diffusive, and the radial drift depends on an ordering between collision  frequency and cyclotron frequency. From a force balance with a collisional drag this gives the radial drift as\cite{Braginskii1965}, 
\begin{equation}
    \vec{v}_{wf}^{(coll.)}\cdot\hat{r} = \frac{\omega_{cj}g_{0}}{\omega_{cj}^{2}+\nu^{2}}
\end{equation}
From considering the above, and the flux of mass out of the plasma due to these drifts, it is evident that there is a substantial challenge in achieving significant mass deposition. The hope is that the core of the pinch will produce fast enough particles in a small enough expanse that the timescale of their traversal will be small enough to avoid collisions. A full treatment of the problem requires studying the collisional transport of a specific configuration outside the asymptotic limits presented here. This is outside the scope of the present work. The purpose of this section is to ideate an application of this technology which potentially has great technological promise in a critical field given the magnetic null on the pinch axis. 

\begin{figure}[ht!]
    \centering
    \includegraphics[width=\linewidth]{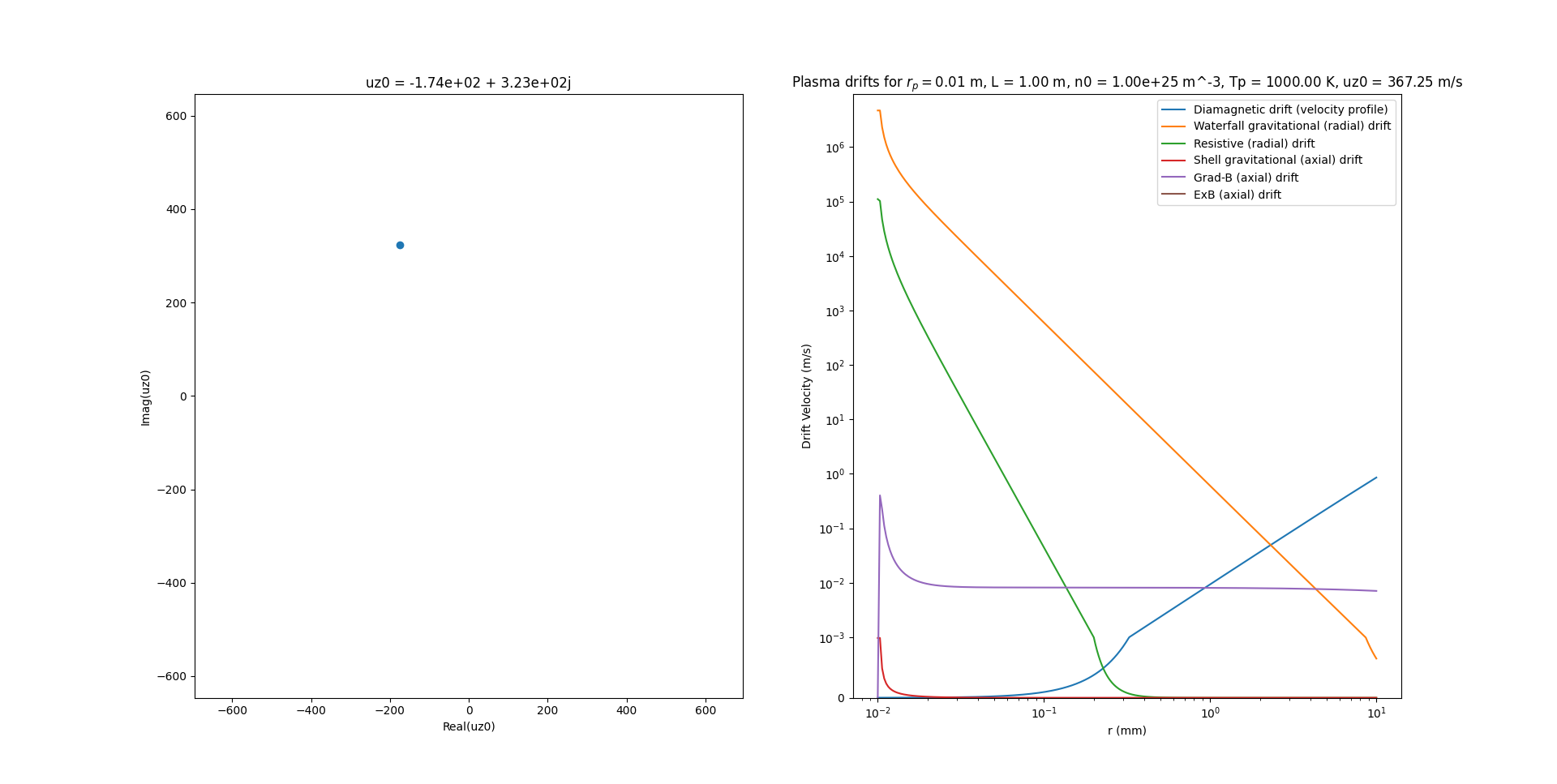}
    \caption{The air plasma drifts for the first root of a pure-flow cubic Bennett vortex based on a meter per second edge speed, and a modest plasma temperature. Due to the similarity in computed flow speeds between the first three roots, being slightly above the speed of sound in air, for these properties the drifts are very similar in structure. Both resistive and waterfall drifts are very strong close to the axis with the plasma needing to reach close to its edge state before its flow dominates.}
    \label{fig:air_tpc_drifts_root1}
\end{figure}

\begin{figure}[ht!]
    \centering
    \includegraphics[width=\linewidth]{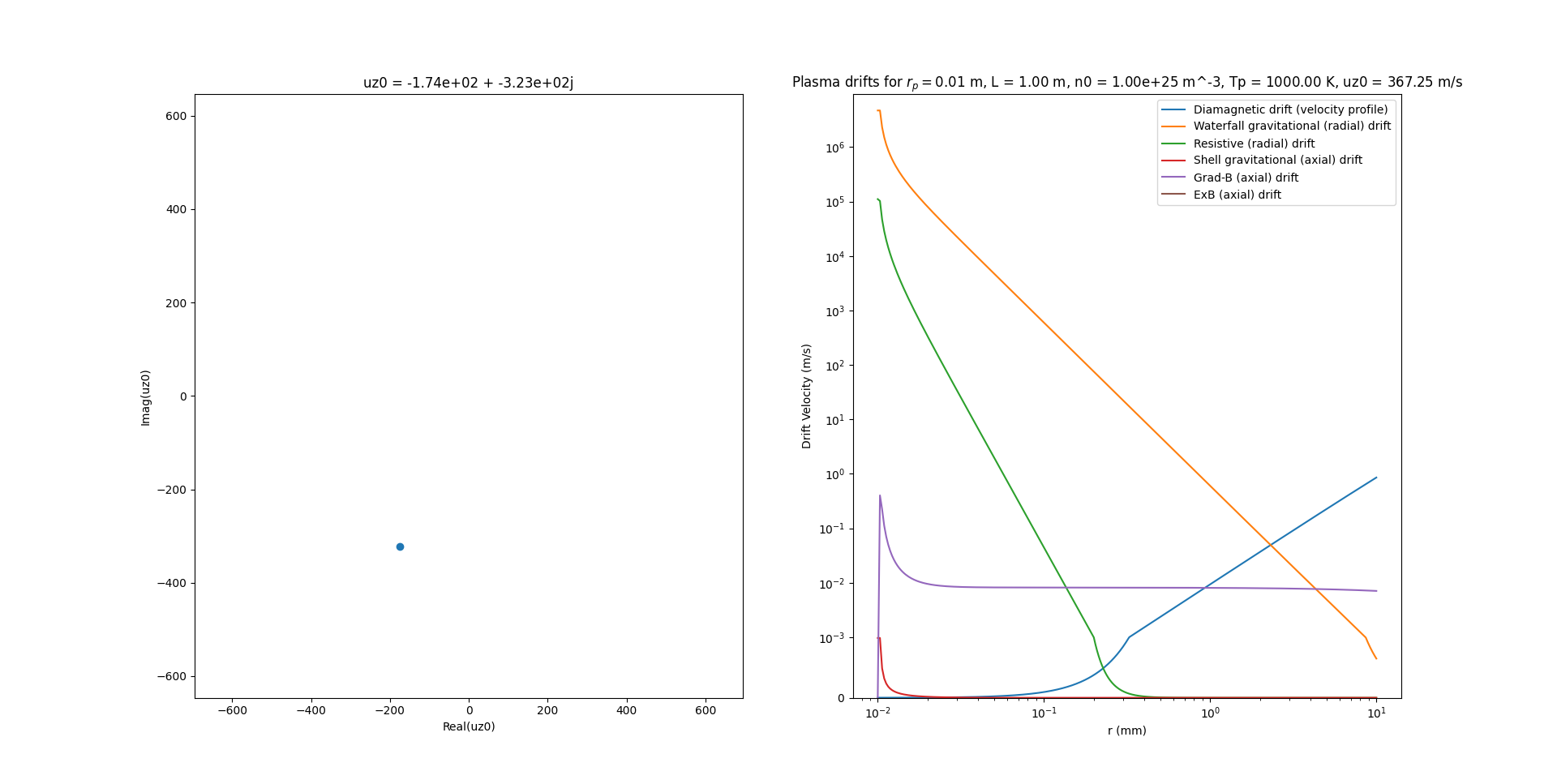}
    \caption{The air plasma drifts for the second root of a pure-flow cubic Bennett vortex based on a meter per second edge speed, and a modest plasma temperature. Due to the similarity in computed flow speeds between the first three roots, being slightly above the speed of sound in air, for these properties the drifts are very similar in structure. Both resistive and waterfall drifts are very strong close to the axis with the plasma needing to reach close to its edge state before its flow dominates.}
    \label{fig:air_tpc_drifts_root2}
\end{figure}

\begin{figure}[ht!]
    \centering
    \includegraphics[width=\linewidth]{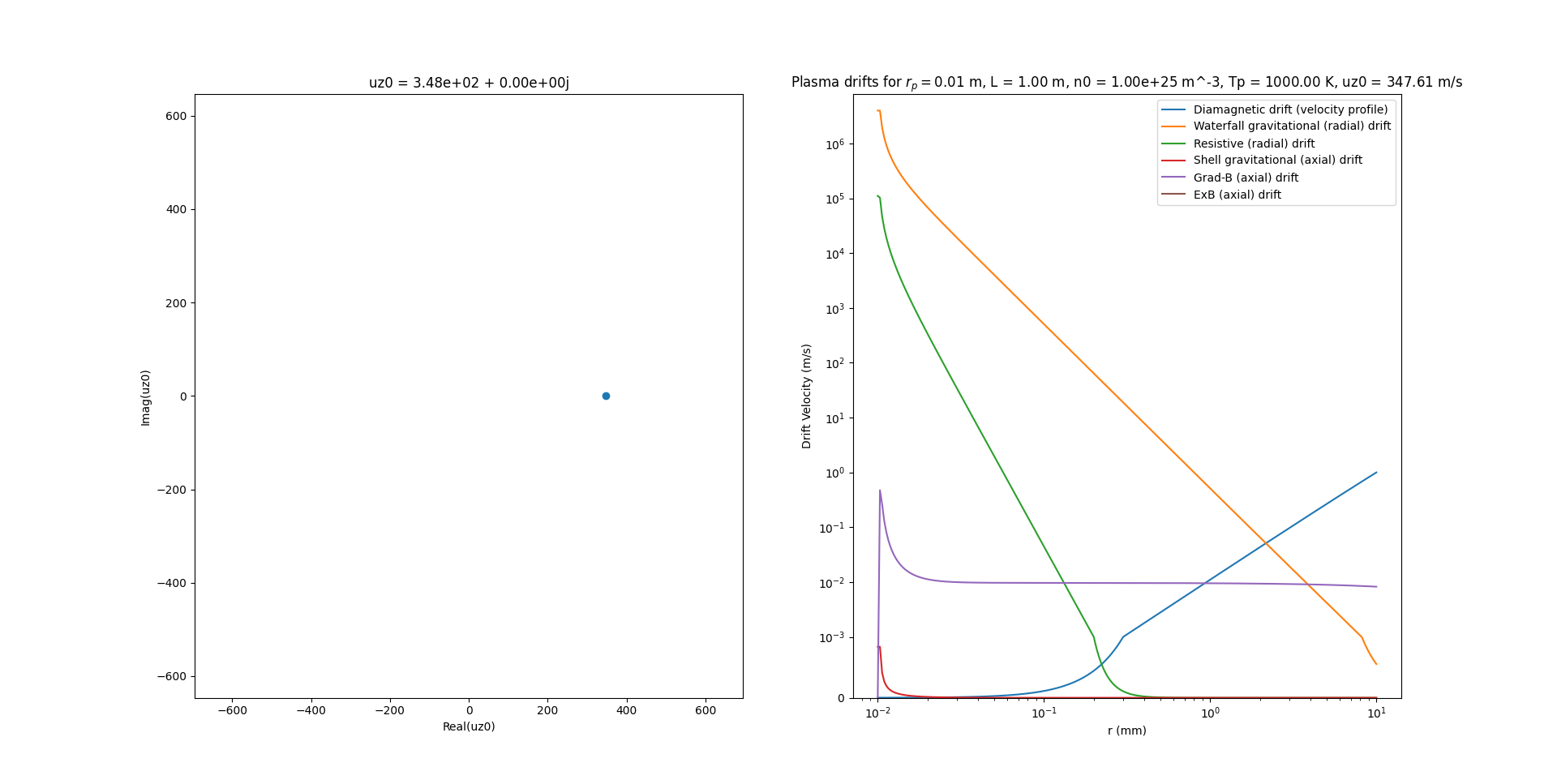}
    \caption{The air plasma drifts for the third root of a pure-flow cubic Bennett vortex based on a meter per second edge speed, and a modest plasma temperature. Due to the similarity in computed flow speeds between the first three roots, being slightly above the speed of sound in air, for these properties the drifts are very similar in structure. Both resistive and waterfall drifts are very strong close to the axis with the plasma needing to reach close to its edge state before its flow dominates.}
    \label{fig:air_tpc_drifts_root3}
\end{figure}

\begin{figure}[ht!]
    \centering
    \includegraphics[width=\linewidth]{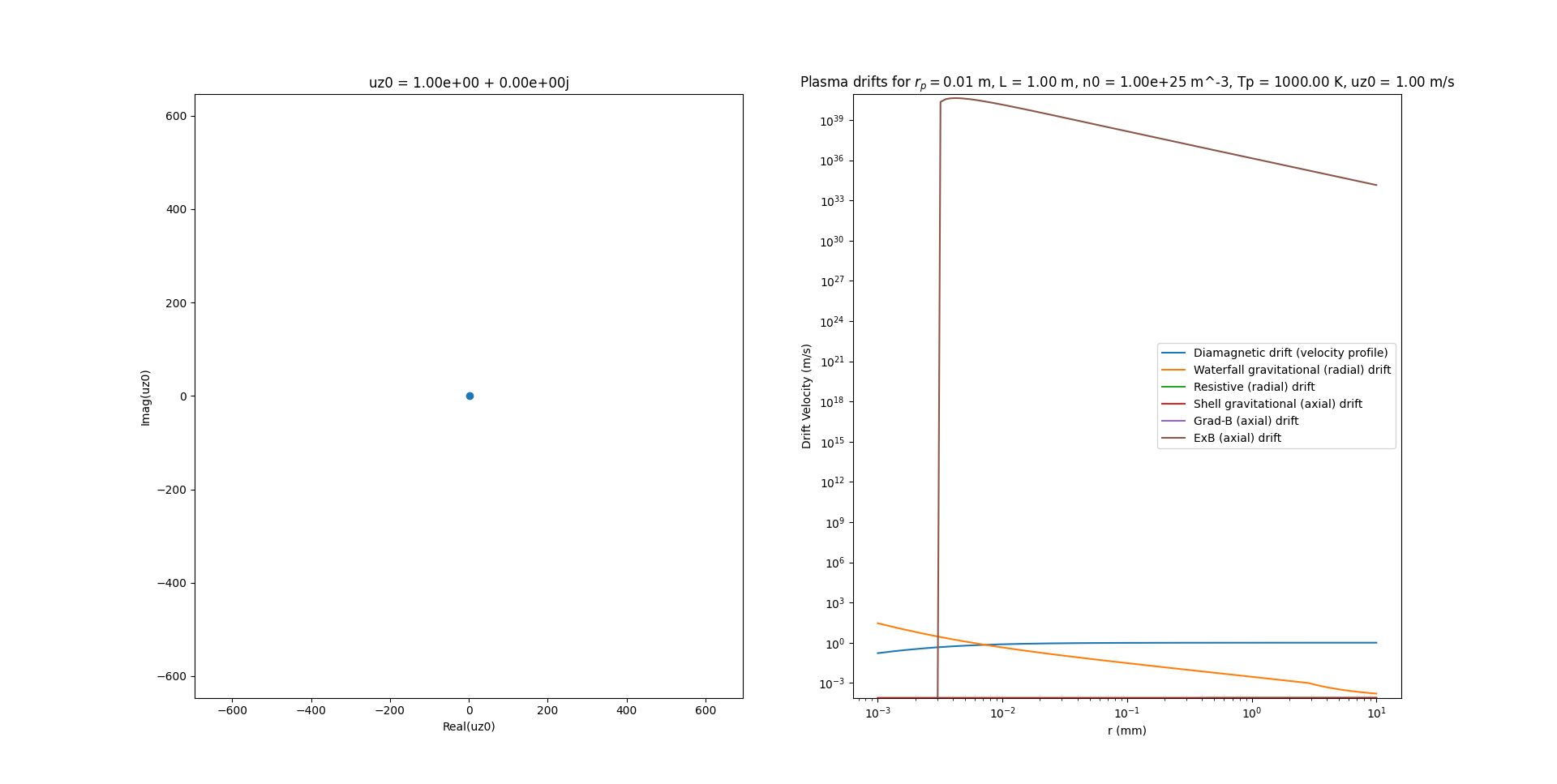}
    \caption{The air plasma drifts for the fourth root of a pure-flow cubic Bennett vortex based on a meter per second edge speed, and a modest plasma temperature. This solution is the one where the flow constant solution is the edge speed, and what is seen is an unphysical growth in the axial ExB drift. Strictly-speaking, this is not the same thing as "electron runaway", but it is a feature which does not disappear when the interior of the plasma disc is increased, and suggests a very sudden acceleration to the plasma travelling down the axis in this solution which indicates a breaking down of the fluid theory in a regime which is kinetic as the unmagnetized core of these pinches involve betatron motion rather than that of a cyclotron.}
    \label{fig:air_tpc_drifts_root4}
\end{figure}

\qquad Analytic solutions for the drifts of a representative air plasma cubic vortex are shown in Figures (\ref{fig:air_tpc_drifts_root1}) - (\ref{fig:air_tpc_drifts_root4}) where the edge speed is taken to be a meter per second. These profiles are constructed with a value of $r_{int} = 10 \ [\mu m]$ for the minimum radius of the disc. Varying this value will change the core velocity, for example, decreasing it to $r_{int} = 1 \ [\mu m]$ with the given plasma properties will formally lead to the presence of superluminal, i.e. divergent, velocities for the radial drifts close to the axis because of issues of numerical instability in the computation of natural logarithms. This numerical artifact stands in contrast to the persistently divergent axial, $\vec{E}\times\vec{B}$ drift velocity that remains in the idealized framework even when the aforementioned unphysical numerical artifacts are gone. This is not a numerical artifact, so much as it is an asymptotic divergence occurring in this idealized classical framework, but it may indicate a deeper structural component related to the asymptotic component of the axial $\vec{E}\times\vec{B}$ drift.

\qquad Another interesting point to note in this data is the occurrence of the flow solutions for the first three roots, namely that they are all clustered tightly around values which are very close to the speed of sound in air. The clustering of the first three roots near these values may indicate an emergent scaling behavior within the ideal framework presented here. 

% \qquad On the similarity between the air and fusion experiments in terms of flow speed, and in the face of the ability to construct relatively accurate solutions for both structures out of representations whose only crime is to suggest the system has a certain temperature profile with absolute zero at the pinch core
 
% \qquad In the warm core of a plasma, which is a soluble condition for a Bennett vortex by applying the weak form of the shear-flow criterion to the structure, then the collisionality will be lowered in comparison to the case of the purely cubic temperature. A further lowering can be possibly achieved by constructing a vortex with a small density near the pinch axis. By maintaining a collisionality that is high enough to result in substantial ionization, but not high enough to result in radial stagnation - a concern that is also addressed by resistive drifts -
% it is ventured that a depository diaspora of pollutant particles may be transported out of the reactor, and a way of achieving significant sequestration of dangerous greenhouse gases may be engineered based on modest temperatures, and standing a shear-flow stabilized Z-pinch upright.  

\subsection*{Deep Neural Network (DNN) Neuron}
\qquad The flow profile discussed in this article is also suitable for implementation as the activation function of a deep neural network\cite{Goodfellow-et-al-2016} because it defines a parametrized, nonlinear form with a bounded nonlinearity, tunable scale, and asymptotic structure. However, the same is true of many functions and this shouldn't be taken to inscribe any unscientific meaning to the form. Instead it is an observation of its mathematical properties that allow it to serve as part of a universal function approximator\cite{cybenko:mcss}, i.e., a DNN. 

\qquad It is worth briefly comparing it in structure to the `sigmoid` function,
\begin{equation}
    f_{sigmoid}(x) = \frac{1}{1 + \exp(-x)}
\end{equation}
as the two nonlinear profiles share similarities. The normalized form of a pureflow cubic vortex can be written as,
\begin{equation}
    f_{cpf}(x) = \frac{1}{(1 + x)^{2}}
\end{equation}
where $x = \frac{C_{B,T}}{r}$. Pointwise correspondence between the two depends on the denominators being identical,
\begin{equation}
    (1 + x)^{2} = 1 + \exp(-x)
\end{equation}
This requires $x$ to be negative-valued, so from the perspective of a real physical system we cannot draw an exact correspondence across the complete activation domain. Mathematically this yields,
\begin{equation}
    -x = \ln(x(x+2))
\end{equation}
which necessitates a negative value for x, and a positive value for the log argument to remain consistent. There is a singularity at $x = -1$ in this correspondence, and $0 > x > -2$ bounds the span of activations in which isolated points can exist at which the profiles coincide exactly to the negative halfspace.

\qquad There are other properties of the nonlinear cubic vortex function that make it interesting to study from the perspective of a deep neural network, and the above is not the only normalized form that can be produced. The radius-normalized coordinate can be inverted and then the shear-normalized form is,
\begin{align}
    f_{cpf}(\phi) &= \frac{\phi^{2}}{(\phi+1)^{2}} \\
    \phi = \frac{r}{C_{B,T}}
\end{align}
For example, the tail-end of the vortex nonlinearity in the denominator is an inverse power-law decay in the $x = \frac{C_{B,T}}{r}$ formulation for large $x$, while still reaching the same asymptotic limiting value as the shear-normalized $\phi$ representation of the vortex.  

\qquad The un-normalized form of a cubic pureflow vortex presents two trainable parameters, the flow solution, $u_{z,0}$, and the shear layer scale parameter, $C_{B,T}$. Further discussion or analysis of this topic is outside the scope of this work. The purpose of this section is only to note that the nonlinear structure of the vortex profile permits its interpretation as a possible activation-function family within a universal function approximator. 

\qquad Grounded caution should be wielded when approaching the idea of treating this profile as the activation function for a DNN. It is merely a statement of fact about this profile's mathematical properties, and how they would fit in amongst the manifold such ones for inferring the targets to highly complex features, and not an ascription of any broader physical interpretation about neural network latent spaces.

\subsection*{Sawteeth}
\qquad Magnetic sawteeth\cite{JAWesson_1986} structures naturally arise in extended chains of this equilibrium. This structure is obscured in a layer-normalized basis,
\begin{equation}
    \phi = \frac{r}{C_{B,T}}
\end{equation}
because this requires the normalized pinch to be placed at a radius, 
\begin{equation}
    \phi_{p} = \frac{r_{p}}{C_{B,T}} = -\frac{1}{2}
\end{equation}
as shown in the Supplement.

\begin{figure}[ht!]
    \centering
    \includegraphics[width=\linewidth]{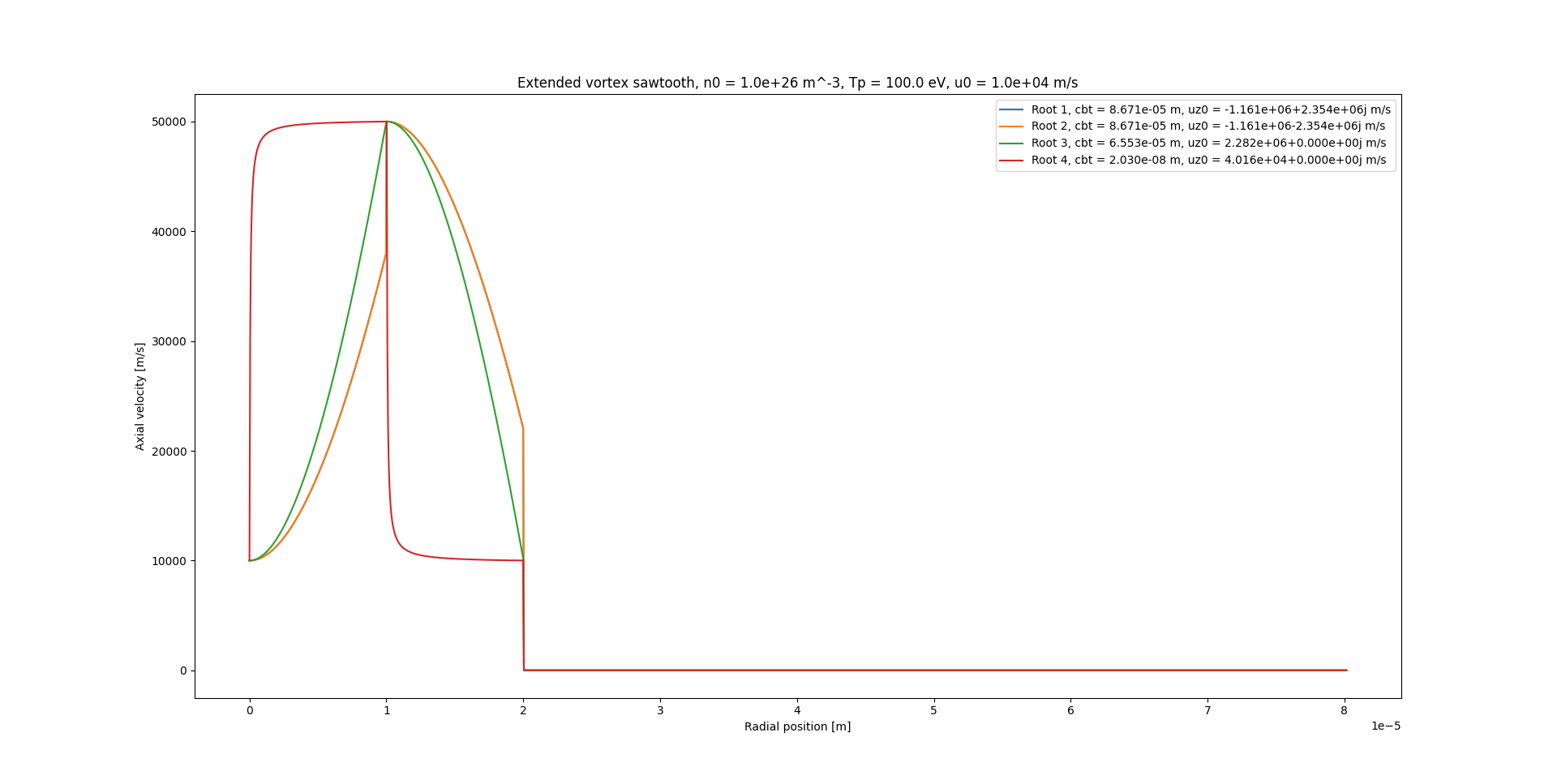}
    \caption{Analytic solutions to a chain of microscale vortices which illustrates the natural manner in which sawteeth can arise from this form. Interestingly, we see the same pattern arise as from previous chains of multiple vortices, where the first and second roots form complex conjugates, the third root is the best fit, and the fourth has a shear layer located much closer to the axis than the other roots do.}
    \label{fig:extended_sawtooth}
\end{figure}

\begin{figure}[ht!]
    \centering
    \includegraphics[width=\linewidth]{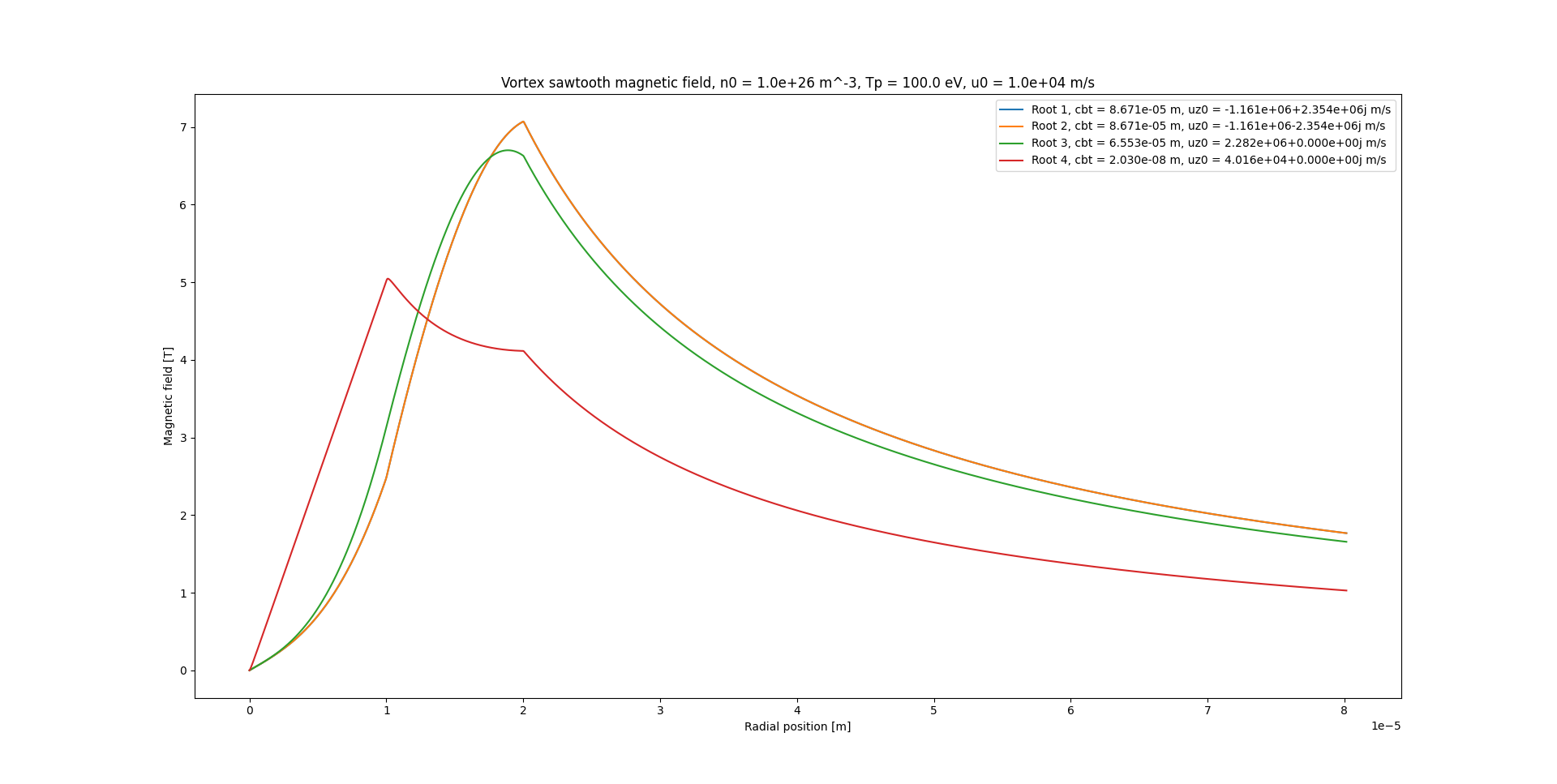}
    \caption{Magnetic field for the microscale vortices from Figure (\ref{fig:extended_sawtooth}) integrated with a cumulative trapezoidal routine. The lack of current in the field-reversal region causes the magnetic field to fall off as $\sim \frac{1}{r}$ there so the duty cycle of this system can be tuned based on the desired falloff to the magnetic field.}
    \label{fig:extended_sawtooth_bfield}
\end{figure}

\qquad Regardless, alternating-sign pureflow vortices may be chained together before the onset of a stagnation region, and then the effective duty cycle of this system, i.e., the relative spacing and width of the vortices, will regulate how far the magnetic field falls. Figures (\ref{fig:extended_sawtooth}), and (\ref{fig:extended_sawtooth_bfield}) illustrate this with a micrometer scale chain of vortices. 

\section*{Limitations \& Discrepancies}
\qquad Unphysical vortices, i.e., solutions which lie outside the range of validity for this ideal nonrelativistic treatment, must be neglected. For example, the non-relativistic thermal interpretation of an electron plasma breaks down once the thermal energy of a plasma particle reaches its rest-energy: $k_{B}T \sim m_{e}c^{2}$. Remaining vortices are understood as being an ideal construct where certain physics related to viscosity, resistivity, relativity, and higher-order kinetic effects, are not incorporated. These structures are understood as emergent features of the axisymmetric shear-flow stabilized Z-pinch theory in a regime where the plasma behaves analogously to a magnetic spring. While the profile reproduces the shape of the experimental data qualitatively, a large source of discrepancy arises when it does not locate the shear layer properly based on the experimental values. These complex solutions to the eigenvalues of the shear-flow stabilized theory can be interpreted as unstable growth modes if mixed complex and oscillatory stable states if imaginary\cite{uri_1995}. 

\qquad This is unsurprising as the ideal model presented here only roughly captures global features related to plasma properties supported by the 1D treatment of a shear-flow stabilized Z-pinch equilibrium. Systems that require a 2D treatment, such as a problem that considers this equilibrium in spherical coordinates, or involve mechanisms beyond the ideal treatment of a shear-flow stabilized Z-pinch, e.g., viscosity, will naturally deviate from the ideal form presented here. 

\begin{figure}[ht!]
    \centering
    \includegraphics[width=\linewidth]{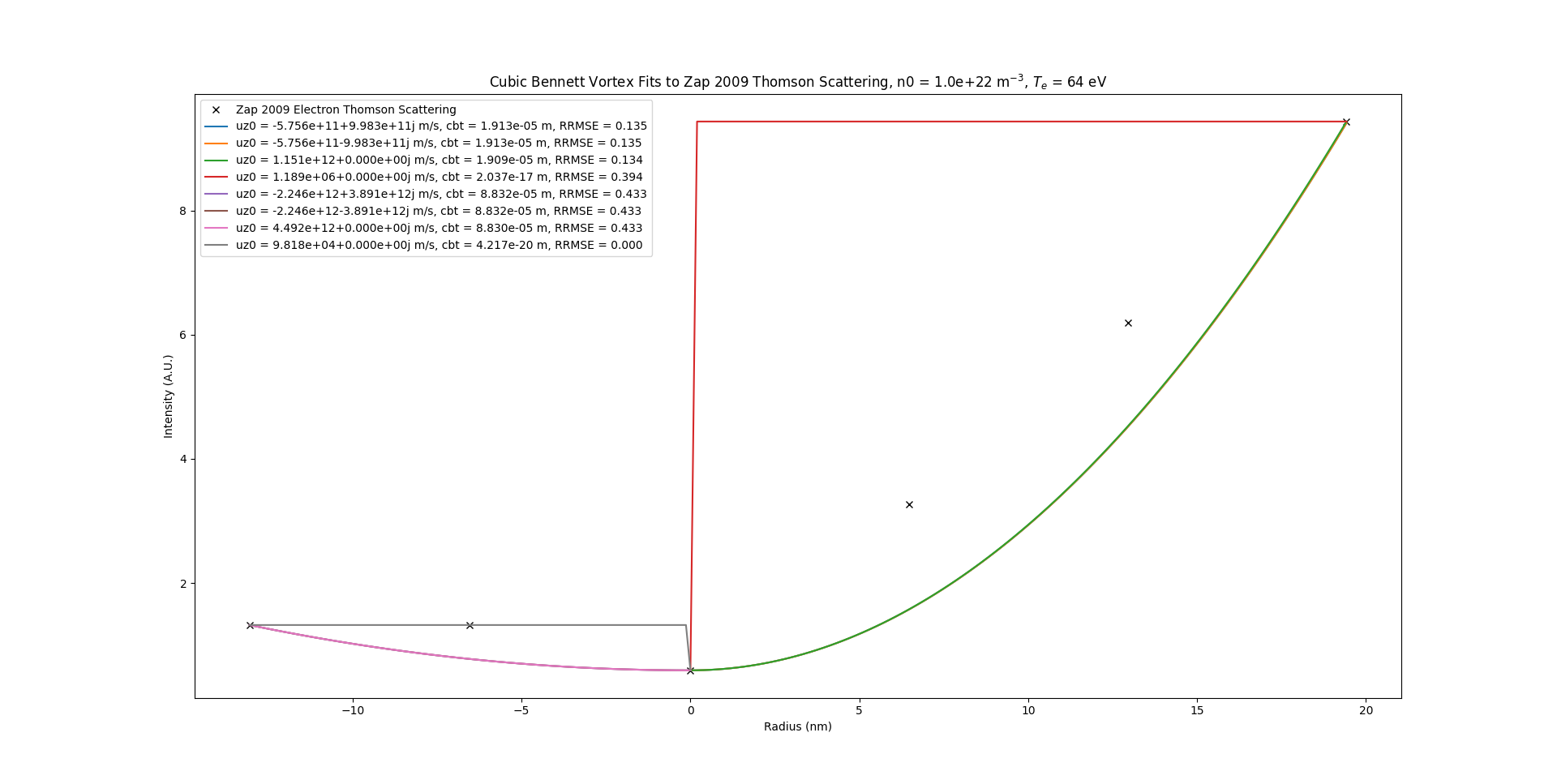}
    \caption{Zap 2009 solutions for the Thomson scattering observable where the wavelength of the pulse is interpreted as a plasma radius as the ordered set of ascending wavelengths probing the plasma gives information regarding the intensity of the scattered light at successive points in the plasma. Accurate solutions to this observable here have unphysical flow velocities. Physical solutions can be seen to have shear layers around the value of the classical electron radius.}
    \label{fig:zap2009_thomson}
\end{figure}

\begin{figure}[ht!]
    \centering
    \includegraphics[width=\linewidth]{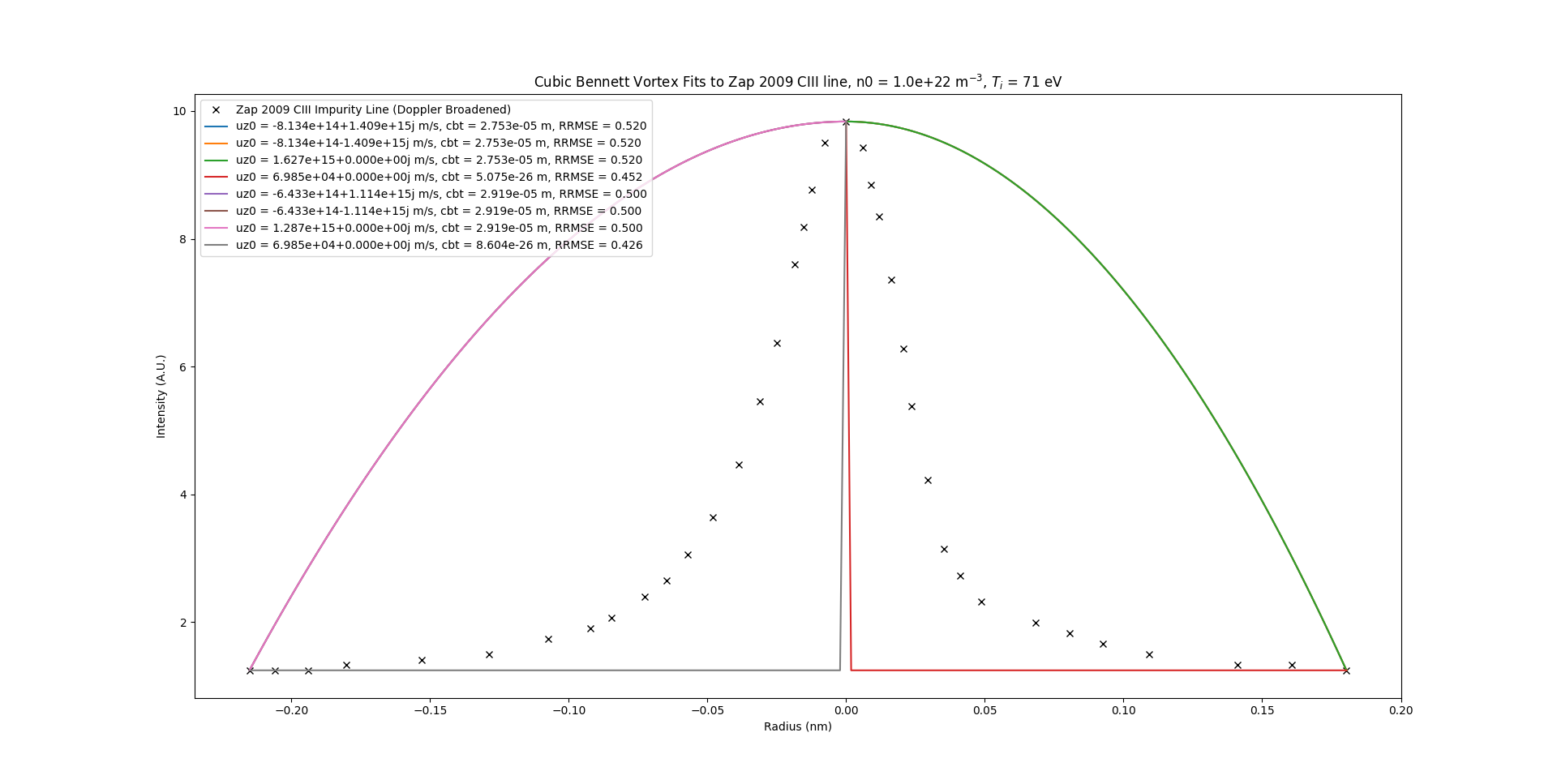}
    \caption{Zap 2009 solutions when interpreted through an effective-radius reconstruction procedure for the ion Doppler spectroscopy intensity observable. The shape of the structure is a Gaussian, and vortex solutions to the ideal model either exhibit a sub-atomic scale shear layer or require unphysical flow speeds. The former leads to inaccurate solutions because the shear layer does not coincide with the observed profile, and the latter because the shear layer is larger than the interpreted pinch radius.}
    \label{fig:zap2009_iondoppler}
\end{figure}

\section*{Discussion}
\qquad The applicability of this equilibrium is evident here in a cross-regime manner, but it remains to be seen how far it extends. It arises here because the observables depend functionally on the current density, which is based purely on the flow as the plasma density follows from it. The flow pattern being shear-flow stabilized, and mathematically the equilibrium possessing no intrinsic cutoff in scale within the ideal theory, suggests that the shear-flow stabilization mechanism may play a broader role in plasma organization across multiple scales.

\qquad There is a strange ability for this equilibrium to get results that are not far off for many disparate plasma discharge systems, and it is suggested that this can be explained under the umbrella of Ideal MHD's intrinsic efficacy as it describes the first-order behavior of the magnetohydrodynamic equilibrium. Unsteady, and diffusive contributions from the magnetic field as well as viscous interactions can be layered on top of this base state to further improve the accuracy of the calculations and the complexity of the physics. 

\qquad Within just the ideal framework, exploring the thermal structure of the temperature perturbations reveals that the solutions are characterized by sums of spatial Bessel functions and temporal eigenfunctions, suggesting a further richness to the model than has been captured by the simplified power-law plasma temperature used in the present calculations. 

\qquad These claims and investigations must be approached with caution and care, because there is a large amount of uncaptured physics in this limited, idealized model, but that studying it alongside this additional physics will give further insight into the multi-scale behavior of plasmas which have the shared Shumlak-Hartmann shear-flow stabilized characteristic. 

\qquad As high-speed directed plasma flows are common across laboratory and astrophysical plasmas, there may exist additional systems in which shear-flow stabilized structures emerge. For example, coronal mass ejections (CMEs) are observed to proceed in three phases where slow magnetic expansion transitions into rapidly accelerated plasma outflow from the solar surface\cite{Russell2023}. 

\qquad The Shumlak-Hartmann criterion can also be integrated to formulate the shear-flow stabilized inequality in terms of the flow velocity at a particular point,
\begin{equation}
    u_{z}(r) > u_{0} + \frac{\pi}{5L\mu_{0}}\int_{0}^{r}\frac{B}{\sqrt{\rho}}dr'
\end{equation}
which can also be written as,
\begin{equation}
    u_{z}\bigg(1 - \frac{\pi}{5}\frac{\sigma}{R_{m}}\int_{0}^{r}\frac{B}{\sqrt{\rho}}dr'\bigg) > u_{0}
\end{equation}
As a form that emphasizes the shear-flow stabilized property is not solely a local property of the velocity shear, but depends on the integrated impact of the magnetic field and inertial MHD properties across the plasma radius. For a Z-pinch the magnetic field is,
\begin{equation}
    B(r) = B_{\theta}(r) = \frac{\mu_{0}}{r}\int_{0}^{r}r'J_{z}(r')dr'
\end{equation}
but not in general. In general, the shear has to contend with the full strength of the magnetic field for calculating the Alfvenic speed. 

\qquad We can study two limits, the "core-pinch", and "edge" limits where
\begin{align}
    \phi &= \frac{r}{C_{B,T}} \\
    \beta_{L} &= \phi^{-1} = \frac{C_{B,T}}{r}
\end{align}
define the transformation of coordinate to obtain, respectively, the forms,
\begin{align}
    \tilde{u}_{z}(\phi) &= \frac{u_{z}}{u_{z,0}} = \phi^{2}(1 + \phi)^{-2} \\\
    \tilde{u}_{z}(\beta_{L}) &= \frac{u_{z}}{u_{z,0}} = (1 + \beta_{L})^{-2}
\end{align}
These forms can be investigated in the case where their associated parameter is small to study the structure of a power-series expansion in this region. In general, the "edge" limit where $\beta_{L} < 1$ yields a MacLaurin expansion,
\begin{equation}
    \tilde{u}_{z}(\beta_{L}) = \sum_{n=0}^{\infty}(-1)^{n}(n+1)\beta_{L}^{n}
\end{equation}
Consider then a plasma flow constructed out of a sum of N vortices,
\begin{equation}
    u_{z} = \sum_{j=1}^{N}u_{z,j}\frac{r^{2}}{(r + C_{j})^{2}}
\end{equation}
where $C_{B,T} \rightarrow C_{j}$. We can rewrite the above as,
\begin{equation}
    u_{z} = \sum_{n=0}^{\infty}(-1)^{n}(n+1)M_{n}r^{-n}
\end{equation}
where,
\begin{equation}
    M_{n} = \sum_{j=1}^{N}u_{z,j}C_{j}^{n}
\end{equation}
with $\beta_{L} < 1$ this ensures we have a sufficiently large enough radius, $r > C_{j}$, that we do not need to worry about any singularities where this approximation is valid. In this region we find that a natural multipole-like description of the axial velocity emerges. By suitable choice of moments, we can match this finite sum of vortices to an arbitrary velocity expansion in this asymptotic basis,
\begin{equation}
    u_{z} = \sum_{n=0}^{\infty}b_{n}r^{-n}
\end{equation}
with,
\begin{equation}
    b_{n} = (-1)^{n}(n+1)M_{n}
\end{equation}
a Vandermonde matrix describes the shear-layer lengths which match the flow speed roots to the desired profile expressed in this "edge" basis,
\begin{equation}
    V_{nj} = C_{j}^{n}
\end{equation}
This can be seen readily from writing out the first few terms of the expansion and comparing it to the necessary moment. This suggests that sufficiently large sums of vortices with tailored properties may be capable of approximating a broad class of plasma velocity profiles within the asymptotic region where the expansion remains valid.

\qquad On the other hand, if we study a power series expansion about the core, i.e., $\phi = \frac{r}{C_{B,T}} < 1$, then we have,
\begin{align}
    u_{z} &= \sum_{j=1}^{N}u_{z,j}\frac{\phi_{j}^{2}}{(\phi_{j}+1)^{2}} \\
    &= \sum_{n=0}^{\infty}\sum_{j=1}^{N}u_{z,j}(-1)^{n}(n+1)\phi_{j}^{n+2} \\
    &= \sum_{n=0}^{\infty}(-1)^{n}(n+1)W_{n}r^{n+2}
\end{align}
where,
\begin{equation}
    W_{n} = \sum_{j=1}^{N}u_{z,j}C_{j}^{-(n+2)}
\end{equation}
describes the moment that needs to be matched with the coefficients of a power-series expansion about $r^{n}$,
\begin{equation}
    u_{z} = \sum_{n=0}^{\infty}d_{n}r^{n}
\end{equation}
Which gives,
\begin{align}
    d_{0} &= d_{1} = 0 \\
    d_{n} &= (-1)^{n}(n-1)W_{n-2} \\
    n&\geq 2
\end{align}
This connection between these moments in an underlying "edge" or "core" approximation and the corresponding power-series representation of the flow is a dual asymptotic moment structure where there are outer moments, $M_{n}$, and inner moments $W_{n}$ to connect the vortex theory to these power-series representations.

\qquad While the present model is limited by its idealized treatment and omission of higher-order kinetic, resistive, viscous, and radiative effects, the repeated emergence of qualitatively similar shear-organized structures across disparate plasma environments suggests that the equilibrium captures a physically meaningful component of the underlying magnetohydrodynamic behavior. Future work incorporating non-ideal transport, time dependence, and kinetic corrections will determine the extent to which these structures persist beyond the ideal limit, and whether the Shumlak-Hartmann stabilized framework can serve as a useful reduced description for a broader class of magnetized plasma flows.

\section*{Acknowledgements}
The author thanks Aria Johansen for constructive feedback and critical comments on the early versions of the manuscript, Ilya Dodin for his feedback, Sander Nijdam for discussions on the application of this equilibrium to air plasma streamer formation, Iman Datta for insightful discussions on Ideal MHD, Jack Coughlin for his feedback, Eric Meier for introducing the topic of the Bennett pinch, Peter Thoreau and James Penna for their support, and Whitney Thomas for her patience. 

\clearpage
\appendix
\section*{Supplemental Appendix}
% \documentclass[11pt]{article}

% \usepackage[T1]{fontenc}
% \usepackage[utf8]{inputenc}
% \usepackage{amsmath,amssymb,amsthm}
% \usepackage{graphicx}
% \usepackage{booktabs}
% \usepackage{geometry}
% \geometry{margin=1in}

% \usepackage{physics}

% \title{Supplementary Information for Bennett Vortices}
% \author{Matt Russell}
% \date{}

% \begin{document}
% \maketitle
\section*{Supplementary Information for Bennett Vortices}
\addcontentsline{toc}{section}{Supplementary Information}
\section*{Weak Form}
Under the exchange of the Bennett nonlinearity from density to flow we have,
\begin{equation}
    u_{z}(r) = \frac{u_{z,0}}{(1 + \xi^{2}r^{2})^{2}}
\end{equation}
where,
\begin{equation}
    \xi^{2} = bn_{0} = \frac{C_{B}n_{0}}{T(r)}
\end{equation}
with,
\begin{equation}
    C_{B} = \frac{\mu_{0}e^{2}u_{z,0}^{2}}{16k_{B}}
\end{equation}
The Shumlak criterion for the shear-flow stabilization of a Z-pinch in vacuum reads,
\begin{equation}
    \label{eqn:shumlak_crit}
    \dv{u_{z}}{r} > 0.1kV_{A} = \frac{\pi}{5L}\frac{B}{\sqrt{\rho\mu_{0}}}
\end{equation}
and the magnetic Reynold's number,
\begin{equation}
    R_{m} = \sigma\mu uL
\end{equation}
For a fixed, and finite, Z-pinch plasma the length of the pinch can be taken as infinite in order to study the consequences of the ideal theory of magnetohydrodynamics for this profile. Then the Shumlak criterion reads,
\begin{equation}
    \dv{u_{z}}{r} > 0
\end{equation}
provided that the magnetic field does not go to infinity, or that the density does not become trivial. What then is the condition on the temperature profile in order to satisfy this criteria and obtain a shear-flow stabilized Z-pinch?
\begin{align}
    \dv{u_{z}}{r} &= \dv{}{r}\bigg[u_{z,0}\frac{T(r)^{2}}{(T(r) + C_{B}n_{0}r^{2})^{2}}\bigg]\\
    &= u_{z,0}\bigg[2TT'(T + C_{B}n_{0}r^{2})^{-2}-2T^{2}(T + C_{B}n_{0}r^{2})^{-3}(T' + 2C_{B}n_{0}r)\bigg] \\
    &= u_{z,0}\bigg[\frac{2TT'(T + C_{B}n_{0}r^{2})}{(T+C_{B}n_{0}r^{2})^{3}} - \frac{2T^{2}(T'+2C_{B}n_{0}r)}{(T + C_{B}n_{0}r^{2})^{3}}\bigg] \\
    &= u_{z,0}\bigg[\frac{2TT'C_{B}n_{0}r^{2} - 4T^{2}C_{B}n_{0}r}{(T + C_{B}n_{0}r^{2})^{3}}\bigg] \\
    &= u_{z,0}\bigg[2TC_{B}n_{0}r\frac{rT' - 2T}{(T + C_{B}n_{0}r^{2})^{3}}\bigg] \\
    &>0
\end{align}
\qquad Restricting to $r > 0$, the denominator is strictly positive, and doe snot affect the sign of the inequality. Even if it is zero it does not matter as the RHS will remain unchanged, but there is the matter that this would result in the fallacious $0/0 = 1$ if $r=0$, and $T(0) = 0$. However, we work around this by turning our attention away from the axis as the plasma state is specified entirely there by our choices, so that we do not have to worry about such a problem in the rest of the pinch where we wish to understand the necessary structure of $T(r)$ for our construction.  
\begin{align}
    \therefore &\bigg[2TC_{B}n_{0}r(rT'-2T)\bigg] > 0 \\
    \rightarrow &(rT'-2T) > 0 
\end{align}
for the same reason that we can throw away the earlier denominator, because we are restricting our attention away from the origin where an absolute zero temperature would pose an issue to our machinations. From here we simply treat this conditional as an integrable differential equation, move infinitesimals around the way physicists do, and arrive at, 
\begin{align}
    &\frac{T'}{T} > \frac{2}{r} \\
    &\rightarrow \frac{dT}{T} > \frac{2}{r}dr \\
    &\therefore \ln(T) > 2\ln(r) \\
    &\therefore e^{\ln(T)} > e^{2\ln(r)} \\
    &\implies T(r) > r^{2}
\end{align}
is sufficient. However, this is the "weak form", and it is called as such because it requires the existence of an integral, and expresses a shear of zero in the pure-flow profile when the temperature is taken to be greater than a parabola, while also being parabolic in form,
\begin{align}
    T(r) &= C_{T}^{(2)}r^{2} = \frac{T_{p}}{r_{p}^{2}}r^{2} \\
    \therefore\dv{u_{z}}{r} &= \dv{}{r}\bigg[\frac{u_{z,0}}{(1 + \frac{C_{B}n_{0}}{C_{T}^{(2)}}\frac{r^{2}}{r^{2}})^{2}}\bigg] = 0
\end{align}
Of course, it is also highly advised to consider,
\begin{equation}
    T(r) = T_{0} \pm C_{T}^{(2,0)}r^{2}
\end{equation}
as a more realistic weak form whose consequences should be investigated since the inverse parabolic form is what is seen in experimental temperature profiles. However, that is outside the scope of this supplement as the results presented in the article do not involve this case explicitly. 

\section*{Strong Form}
\qquad The result of the weak form invites a consideration of what happens when a power-law is adopted for the temperature profile,
\begin{equation}
    T(r) = C_{T}^{(n)}r^{n} = \frac{T_{p}}{r_{p}^{n}}r^{n}
\end{equation}
and this is termed the "strong form" because it does not hinge on the existence of an integral, and it expresses a non-trivial shear. Skipping back to the end of the Shumlak criterion for this new case, we have,
\begin{align}
    &(rT' - 2T) > 0 \\
    &\therefore rT' > 2T \\
    &\rightarrow rC_{T}^{(n)}nr^{n-1} > 2C_{T}^{(n)}r^{n} \\
    &\implies n > 2
\end{align}

\section*{Cubic Pure-flow Profile}
\qquad Since the $n = 2$ case is a weak satisfaction expressing a trivial shear unless there is a core temperature added into the expression, $n = 3$ is the first case to check for whether there is actually a soluble shear-flow stabilized equilibrium which can be curtailed from the base flow profile,
\begin{align}
    u^{(2,3)}_{z}(r) &= \frac{u_{z,0}}{(1 + \frac{C_{B}n_{0}}{C_{T}^{(3)}}\frac{r^{2}}{r^{3}})^{2}} \\ 
    &= \frac{u_{z,0}}{(1 + \frac{C_{B,T}^{(3)}}{r})^{2}} \\ 
    &= u_{z,0}\frac{r^{2}}{(r + C_{B,T}^{(3)})^{2}}
\end{align}
where it was convenient to lump all the plasma properties and constants into,
\begin{equation}
    C_{B,T}^{(3)} = C_{B}n_{0}\frac{r_{p}^{3}}{T_{p}}
\end{equation}
which is the origin of the parenthetical "3" superscript, i.e., the cubic temperature profile supporting this structure, with the "2" being present for consistency with the article notation where it is present to indicate this is the $\chi = 2$ member of an entire family predicated on this fundamental act of exchange whose existence is being shown in this supplement.

\qquad The shear of this profile is,
\begin{align}
    \dv{u_{z}^{(2,3)}}{r} &= \dv{}{r}\bigg[u_{z,0}\frac{r^{2}}{(r + C_{B,T}^{(3)})^{2}}\bigg] \\
    &= u_{z,0}\bigg[2r(r+C_{B,T})^{-2} - 2r^{2}(r + C_{B,T})^{-3}\bigg] \\
    &= u_{z,0}\bigg[\frac{2r(r+C_{B,T})-2r^{2}}{(r + C_{B,T})^{3}}\bigg] \\
    &= u_{z,0}\frac{2rC_{B,T}}{(r + C_{B,T})^{3}}
\end{align}
where the parenthetical superscript was dropped for brevity in the $C_{B,T}$ shear-layer parameter. The shear at the origin is well-defined, and can be evaluated by inspection to dispel concerns about our projection of focus away from the pinch axis when deriving this profile,
\begin{equation}
    \dv{u_{z}}{r}\bigg|_{r=0} = 0
\end{equation}
Points of extremal shear occur when the derivative of the shear is itself 0,
\begin{align}
    u_{z}'' &= 2C_{B,T}u_{z,0}\dv{}{r}\bigg[r(r+C_{B,T})^{-3}\bigg] \\
    &= 2C_{B,T}u_{z,0}\bigg[(r+ C_{B,T})^{-3} - 3r(r+C_{B,T})^{-4})] \\
    &= 2C_{B,T}u_{z,0}\bigg[\frac{C_{B,T}-2r}{(r + C_{B,T})^{4}}\bigg]
\end{align}
Setting the above equal to zero yields, 
\begin{equation}
    r_{max} = \frac{C_{B,T}}{2}
\end{equation}
The shear at this location is,
\begin{align}
    u_{z,max}' &= \frac{2\frac{C_{B,T}}{2}C_{B,T}u_{z,0}}{(\frac{C_{B,T}}{2} + C_{B,T})^{3}} \\
    &= \frac{C_{B,T}^{2}u_{z,0}}{(\frac{3}{2}C_{B,T})^{3}} \\
    &= \frac{8}{27}\frac{u_{z,0}}{C_{B,T}}
\end{align}
With these basic properties established, our next step is to evaluate the equilibrium as it proceeds entirely from this pure-flow profile. There is no \textit{a priori} guarantee that this profile will integrate to produce a magnetic field, or that this magnetic field will further integrate to produce a salient plasma pressure. Obtaining the plasma pressure via integration of the momentum equation is preferred to the ideal gas law closure because Noether's Theorem, which states that conservation laws, like that of a momentum balance, are a consequence of symmetries, like that of the translational invariance of the laws of physics, is the stronger support compared to an \textit{ad hoc} closure relation based on a rarefied gas picture.  

\qquad The magnetic field is obtained from Ampere's Law in ideal magnetohydrodynamics, and for a cylindrical Z-pinch equilibrium this field only has an azimuthal component,
\begin{align}
    B_{\theta}(r) &= \frac{\mu_{0}}{r}\int_{0}^{r} r'J_{z}(r')dr' \\
    &= \frac{-en_{0}u_{z,0}\mu_{0}}{r}\int_{0}^{r} \frac{r'^{3}}{(r' + C_{B,T})^{2}}dr'
\end{align}
This integral can be evaluated analytically with a u-substitution,
\begin{equation}
    u = r' + C_{B,T}
\end{equation}
\begin{align}
    \therefore B_{\theta}(r) &= -\frac{en_{0}\mu_{0}u_{z,0}}{r}\int_{u=C_{B,T}}^{u=r+C_{B,T}}\frac{(u - C_{B,T})^{3}}{u^{2}}du \\
    &= -\frac{en_{0}\mu_{0}u_{z,0}}{r}\int_{u=C_{B,T}}^{u=r+C_{B,T}}\frac{u^{3} - 3u^{2}C_{B,T} + 3uC_{B,T}^{2}-C_{B,T}^{3}}{u^{2}}du \\
    &= -\frac{en_{0}\mu_{0}u_{z,0}}{r}\int_{u=C_{B,T}}^{u=r+C_{B,T}}u - 3C_{B,T} + 3\frac{C_{B,T}^{2}}{u}-\frac{C_{B,T}^{3}}{u^{2}}du \\
    &= -\frac{en_{0}\mu_{0}u_{z,0}}{r}\bigg[\frac{u^{2}}{2}-3C_{B,T}u+3C_{B,T}^{2}\ln(u)+\frac{C_{B,T}^{3}}{u}\bigg|_{u=C_{B,T}}^{u=r+C_{B,T}} \\ 
    &= -\frac{en_{0}\mu_{0}u_{z,0}}{r}\bigg[\frac{(r+C_{B,T})^{2}}{2} - 3C_{B,T}(r + C_{B,T}) + 3C_{B,T}^{2}\ln(r+C_{B,T}) + \frac{C_{B,T}^{3}}{r+C_{B,T}} \\
    &\qquad \qquad - \frac{C_{B,T}^{2}}{2} + 3C_{B,T}^{2} - 3C_{B,T}^{2}\ln(C_{B,T}) - C_{B,T}^{2}\bigg]
\end{align}
or with Wolfram Mathematica to yield the form presented in the article. Evaluating via u-sub is equivalent after some algebra,
\begin{align}
    B_{\theta}(r) &= -en_{0}u_{z,0}\mu_{0}\bigg(\frac{r^{2}-3rC_{B,T}-6C_{B,T}^{2}}{2(r+C_{B,T})} + \frac{3C_{B,T}^{2}\ln(\frac{r + C_{B,T}}{C_{B,T}})}{r}\bigg) \\
    &= -en_{0}u_{z,0}\mu_{0}\frac{r^{3}-3r^{2}C_{B,T}-6rC_{B,T}^{2}+6(r+C_{B,T})C_{B,T}^{2}\ln(\frac{r+C_{B,T}}{C_{B,T}})}{2r(r+C_{B,T})} \\
    &= \frac{-\mu_{0}en_{0}u_{z,0}}{2r(r+C_{B,T})}f(r,C_{B,T})
\end{align}
which will require the application of the log property,
\begin{equation}
    \ln(a / b) = \ln(a) - \ln(b) = -\ln(b/a)
\end{equation} 

\qquad The plasma pressure kernel is determined by this magnetic field, and the structure of the B-field is crucial to the validity of this equilibrium in the context of the Shumlak criterion, so we will integrate this plasma pressure kernel first to see if a full equilibrium is produced, 
\begin{align}
    p(r) &= p_{0} - \int_{0}^{r}J_{z}B_{\theta}dr' \\
    p(r_{p}) &= 0 \implies p_{0}=\int_{0}^{r_{p}}J_{z}B_{\theta}dr \\
    \therefore p(r) &= p_{0} - \int_{0}^{r}-en_{0}u_{z,0}\frac{r'^{2}}{(r' + C_{B,T})^{2}}(-\frac{\mu_{0}en_{0}u_{z,0}}{2r'(r'+C_{B,T})} f(r',C_{B,T}))dr' \\
    &= p_{0} - \frac{\mu_{0}}{2}(en_{0}u_{z,0})^{2}\int_{0}^{r}\frac{r'}{(r' + C_{B,T})^{3}}f(r', C_{B,T})dr' \\
    &= p_{0} - \frac{\mu_{0}}{2}(en_{0}u_{z,0})^{2}\int_{0}^{r}\frac{r'}{(r' + C_{B,T})^{3}}(f_{1}+f_{2}+f_{3}+f_{4})dr'
\end{align}
Despite its appearance the above integral is tractable as it splits into four pieces,
\begin{equation}
    \int_{0}^{r}\frac{r'}{(r' + C_{B,T})^{3}}(f_{1}+f_{2}+f_{3}+f_{4})dr' = P_{1} + P_{2} + P_{3} + P_{4} 
\end{equation}
\begin{align}
    P_{1} &= \int_{0}^{r}\frac{r'^{4}}{(r' + C_{B,T})^{3}}dr' \\
    P_{2} &=-3C_{B,T}\int_{0}^{r}\frac{r'^{3}}{(r' + C_{B,T})^{3}}dr' \\ 
    P_{3} &= -6C_{B,T}^{2}\int_{0}^{r}\frac{r'^{2}}{(r' + C_{B,T})^{3}}\bigg(1 + \ln(\frac{C_{B,T}}{r' + C_{B,T}})\bigg)dr' \\
    P_{4} &= -6C_{B,T}^{3}\int_{0}^{r}\frac{r'}{(r' + C_{B,T})^{3}}\ln(\frac{C_{B,T}}{r' + C_{B,T}})dr'
\end{align}
which are either rational functions amenable to u-substitution or the product of a rational function and a logarithmic term that can be integrated by parts after appropriate substitution.

\section*{Bulk Profiles \& Mixed Vorticity}
\qquad The addition of a uniform background flow can be used to extend the cubic pure-flow profile to more realistic situations where the pinch axis will not be in stagnation. There is a deep consequence of this fact, namely that there is a \textit{mixed} nature at the heart of the fundamental character of this result which can be seen from considering the inclusion of a bulk profile in the electron plasma current density, 
\begin{equation}
    \vec{J} = J_{z}\hat{z} = -en(r)(u_{0}\pm u_{z,0}\frac{r^{2}}{(r + C_{B,T})^{2}})
\end{equation}
If the $n(r)$ in the above is selected appropriately so that an analytic plasma current density remains then the combination of flow and density is naturally tractable. The ability to distribute the Bennett nonlinearity reflects this notion, as exponents can be attached to the density and flow to describe how much of the original nonlinearity is exchanged,
\begin{align}
    n^{(\nu)}(r) &= n_{0}\frac{T^{\nu}}{(T + C_{B}n_{0}r^{2})^{\nu}} \\
    u_{z}^{(\chi)}(r) &= u_{z,0}\frac{T^{\chi}}{(T + C_{B}n_{0}r^{2})^{\chi}}
\end{align}
where the constraint that,
\begin{equation}
    \chi + \nu = 2
\end{equation}
exists so that we arrive at the basic form of the analytic current density we obtained,
\begin{equation}
    \vec{J} = -en_{0}u_{z,0}\frac{T(r)^{2}}{(T(r) + C_{B}n_{0}r^{2})^{2}}\hat{z}
\end{equation}
Taking this a step further, we know that a cubic temperature profile in the above leads to an analytic Z-pinch equilibrium based on the plasma current density,
\begin{equation}
    \vec{J} = -en_{0}u_{z,0}\frac{r^{2}}{(r + C_{B,T}^{(3)})^{2}}\hat{z}
\end{equation}
Consequently, any combination of flow and density which produces this current density will also result in the same analytic Z-pinch equilibrium as before,
\begin{equation}
    \vec{J} = -en(r)u_{z}(r)\hat{z}
\end{equation}
Another point to note is that the inclusion of a uniform background current density is also tractable, 
\begin{equation}
    \vec{J} = \vec{J}_{0} - en_{0}u_{z,0}\frac{r^{2}}{(r + C_{B,T}^{(3)})^{2}}\hat{z}
\end{equation}
This can be seen trivially because the uniform background integrates in the aforementioned manner to produce a tractable magnetic field. One note to make is that in practice, the analytic magnetic field of a cubic vortex contains logarithmic terms which introduce numerical oscillations into the computation of any equilibrium properties. It is arguably better then, in practice, to integrate the magnetic field numerically when these numerical oscillations would otherwise be present in the analytic form. 

\qquad To close the section, let us consider a plasma current density representing the cubic, pure-flow ($\chi = 2$), Bennett vortex but which is described by a flow that includes a uniform background,
\begin{equation}
    u_{z}(r) = u_{0} + u_{z,0}\frac{r^{2}}{(r + C_{B,T})^{2}}
\end{equation}
so that the plasma current density is,
\begin{equation}
    J_{z}(r) = -en_{0}(u_{0} + u_{z,0}\frac{r^{2}}{(r + C_{B,T})^{2}})
\end{equation}
It is worth noting that the above form could be also constructed out of a non-uniform density and a cubic pure-flow profile and the results would be the same, albeit with a complicated non-uniformity describing the density. The exchange of sign from positive to negative has no influence on the tractability of the equations, and will only introduce a change of sign into the relevant parts of the result so the positive is taken here without loss of generality.

\qquad The magnetic field clearly integrates as the uniform background flow becomes a uniform background plasma current when the density is considered uniform, yielding,
\begin{equation}
    B_{\theta}(r) = -en_{0}\mu_{0}\bigg(\frac{u_{0}}{2}r + u_{z,0}\frac{f(r)}{2r(r+C_{B,T})}\bigg)
\end{equation}
Incredibly, the pressure integrates in this case as well, being done here with the Wolfram Mathematica CAS due to the complexity of the expressions,
\begin{align}
    p(r) - p_{0} &= -\mu_{0}e^{2}n_{0}^{2}\bigg(P_{1} + P_{2} + P_{3} + P_{4}\bigg) \\
    P_{1} &= \int_{0}^{r}\frac{u_{0}^{2}}{2}r'dr' \\
    P_{2} &= \int_{0}^{r}u_{0}u_{z,0}\frac{f(r', C_{B,T})}{2r'(r'+C_{B,T})}dr' \\
    P_{3} &= \int_{0}^{r}u_{0}u_{z,0}\frac{r'^{3}}{2(r' + C_{B,T})^{2}}dr' \\
    P_{4} &= \int_{0}^{r}\frac{u_{z,0}^{2}}{2}\frac{r'f(r', C_{B,T})}{(r' + C_{B,T})^{3}}dr'
\end{align}
$P_{1}$ is trivial, and $P_{3}$ is a re-scaled version of the previous cubic integral giving the functional form for the magnetic field of a pure-flow vortex. $P_{2}$, and $P_{4}$ are substantially more complicated. Interestingly, $P_{2}$ introduces a complex pressure, and a Jonquiere function which is typically seen only in quantum statistics rather than classical plasma physics, 
\begin{align}
    P_{2} &= \frac{1}{4} u_0 u_{z0} \Bigg\{ r^2 - 8 r C_{B,T} \\ 
    &\quad - 2 C_{B,T}^2 \Bigg[ \pi^2 - 6 \operatorname{arctanh}\left(\frac{r}{r + 2 C_{B,T}}\right) \nonumber \\
    &\quad + 5 \ln\left(1 + \frac{r}{C_{B,T}}\right) + 6 \ln\left( r \left(1 + \frac{r}{C_{B,T}}\right) \right) \ln(C_{B,T}) \nonumber \\
    &\quad + 6 i \pi \ln\left(\frac{C_{B,T}}{r + C_{B,T}}\right) - 6 \ln(r) \ln(r + C_{B,T}) \nonumber \\
    &\quad - 6 \operatorname{Li}_2\left(1 + \frac{r}{C_{B,T}}\right) \Bigg] \Bigg\} \nonumber
\end{align}
where $Li_{2}(1 + \frac{r}{C_{B,T}})$ is the second-order Jonquiere function (polylogarithm) typically seen in quantum statistics when integrating Fermi-Dirac, or Bose-Einstein, distributions. Its emergence here in a classical, ideal, MHD context is suggestive of a deeper structural connection.

% \section*{Mixed Vorticity}
% \qquad The existence of an equilibrium for the cubic Bennett plasma current implies that the Bennett nonlinearity may be distributed in any way that leaves an analytic current behind. 

\section*{Flow Boundary Conditions}
\qquad Before continuing on past this section to evaluate the shear-flow stabilized character with the Shumlak criterion, Equation (\ref{eqn:shumlak_crit}), we must take a minute to discuss the boundary conditions of these cases. In the cubic, pure-flow ($\chi = 2$) case we have,
\begin{equation}
    u_{edge} = u_{z}(r_{p}) = u_{z,0}\frac{r_{p}^{2}}{(r_{p} + C_{B,T})^{2}}
\end{equation}
The primary wrinkle in the above is the existence of an additional quadratic factor of $u_{z,0}$ inside of the shear layer placement $C_{B,T}$. If this value is sufficiently small, meaning, 
\begin{equation}
    C_{B,T} << r_{p}
\end{equation}
then the flow constant root becomes degenerate,
\begin{equation}
    u_{edge} \simeq u_{z,0}
\end{equation}
Otherwise the boundary condition must be translated into an algebraic system,
\begin{align}
    &u_{edge}(r_{p} + C_{B,T})^{2} - u_{z,0}r_{p}^{2} = 0 \\
    &\therefore u_{edge}(r_{p}^{2} + 2r_{p}C_{B,T} + C_{B,T}^{2}) - u_{z,0}r_{p}^{2} = 0 \\ 
    &\implies Au_{edge}u_{z,0}^{4} + 2Bu_{edge}u_{z,0}^{2} - r_{p}^{2}u_{z,0} + u_{edge}r_{p}^{2} = 0
\end{align}
where
\begin{align}
    A &= \frac{\mu_{0}^{2}e^{4}n_{0}^{2}r_{p}^{6}}{(16k_{B}T_{p})^{2}} \\
    B &= \frac{\mu_{0}n_{0}e^{2}r_{p}^{4}}{16k_{B}T_{p}}
\end{align}
that can technically be solved as all quartic equations can to give four different solutions for $u_{z,0}$. Nothing requires these roots to be real, and a pattern of complex roots shows up in the results presented in the article. The existence of complex roots relates to the structure of these quartic solutions which are based on closed form expression obtained in the 1800s by the Italian mathematician Ruffini\cite{Ruffini1813}. An in-depth exploration of this topic is beyond the scope of this supplement. 

\qquad In the bulk case we have,
\begin{align}
    &u_{edge} = u_{0} \pm u_{z,0}\frac{r_{p}^{2}}{(r_{p} + C_{B,T})^{2}} \\
    &\therefore u_{edge} - u_{0} = \pm u_{z,0}\frac{r_{p}^{2}}{(r_{p} + C_{B,T})^{2}} 
\end{align}
so that the structure of the algebraic system remains largely unchanged except for the substitution implied by the above. Of course, a subtlety is introduced by this, for if $u_{edge} = u_{0}$, then only trivial solutions to the flow roots exist. Evidently, within this equilibrium family it is inadmissible for there to be a symmetry between the edge and core flows, because such a uniformity would cause the shear to collapse between the span of the column as a consequence of the flow stagnation. 

\section*{Minimum Pinch Lengths}
\qquad We must evaluate the Shumlak criterion, Equation (\ref{eqn:shumlak_crit}) for these profiles to determine the minimum length required for the Z-pinch equilibrium to be shear-flow stabilized. This analysis will consider the limit as the plasma radius goes to zero so that the behavior of this property in an arbitrarily small space will be elucidated.  

\qquad There are four primary forms to evaluate this criterion for in this supplement, namely, the cubic, n-form, and their bulk forms. Specifically, we are evaluating the expression,
\begin{equation}
    L > \frac{\pi}{5}(\rho\mu_{0})^{-1/2}\bigg(\dv{u_{z}}{r}\bigg)^{-1}B_{\theta}
\end{equation}
for its value as $\lim_{r\rightarrow 0}$ is taken. This expression can be simplified by considering a length that is normalized by the uniform values on the RHS, and this can include the mass density if we restrict our study to situations which admit the treatment of a uniform density. In principle, this means any plasma current density that describes a Bennett vortex without loss of generality because a vortex with a non-uniform density can be treated as isomorphic to the case of a uniform density and the flow patterns investigated in this article. 

\qquad Then, we study
\begin{align}
    \lim_{r\rightarrow 0} L^{\ast} > \bigg(\dv{u_{z}}{r}\bigg)^{-1}B_{\theta}  
    \\
    L^{\ast} = L / \bigg(\frac{\pi}{5}(\rho\mu_{0})^{-1/2}\bigg)
\end{align}
The influence of a bulk flow will only impact the magnetic field form so we begin with the non-bulk cases, and first amongst them the cubic case,  
% COMPLETE
\begin{align}
    \lim_{r\rightarrow 0}L^{\ast}_{3} &> -\frac{en_{0}\mu_{0}}{u_{z,0}C_{B,T}}\lim_{r\rightarrow 0}\frac{(r + C_{B,T})^{3}}{r}\frac{f(r)}{2r(r+C_{B,T})} \\
    &> -\frac{en_{0}\mu_{0}}{2u_{z,0}C_{B,T}}\lim_{r\rightarrow 0}\frac{(r + C_{B,T})^{2}}{r^{2}}f(r)
\end{align}
Two applications of L'Hopitals rule are required to lift the singularity in the denominator. The limit we wish to evaluate then becomes,
\begin{align}
    \lim_{r\rightarrow 0}2f + 4(r + C_{B,T})f' + (r+C_{B,T})^{2}f''
    = 2f(0) + 4C_{B,T}f'(0) + C_{B,T}^{2}f''(0)
\end{align}
The derivatives that need to be evaluated are,
\begin{align}
    f'(r) &= -6C_{B,T}r + 3r^{2} + \frac{6C_{B,T}^{3}}{r + C_{B,T}} + \frac{6C_{B,T}^{2}r}{r + C_{B,T}} - 6C_{B,T}^{2}\bigg(1 + \ln(\frac{C_{B,T}}{r + C_{B,T}})\bigg) \\
    f''(r) &= -6C_{B,T} + 6r - \frac{6C_{B,T}^{3}}{(r + C_{B,T})^{2}} - \frac{6C_{B,T}^{2}r}{(r + C_{B,T})^{2}} + \frac{12C_{B,T}^{2}}{r + C_{B,T}}
\end{align}
At the origin they become, alongside $f(r)$,
\begin{align}
    f(0) &= 0 - 0 - 6C_{B,T}^{3}\ln(1) - 6\bigg(1 + \ln(1)\bigg)*0*C_{B,T}^{2} = 0 \\ 
    f'(0) &= 0 + 0 + 6C_{B,T}^{2} + 0 - 6C_{B,T}^{2}\bigg(1 + \ln(1)\bigg) = 0\\
    f''(0) &= -6C_{B,T} + 0 -6C_{B,T} + 0 + 12C_{B,T} = 0 
\end{align}
evidently, we find,
\begin{equation}
    \lim_{r\rightarrow 0} L^{\ast}_{3} > 0 
\end{equation}
This limit suggests a divergence analogous to an ultraviolet-type pathology, as the only non-trivial singularity this result can suffer from in the remaining coupling is when $T_{p}\rightarrow \infty$ since this causes the shear structure to collapse, shown here by an explosion of the minimum length required for this equilibrium to assume a shear-flow stabilized state in an arbitrarily small space towards $\infty$. 

\qquad The addition of a bulk flow does not change the structure of the shear but it does change the form of the magnetic field on the RHS of the Shumlak criterion,
\begin{align}
    \lim_{r\rightarrow 0}L^{\ast}_{3,0} &> -\frac{en_{0}\mu_{0}}{u_{z,0}C_{B,T}}\lim_{r\rightarrow 0}\frac{(r + C_{B,T})^{3}}{r}\bigg(\frac{u_{0}}{2}r\pm\frac{f(r)}{2r(r+C_{B,T})}\bigg) \\
    &> -\frac{en_{0}\mu_{0}}{2u_{z,0}}u_{0}C_{B,T}^{2}
\end{align}
This is negative-valued, and when the pinch radius goes to zero it does as well because of the $\sim r_{p}^{6}$ scaling of the RHS. 

\qquad The last cases to consider involve when the power-law of the temperature profile is defined by an arbitrary integer, $n$, 
\begin{equation}
    T(r) = C_{T}^{(n)}r^{n} = \frac{T_{p}}{r_{p}^{n}}r^{n}
\end{equation}
The magnetic field in this case is,
\begin{align*}
    B_{\theta;2,n} &= -\frac{en_{0}\mu_{0}u_{z,0}}{(n-2)^{2}r}\bigg(\frac{(n-2)r^{2n}}{C_{B,T}^{(n)}(C_{B,T}^{(n)}r^{2} + r^{n})} \\
    &\qquad - (-\frac{1}{C_{B,T}^{(n)}})^{-\frac{2}{n-2}}n\beta_{I}(-\frac{r^{n-2}}{C_{B,T}^{(n)}},2+\frac{2}{n-2},0)\bigg)
\end{align*}
where,
\begin{equation}
    \beta_{I}(z;a,b)=\int_{0}^{z}u^{a-1}(1-u)^{b-1}du
\end{equation}
is the incomplete beta function. A plasma pressure was not found when an attempt was made, but this is enough to evaluate the minimum pinch length necesssary for an arbitrarily small Z-pinch equilibrium of this kind to form a shear-flow stabilized state,

\qquad The shear of a pure-flow n-vortex is,
\begin{align}
    u_{z}(r) &= u_{z,0}\frac{T^{2}}{(T + C_{B}n_{0}r^{2})^{2}} = u_{z,0}\frac{1}{(1 + C_{B,T}^{(n)}r^{2-n})^{2}} \\
    \dv{u_{z}}{r}\bigg|_{2,n} &= -2u_{z,0}(2-n)\frac{r^{1-n}}{(1 + C_{B,T}^{(n)}r^{2-n})^{3}}
\end{align}
so we have,
\begin{equation}
    L_{n}(0) > \lim_{r\rightarrow 0} -\frac{\pi}{10}(\rho\mu_{0})^{-1/2}\frac{(1 + C_{B,T}^{(n)}r^{2-n})^{3}}{(2-n)u_{z,0}}r^{n-1}B_{\theta;2,n}(r)  
\end{equation}
expanding out the magnetic field obtains,
\begin{align}
    L_{n}(0) &> -\frac{\pi}{10}(\rho\mu_{0})^{-1/2}\frac{en_{0}\mu_{0}}{C_{B,T}^{(n)}}\lim_{r\rightarrow 0} r^{n-2}\frac{(1 + C_{B,T}^{(n)}r^{2-n})^{3}}{(2-n)(n-2)^{2}}\bigg[\frac{(n-2)r^{2n}}{C_{B,T}^{(n)}r^{2}(C_{B,T}^{(n)}+r^{n-2})} \\
    &\qquad - (-\frac{1}{C_{B,T}^{(n)}})^{-\frac{2}{n-2}}n\beta_{I}(-\frac{r^{n-2}}{C_{B,T}^{(n)}},2 + \frac{2}{n-2},0)\bigg]
\end{align}
In the given limit the beta function will go to zero because the size of the space its kernel is being integrated across does. Importantly, this kernel remains well-defined here because $a > 1$, so there is no objection to be made on the grounds of the kernel losing its regularity. The remaining portion then becomes,
\begin{align}
    L^{\ast}_{n}(0) &> \lim_{r\rightarrow 0} r^{n-2}(1 + C_{B,T}^{(n)}r^{2-n})^{3}\frac{(n-2)r^{2n}}{C_{B,T}^{(n)}(C_{B,T}^{(n)}r^{2}+r^{n})} \\
    &> \frac{n-2}{C_{B,T}^{(n)}}\lim_{r\rightarrow 0}\frac{(C_{B,T}^{(n)}r^{2} + r^{n})^{2}}{r^{2}}\\
    &> 0
\end{align}
which can be evaluated by applying L'Hopitals Rule twice, and where,
\begin{align}
    L^{\ast}_{n} &= \frac{L_{n}}{C_{L,2,n}} \\
    C_{L,2,n} &= -\frac{\pi}{10}(\rho\mu_{0})^{-1/2}\frac{en_{0}\mu_{0}}{C_{B,T}^{(n)}}\frac{1}{(2-n)(n-2)^{2}}
\end{align}
Evidently, every member of this part of the family of shear-flow stabilized Bennett vortices can form in this crucially stable state for arbitrarily small spaces. Note that the above also implies that the weak form of $n = 2$ will also have an arbitrarily small requirement for pinch length. 

\qquad The only remaining case to consider from the original four discussed is when a bulk flow is added to that of an n-vortex. The shear remains unchanged, and as in the previous bulk case we have already shown that the pure-flow contribution will be nothing. That leaves us with,
\begin{equation}
    L_{n,0}(0) > -\frac{e\pi n_{0}\mu_{0}}{5}(\rho\mu_{0})^{-1/2}\lim_{r\rightarrow 0} \bigg(\dv{u_{z}}{r}\bigg)^{-1}\bigg(\frac{u_{0}}{2}r \bigg)
\end{equation}
which boils down to,
\begin{equation}
    \tilde{L}_{n,0}(0) > \lim_{r\rightarrow 0}\frac{(r^{n}+C_{B,T}^{(n)}r^{2})^{3}}{r^{2n}}
\end{equation}
This limit is indeterminate for general n, and shear layer. However, as can be seen previously from the cubic case this limit will evaluate for certain values of $n$, e.g., $n = 3$. In this specific case, it can also be seen that an arbitrarily small pinch length is sufficient for such an equilibrium to be shear-flow stabilized when the pinch radius is arbitrarily small as this will kill the quadratic term in the above and leave a single factor of $r^{n}$ on the RHS. 

\qquad A trivial shear layer also corresponds to the ultraviolet limit, and not just an infinitesimally vanishing pinch radius, so that is another aspect to consider because it is suggestive of the possibility that this equilibrium could possibly explain how the aftermath of the Big Bang stabilized into a quark-gluon plasma that then cooled and eventually condensed into ordinary matter. Letting the temperature run to infinity admits the possibility that an arbitrarily small pattern of this kind of flow could establish itself with a thermal lifetime and persist to interact with its fellows if there were a plasma state setting the stage for these events. What is interesting to close this section with as an observation, is that the same is not true for the cubic case since there the ultraviolet limit will cause the length to become indeterminate so that some relaxation process would first be necessary for the universe to cool off before these lower-order vortices could form. 

\section*{Thermal Structure}
\qquad The classic Bennett pinch treats a uniform plasma temperature and therefore matters of heat flux, thermal confinement time, and thermal energy go trivially unanswered. By itself a cubic Bennett vortex represents a steady-state heat flux distribution that can be treated as producing a net volumetric heating through its divergence,
\begin{equation}
    \nabla\cdot\vec{q} = Q 
\end{equation}
A Braginskii type closure\cite{Braginskii1965} will study the dressed thermal transport properties in greater detail than is required at a first pass to understand the basic thermal structure of the cubic vortex and how it responds to unsteady perturbations. For this first pass we consider a uniform perpendicular conductivity,
\begin{equation}
    \vec{q} = -\kappa_{\perp}\nabla T
\end{equation}
The cross-field, Righi-Leduc term,
\begin{equation}
    \vec{q}_{\wedge} = -\kappa_{\wedge}\hat{b}\times\nabla T = \kappa_{\wedge}\dv{T}{r}\hat{z}
\end{equation}
shows that an axial flux of heat will develop, but this heat will remain confined to the axis if the system is axially symmetric, and its transport properties uniform, because then there will be no divergence of the heat. 

\qquad With a uniform thermal conductivity we have in steady state,
\begin{equation}
    -\kappa_{\perp}\nabla^{2}T = Q
\end{equation}
The steady volumetric heating provided by an n-vortex,
\begin{equation}
    T_{n}(r) = \frac{T_{p}}{r_{p}^{n}}r^{n} = C_{T}^{(n)}r^{n}
\end{equation}
can be evaluated to give,
\begin{align}
    Q_{n}(r) &= -\frac{\kappa_{\perp}}{r}\dv{}{r}\bigg(r\dv{T_{n}}{r}\bigg) \\
    &=-\frac{\kappa_{\perp}}{r}\bigg(C_{T}^{(n)}nr^{n-1} + C_{T}^{(n)}n(n-1)r^{n-1}\bigg) \\
    &= -\kappa_{\perp}\frac{T_{p}}{r_{p}^{n}}r^{n-2}\bigg(n + n(n-1)\bigg) \\
    &= -\kappa_{\perp}\frac{T_{p}}{r_{p}^{n}}n^{2}r^{n-2}
\end{align}
Integrating over the entire plasma volume gives the thermal power of the vortex,
\begin{equation}
    -\int_{V_{p}}\kappa_{\perp}\nabla^{2}T_{n}dV = \int_{V_{p}}Q_{n}dV = S_{n}
\end{equation}
For an n-vortex we have,
\begin{align}
    S_{n} &= -\kappa_{\perp}2\pi L\int_{0}^{r_{p}}\dv{}{r}\bigg(r\dv{T_{n}}{r}\bigg)dr \\
    &= -2\pi L\kappa_{\perp}T_{p}\frac{1}{r_{p}^{n}}\int_{0}^{r_{p}}n^{2}r^{n-1}dr \\
    &= -2 \pi L\kappa_{\perp}T_{p}n^{2}\frac{1}{r_{p}^{n}}\frac{r_{p}^{n}}{n}\\
    &= - 2\pi nL\kappa_{\perp}T_{p}
\end{align}
after the integration is performed. Amongst the strong-form solutions the total thermal power is then minimized for the cubic temperature case when $n = 3$ and the system is axially symmetric with a uniform thermal conductivity. Parabolic temperatures have less thermal power, but they are only weakly shear-flow stabilized and have mathematically zero shear.

\qquad The energy confinement time\cite{Freidberg_2007} is defined according to the pressure and heat flux at the pinch boundary,
\begin{equation}
    \frac{3}{2}\frac{p_{0}}{\tau_{E}} = \frac{1}{V_{p}}\int_{A_{L}}\vec{q}\cdot d\vec{A}_{L}
\end{equation}
For a cubic vortex it can be found to be,
\begin{equation}
    \label{eqn:cubic_tauE}
    \tau_{E} = \frac{1}{12}\frac{p_{0}}{\kappa_{\perp}(r_{p}) T_{p}}r_{p}^{2}
\end{equation}
where the plasma pressure at the core of the pinch can be solved for by taking the pressure at the boundary to be equal to the vacuum pressure. In the pure-flow, $\chi = 2$ case with a pure vacuum this yields, 
\begin{align}
    p_{0} &= \frac{\mu_{0}(en_{0}u_{z,0})^{2}}{2}\int_{0}^{r_{p}}\frac{r}{(r+C_{B,T})^{3}}f(r, C_{B,T})dr \\
    &= \frac{\mu_{0}(en_{0}u_{z,0})^{2}}{2(r_{p} + C_{B,T})^{2}}\bigg(r_{p}(2C_{B,T} + r_{p})(r_{p}^{2} - 12C_{B,T}r_{p} - 15C_{B,T}^{2}) \\
    &\qquad +6C_{B,T}^{2}(C_{B,T}+r_{p})\ln(\frac{C_{B,T}}{r_{p}+C_{B,T}})((C_{B,T} + r_{p})\ln(\frac{C_{B,T}}{r_{p}+C_{B,T})}) \\
    &\qquad -3r_{p}-5C_{B,T})\bigg) \nonumber
\end{align}
at the plasma edge the (perpendicular) thermal conductivity is\cite{Braginskii1965},
\begin{align}
    \kappa_{\perp} &= 4.7\frac{n_{0}k_{B}T_{e}}{m_{e}\omega_{ce}^{2}\tau_{e}}\\ 
    % &= 4.7\frac{\ln(\Lambda_{c})\sqrt{m_{e}}e^{2}}{6\sqrt{2}\pi^{3/2}\epsilon_{0}^{2}}\frac{n_{0}^{2}}{B_{max}^{2}T_{p}^{1/2}}
    % &= 4.7 \frac{\Lambda_{C}\sqrt{m_{e}}e^{2}4\sqrt{2\pi}}{3}\frac{n_{0}^{2}}{B(r_{p})^{2}\sqrt{k_{B}T_{p}}}
\end{align}
and the maximum magnetic field, which occurs at the edge of the plasma, is,
\begin{align}
    B_{\theta}(r_{p}) &= B_{max} \\
    &= \frac{\mu_{0}en_{0}u_{z,0}}{2r_{p}(r_{p}+C_{B,T})}f(r_{p}, C_{B,T})
\end{align}

\qquad Because of the linear nature of the source and sink terms in the power balance of the plasma, any collection of sources and sinks, e.g., fusion power, Ohmic heating, Bremsstrahlung losses, etc. can be represented by an n-vortex whose temperature defines the natural scale of the net conductive flux of heat at the plasma edge. Equating the net volumetric power output of these sources and sinks with the net flux of conductive thermal power from the plasma at the pinch radius, we have,
\begin{align}
    &\int_{A_{L}}\vec{q}\cdot d\vec{A}_{L} = V_{p}\sum_{k}S_{k} \\
    &\rightarrow 2\pi Lr_{p}\kappa_{\perp}(r_{p})\frac{T_{v}}{r_{p}^{n}}nr_{p}^{n-1} = \pi r_{p}^{2}L\sum_{k}S_{k} \\
    &\therefore T_{v} = \frac{r_{p}^2}{2n\kappa_{\perp}(r_{p})}\sum_{k}S_{k}
\end{align}
where the volumetric power density of the $kth$ source or sink is $S_{k}$, and the temperature gradient at the plasma edge is,
\begin{equation}
    \dv{T}{r}\bigg|_{r=r_{p}} = \frac{T_{v}}{r_{p}^{n}}nr_{p}^{n-1} = \frac{T_{v}}{r_{p}}n
\end{equation}

\section*{Plasma Filament Formation}
The plasma being a circular disk of uniform density, the impact of its gravitational forces measured along the positive half-chord of the pinch cross-section can be calculated analytically by treating the gravitational forces along the cylinder of mass in the classical sense, i.e., as if all the mass in a given shell were concentrated at the center of the plane, and a central-force was pulling on the test mass, $m$, of plasma at some point, $r$. The mass $M$ of the shell is given by,
\begin{align}
    M &= 2\pi L\int_{0}^{r} r'\rho dr' \\
    &=L\pi r^{2}mn_{0} 
\end{align}
and the gravitational force on the test mass, $m$, in the lab frame, 
\begin{equation}
    \vec{F}_{G} = -\frac{GM(r)m}{r^{2}}\hat{r}
\end{equation}
The ratio of the gravitational force to the electromagnetic forces felt by the test plasma mass in the lab frame can be calculated for the case $T(r) = C_{T}r^{3}$ since the entire equilibrium of a Bennett Vortex can be for this situation. The radial electric field felt by the test plasma mass in its rest frame can necessarily be obtained as well, from the ideal version of Ohm's Law,
\begin{equation}
    \vec{E} = -\vec{u}\times\vec{B} = E_{r}\hat{r} = \frac{r}{(r + C_{B,T})^{3}}\mu_{0}en_{0}u_{z,0}^{2}f(r, C_{B,T})    
\end{equation}
If we use this electric field in a naive attempt to calculate the ratio of these two forces we will find that the electromagnetic force is considered to have no impact on the dynamics of the plasma mass, and this is by definition. Instead, to restore the impact of the electromagnetic force we must boost to a lab frame travelling at a velocity, $\vec{u}_{lab}$, 
\begin{align}
    &\vec{u}_{lab} \times \vec{B} = -2\vec{u}\times\vec{B} \\
    &\therefore \vec{u}_{lab} = -2\vec{u}
\end{align}
Doing so amounts to finding a frame of reference where the electric force adds constructively with the magnetic force instead of cancelling it out fully, as is done in the rest frame of the plasma. Critically, the system remains axisymmetric from this new frame of reference,
\begin{align}
    z_{lab} &= z - 2u_{z}(r)t \\
    \therefore \pdv{q(r)}{z_{lab}} &= \pdv{z}{z_{lab}}\pdv{q}{z} = \pdv{q}{z} = 0
\end{align}
and furthermore, this frame of reference is also inertial as the plasma flow is steady.

\qquad In principle any scaling factor could be coupled to the boost for performing this calculation, subsequently changing the value of the electromagnetic forces seen by the lab observer, but doing so would cause the amplitude of the lab electric field to grow or shrink from its value in the rest frame of the plasma. Furthermore, applying too large a scaling factor, or consider too large an edge flow speed, and the relativistic effects on the observed mass from the lab frame will need to be addressed.
\\ 
\indent 
When measured in this frame, the electric field experienced by the plasma mass becomes,
\begin{align}
    \vec{E}_{lab}=\vec{u}\times\vec{B}
\end{align}
and the electromagnetic forces in the lab frame become,
\begin{align}
    \vec{F}_{L,lab} &= e(\vec{E}_{lab} + \vec{u}\times\vec{B}) \\ 
    &= 2e(\vec{u}\times\vec{B}) \\
    &= F_{L,lab,r}\hat{r} \\
    &= -2eu_{z}B_{\theta}\hat{r}
\end{align}
Defining the ratio of the gravitational force to this lab electromagnetic force as,
\begin{equation}
    \eta_{GL} = \frac{\int_{V}|\vec{F}_{G}(r)|dV}{\int_{V}|\vec{F}_{L,lab}(r)|dV} = \frac{I_{G}}{I_{EM}} 
    % \label{eqn:eta_GL}
\end{equation}
we have, first, the electromagnetic integral,
\begin{align}
    I_{EM} &= 2\pi Le^{2}u_{z,0}^{2}n_{0}\mu_{0}\int_{0}^{r}\frac{f(r', C_{B,T})}{(1 + \xi^{2}r'^{2})^{4}(r' + C_{B,T})}dr' \notag \\
    &= 2\pi Le^{2}u_{z,0}^{2}n_{0}\mu_{0}(L_{1} + L_{2} + L_{3} + L_{4}) 
\end{align}
where,
\begin{equation}
    L_{1} = \int_{0}^{r}\frac{f_{1}(r', C_{B,T})}{(1+\xi^{2}r'^{2})^{4}(r' + C_{B,T})}dr' \label{eqn:Iem_L1}
\end{equation}
\begin{equation}
    L_{2} = \int_{0}^{r}\frac{f_{2}(r', C_{B,T})}{(1+\xi^{2}r'^{2})^{4}(r' + C_{B,T})}dr' \label{eqn:Iem_L2}
\end{equation}
\begin{equation}
    L_{3} = \int_{0}^{r}\frac{f_{3}(r', C_{B,T})}{(1+\xi^{2}r'^{2})^{4}(r' + C_{B,T})}dr' \label{eqn:Iem_L3}
\end{equation}
\begin{equation}
    L_{4} = \int_{0}^{r}\frac{f_{4}(r', C_{B,T})}{(1+\xi^{2}r'^{2})^{4}(r' + C_{B,T})}dr' \label{eqn:Iem_L4}
\end{equation}
These integrals, Equations (\ref{eqn:Iem_L1}) - (\ref{eqn:Iem_L4}), are solved using the Wolfram Mathematica CAS. The same techniques can be used for this as were used for evaluating the pressure integrals, u-substitution and integration by parts for when the rational functions are coupled to a logarithmic term. What is important is their leading order when added together,
\begin{equation}
    I_{EM} \sim \frac{r^{7}}{(r+C_{B,T})^{4}} \sim r^{3}
\end{equation}
and the leading order of the gravitational integral,
\begin{align}
    I_{G} &= 2\pi LGm\int_{0}^{r}\frac{1}{r'}M(r')dr' \\
    &= 2\pi^{2} L^{2}Gm^{2}n_{0}\int_{0}^{r}r'dr' \\
    &= (\pi^{2} L^{2}Gm^{2}n_{0})r^{2}
\end{align}
Together we find that the ratio, Equation (\ref{eqn:eta_GL}) is then given by,
\begin{equation}
    \eta_{GL} \sim \frac{L^{2}r^{2}}{Lr^{3}} \sim \frac{L}{r}
\end{equation}
for this particular, non-relativistic, case where $T = C_{T}r^{3}$.

\qquad For a relativistic flow in a flat spacetime the pinch length, but not radius because there is no radial flow treated in this frame, contracts in the flow frame to a new length, $L'$, 
\begin{equation}
    L' = \frac{1}{\gamma(v)}L = L\sqrt{1 - \frac{v^{2}}{c^{2}}}
\end{equation}
In the ultra-relativistic limit, $v \approx c$, which is naturally entered for sufficiently high energy particles or when the current grows sufficiently large in a plasma whose electrons are carrying it, then this length goes to zero. Ultra-relativistic particles in this equilibrium then would not see a pinch at all as they simply "jump" to the end of it once borne. This suggests a mechanism by which the structure can form, namely, high-energy electrons which propagate on the fastest timescale in a plasma. If fast electrons were to assume the equilibrium, then the rest of the plasma will have no choice but to move collectively with them for the proper duration of the structure's thermal lifetime. 

\qquad The collapse of the pinch length in the fast electron frame suggests that a flat spacetime cannot support a continuous axial structure of this kind, indicating a preference for localized growth of filaments as the current density is amplified in a very narrow region by the shear structure of this equilibrium. When such amplification occurs, which can happen freely across scales ranging from small laboratory plasmas to very large astrophysical systems depending on the characteristics of the shear layer, this provides a unifying mechanism for the appearance of filaments. 

\qquad Where this is challenged is if the length of the structure in the slow frame is very long as then accounting for the strict difference between the local speed of light and fast electron speed will give a finite pinch length in the fast frame. This also highlights the importance of accounting for the gravitational forces instead of neglecting them as otherwise it would be unclear what impact this length has on the overall impact a test plasma mass experiences. What meets these challenges is both the minimal energy of this Z-pinch equilibrium, and its shear-flow stabilized condition, as these force the growth of the electron plasma current into channels which can be arbitrarily narrow depending on the local state of the plasma. 

\section*{Spherical Bennett Vorticity}
\qquad It is valuable to study what form this flow takes in a spherical basis. Transforming from a cylindrical to spherical basis results in,
\begin{align}
    u_{\rho} &= u_{z}(\rho, \phi)\cos(\phi) \\
    u_{\theta} &= 0 \\
    u_{\phi} &= -u_{z}(\rho, \phi)\sin(\phi)
\end{align}
where the relationship,
\begin{align}
    &\rho^{2} = r^{2} + z^{2} = r^{2} + \rho^{2}\cos^{2}(\phi) \\
    &\therefore r^{2} = \rho^{2}(1 - \cos^{2}(\phi))
\end{align}
reveals that the axisymmetric cylindrical current density we have been studying suffers a broken symmetry in the spherical basis that is identified with the polar dependence necessary for this 2D flow to transform back into the 1D axial model. Interestingly, we see that the azimuthal symmetry is preserved. This is suggestive of the reason why the axial symmetry of the equilibrium is relaxable in the case of a spherical air plasma streamer head, because this axial symmetry is a phantom symmetry that breaks when the cylindrical basis is transformed into spherical.  

\qquad The current density then stands as,
\begin{equation}
    \vec{J} = J_{\rho}(\rho,\phi)\hat{\rho} + J_{\phi}(\rho, \phi)\hat{\phi}
\end{equation}
which is the basic structure on which the magnetostatic field rests,
\begin{align}
    &\curl\vec{B} = \mu_{0}\vec{J} \\
    &\implies \frac{1}{\rho\sin\theta}\bigg(\pdv{(\sin\theta B_{\phi})}{\theta} - \pdv{B_{\theta}}{\phi}\bigg)\hat{\rho}
    + \frac{1}{\rho}\bigg(\pdv{(\rho B_{\theta})}{\rho} - \pdv{B_{\rho}}{\theta}\bigg)\hat{\phi} \\
    &\therefore \bigg(\curl\vec{B}\bigg)_{\theta} = 0 \implies \frac{1}{\rho\sin\theta}\pdv{B_{\rho}}{\phi} = \frac{1}{\rho}\pdv{(\rho B_{\phi})}{\rho} 
\end{align}
The above illustrates clearly that the azimuthal magnetic field splits into a radial and polar component with fully broken symmetries while the lack of azimuthal current in the magnetostatic case results in a coupling between the radial and polar magnetic field gradients. A full 3D treatment of the magnetic field is then required to resolve it from this perspective. 

\section*{Resistive Bennett Vorticity}
\qquad The incorporation of a finite resistivity is seamless into the power balance, as the electrical power of a uniform cylindrical ideal plasma is expressable as,
\begin{equation}
    P = I_{encl}^{2}R = I_{encl}^{2}\frac{L}{\pi r_{p}^{2}}\eta 
\end{equation}
which is calculable for an arbitrary pure-flow vortex with pinch radius, $r_{p}$,
\begin{equation}
    I_{encl} = \int_{A_{p}}\vec{J}\cdot\vec{dA_{p}}
\end{equation}
and usage of the Spitzer resistivity\cite{Spitzer1950}, 
\begin{equation}
    \eta_{sp} = \frac{4\sqrt{2\pi}}{3}\frac{Z_{eff}e^{2}m_{}^{1/2}\ln(\Lambda_{c})}{(4\pi\epsilon_{0})^{2}(k_{B}T_{e})^{3/2}}
\end{equation}
where
\begin{equation}
    Z_{eff} = \frac{\sum_{j}Z_{j}^{2}n_{j}}{n_{e}}
\end{equation}
and $Z_{j}$ is the charge (ionization) state of the $j$-th ionic species.
\\
\indent 
The energy confinement time\cite{Freidberg_2007} is defined according to the pressure and heat flux at the pinch boundary,
\begin{equation}
    \frac{3}{2}\frac{p_{0}}{\tau_{E}} = \frac{1}{V_{p}}\int_{A_{L}}\vec{q}\cdot d\vec{A}_{L}
\end{equation}
For a cubic vortex it can be found to be Equation (\ref{eqn:cubic_tauE}),
\begin{equation}
    % \label{eqn:cubic_tauE}
    \tau_{E} = \frac{1}{12}\frac{p_{0}}{\kappa(r_{p}) T_{p}}r_{p}^{2}
\end{equation}
where the plasma pressure at the core of the pinch can be solved for by taking the pressure at the boundary to be equal to the vacuum pressure. In the pure-flow, $\chi = 2$ case with a pure vacuum this yields, 
\begin{align}
    p_{0} &= \frac{\mu_{0}(en_{0}u_{z,0})^{2}}{2}\int_{0}^{r_{p}}\frac{r}{(r+C_{B,T})^{3}}f(r, C_{B,T})dr \\
    &= \frac{\mu_{0}(en_{0}u_{z,0})^{2}}{2(r_{p} + C_{B,T})^{2}}\bigg(r_{p}(2C_{B,T} + r_{p})(r_{p}^{2} - 12C_{B,T}r_{p} - 15C_{B,T}^{2}) \\
    &\qquad +6C_{B,T}^{2}(C_{B,T}+r_{p})\ln(\frac{C_{B,T}}{r_{p}+C_{B,T}})((C_{B,T} + r_{p})\ln(\frac{C_{B,T}}{r_{p}+C_{B,T})}) \\
    &\qquad -3r_{p}-5C_{B,T})\bigg) \nonumber
\end{align}
at the plasma edge the (perpendicular) thermal conductivity is\cite{Braginskii1965},
\begin{align}
    \kappa_{\perp} &= 4.7\frac{n_{0}k_{B}T_{e}}{m_{e}\Omega_{e}^{2}\tau_{ee}}\\ 
    % &= 4.7\frac{\ln(\Lambda_{c})\sqrt{m_{e}}e^{2}}{6\sqrt{2}\pi^{3/2}\epsilon_{0}^{2}}\frac{n_{0}^{2}}{B_{max}^{2}T_{p}^{1/2}}
    &= 4.7 \frac{\Lambda_{C}\sqrt{m_{e}}e^{2}4\sqrt{2\pi}}{3}\frac{n_{0}^{2}}{B(r_{p})^{2}\sqrt{k_{B}T_{p}}}
\end{align}
and the maximum magnetic field, which occurs at the edge of the plasma, is,
\begin{align}
    B_{\theta}(r_{p}) &= B_{max} \\
    &= \frac{\mu_{0}en_{0}u_{z,0,}}{2r_{p}(r_{p}+C_{B,T})}f(r_{p}, C_{B,T})
\end{align}
\indent The Alfven time is another timescale which is of great importance to this kind of equilibrium as it describes the time it takes for magnetic energy, which can be of arbitrary shape, to be transported across a given lengthscale,
\begin{equation}
    \tau_{A} = \frac{L^{\ast}}{V_{A}}
\end{equation}
A plasma pinch has two natural lengthscales, the first being the pinch radius, $r_{p}$, and the second being the pinch length, $L$, as it is measured in a frame at rest with respect to the pinch. This suggests two Alfvenic timescales to consider,
\begin{equation}
    \tau_{A;p} = \frac{r_{p}}{V_{A}}
\end{equation}
and,
\begin{equation}
    \tau_{A;L} = \frac{L}{V_{A}}
\end{equation}
where the Alfven speed is of course,
\begin{equation}
    V_{A} = \frac{B}{\sqrt{\rho\mu_{0}}}
\end{equation}
We should consider the ratio of the confinement time against this time,
\begin{equation}
    R_{EA} = \frac{\tau_{E}}{\tau_{A}}
\end{equation} 
% COMPLETE

\section*{Viscous (Incompressible) Bennett Vorticity}
\qquad A Louiville equation shows how the impact of viscosity will influence the structure of the plasma dynamics in general, and with the full momentum equation being for a plasma species $s$ interacting visco-collisionally with a background $s'$,
\begin{equation}
    m_{s}n_{s}\bigg(\pdv{\vec{u}}{t} + \vec{u}\cdot\nabla\vec{u}\bigg) = \vec{J}\times\vec{B} + \nabla\cdot\uuline{\Pi}+ \sum_{s'}m_{s}n_{s}\nu_{ss'}(\vec{u}_{s'}-\vec{u}_{s})
\end{equation}
In the case of the shear-flow stabilized Z-pinch equilibrium, the flow is purely axial,
\begin{equation}
    \vec{u} = u_{z}(r)\hat{z}
\end{equation}
so for incompressible flow where the viscous stress constitutes a divergence with an isotropic viscosity,
\begin{equation}
    (\nabla\cdot\uuline{\Pi})_{z} = \partial_{j}\Pi_{zj} = \mu_{v}\partial_{j}\partial_{j}u_{z} = \mu_{v}\frac{1}{r}\dv{(r\pdv{u_{z}}{r})}{r}
\end{equation}
then a first-pass Newtonian closure for a singly-ionized plasma presents a buildup of electric field on the z-axis, balanced by viscous momentum transfer in the absence of collisions, 
\begin{align}
0 &= q_{s}n_{0}E_{z} + (\nabla\cdot\uuline{\Pi})_{z} \\
&= q_{s}n_{0}E_{z} + \mu_{v}\frac{1}{r}\dv{(r\pdv{u_{z}}{r})}{r}
\end{align}

\section*{Drifts}
\qquad The very serious question of plasma drifts must also be considered as magnetization of the plasma brings on cyclotron emission, and leads to guiding-center drifts orthogonal to both the perpendicular electric, and magnetic fields of the plasma by introducing a finite ordering to the guiding-center motion based on the cyclotron frequency. Additionally, these equilibria have a magnetic null at the axis of the pinch so that there will be a region in which the magnetized cyclotron orbits transition to betatron orbits where the increasing nature of the magnetic field is what is important, rather than its amplitude. Meaning that, in general, a greater range of drifts must be considered than just those resulting from magnetic effects. 

\qquad Drifts resulting from the isotropic pressure gradient, the non-uniform electromagnetic field of the vortex, and a possible gravitational field will be considered. No appreciable curvature drift will be present in the idealized equilibrium because there is no plasma flow considered to be occurring parallel to the magnetic field. Likewise there will be no polarization drift because the electric field is considered to be steady, although in the presence of a growth mode for a strong accelerating electric field this would not be the case. For axial drifts this leaves the grad-B drift, 
\begin{align}
    \vec{v}_{\nabla B, j} &= \pm\frac{1}{2}v_{\perp}\rho_{L}\frac{\vec{B}\times\vec{\nabla} B}{B^{2}} \\
    &= \pm\frac{1}{2}v_{\perp}^{2}\frac{1}{\omega_{cj}}\frac{B_{\theta}\hat{\theta}\times B'_{\theta}(r)\hat{r}}{B^{2}} \\
    &= \mp\frac{1}{2}m_{j}v_{\perp}^{2}\frac{1}{|q_{j}|}\frac{B'_{\theta}}{B^{2}}\hat{z} \\
    &= \mp\frac{K_{\perp,j}}{|q_{j}|B^{2}}B_{\theta}'\hat{z} \\
\end{align}
the ExB drift with a non-uniform electric field,
\begin{align}
    \vec{v}_{E} &= (1 + \frac{1}{4}\rho_{L}^{2}\nabla^{2})\frac{\vec{E}\times\vec{B}}{B^{2}} \\
    &= (1 + \frac{\rho_{L}^{2}}{4}\nabla^{2})\frac{(-u_{z}\hat{z}\times B_{\theta}\hat{\theta})\times B_{\theta}\hat{\theta}}{B^{2}} \\
    &=  (1 + \frac{\rho_{L}^{2}}{4}\nabla^{2})\frac{u_{z}B_{\theta}\hat{r}\times B_{\theta}\hat{\theta}}{B^{2}} \\
    &= (1 + \frac{\rho_{L}^{2}}{4}\nabla^{2})u_{z}\hat{z} \\
\end{align}
and the diamagnetic drift,
\begin{align}
    \vec{v}_{D} &= \frac{1}{|q_{j}|n}\frac{-\dv{p}{r}\hat{r}\times B_{\theta}\hat{\theta}}{B^{2}} \\
    &= \frac{1}{|q_{j}|n}\frac{J_{z}B_{\theta}\hat{r}\times B_{\theta}\hat{\theta}}{B^{2}} \\
    &= -\frac{1}{ne}enu_{z}\hat{z} \\
    &= -u_{z}\hat{z} \\
\end{align}
where,
\begin{align}
    K_{\perp} &= \frac{1}{2}mv_{\perp}^{2} \\
    \rho_{L,j} &= \frac{v_{\perp,j}}{\omega_{cj}} \\
    \vec{E} &= -\vec{u}\times\vec{B}
\end{align}
and $\rho_{L,j}$ is the Larmor radius of the $jth$ plasma species. If this scale is small compared to the length-scale over which gradients in the plasma exist, then for a singly-ionized plasma with a current carried by the electrons we find the electric drift and diamagnetic drifts cancel in the present equilibrium, leaving the behavior in this regime dependent on the remaining drifts.

\subsection*{Waterfall Drifts}
\qquad If gravity is considered then we must first, while recognizing our discussion remains firmly rooted in a flat spacetime, choose its setting. The most important aspect of this choice is its situation relative to the pinch axis, and the source of the field. If the field is parallel to the pinch axis, and uniform, so that the pinch is thought of as being located on the surface of a massive body which the gravitational field is due to, then we can say this configuration looks something like a waterfall, and the drift is,
\begin{align}
    \vec{v}_{wf} &= \frac{m_{j}}{q_{j}}\frac{g_{0}\hat{z}\times B_{\theta}\hat{\theta}}{B^{2}} \\
    &= -\frac{g_{0}}{\omega_{cj}}\hat{r}
\end{align}
the weakly-magnetized interior of these vortices will result in a large such drift then near the core as indicated in the main article, and additionally because their magnetic field points in the opposite direction to $\hat{\theta}$ this drift will travel outwards rather than inwards as suggested above,
\begin{equation}
    \vec{v}_{wf} = \frac{g_{0}}{\omega_{cj}}\hat{r}
\end{equation}
Of course, this is the collisionless picture, and the weak magnetic field in the core of these shear-flow stabilized Z-pinch equilibria suggests that if the pinch axis of the discharge was aligned with the local gravitational field of a massive body on its surface, then significant radial drifts would be excited in these regions to transport matter radially in the plasma from the gravitational influence of the massive body. In a thermal plasma reactor where excited electrons are available to react with pollutant ions from a background discharge plasma, this process could be used to ionize, and transport industrial effluence out of the impurity gas for deposition on a material substrate as part of a sequestration pipeline. 

\qquad The primary issue to address is how collisions influence this picture. In the absence of viscosity, or unsteady Eulerian effects, then concentrating on a plasma velocity that is made up purely of a radial "waterfall" component and the axial vortex flow presented in this article,
\begin{equation}
    \vec{u} = u_{wf}(r)\hat{r} + u_{z}(r)\hat{z}
\end{equation}
we have to contend with the existence of convective nonlinearities at the fluid scale. However, considering the force balance on a single particle where the effect of collisions are described with a drag term we have,
\begin{equation}
    0 = q_{s}(\vec{E} + \vec{u}\times\vec{B}) + m_{s}g_{0}\hat{z} - m_{s}\nu\vec{v}
\end{equation}
Taking the electric field to be zero in the above as a consequence of plasma screening we have,
\begin{align}
    0 &= -q_{s}u_{z}B_{\theta} - m_{s}\nu u_{r} \\
    0 &= q_{s}u_{r}B_{\theta} + m_{s}g_{0} - m_{s}\nu u_{z}
\end{align}
From here we can obtain,
\begin{equation}
    u_{r}\bigg(1 + (\frac{m_{s}\nu}{q_{s}B_{\theta}})^{2}\bigg) = -\frac{g_{0}}{\omega_{cj}}
\end{equation}
However, recall that the magnetic field of this kind of equilibrium points in the $-\hat{\theta}$ direction instead of the $+$ direction as indicated in the above. Incorporating this additional negative sign we then arrive at,
\begin{equation}
    u_{r} = g_{0}\frac{\omega_{cj}}{\nu^{2} + \omega_{cj}^{2}} = v_{wf}^{(coll.)}(r)
\end{equation}
From the above it is apparent that,
\begin{align}
    \nu >> \omega_{cj} &\implies u_{r}\rightarrow 0\\
    \nu << \omega_{cj} &\implies u_{r} \implies\frac{g_{0}}{\omega_{cj}}
\end{align}
Leaving aside the exact form of $\nu$ for the moment from this over-simplified picture we can imagine a potentially viable scenario for this application whereby highly-collisional regions in a plasma made from a broken-down pollutant gas stream give rise to appreciable quantities of pollutant ions that remain confined in the plasma as long as they continue to react with the background. However, as these pollutant ions travel down the length of the pinch in a collisionless manner, i.e., on a timescale that is shorter than the collisional one, then they will find themselves travelling outwards at significantly large speeds as shown quantitatively in the article, and this permits the possibility of escape from the plasma if they are travelling fast enough for a small enough pinch.  
% COMPLETE

\subsection*{Resistive Drifts}
\qquad Waterfall drifts are not the only radial drifts possible. When a finite resistivity is introduced then an Ohmic electrostatic field arises from the axial plasma current density,
\begin{equation}
    E_{z} = \eta_{\perp}J_{z}
\end{equation}
which leads to a plasma drift of the form,
\begin{equation}
    \vec{v}_{\eta} = \frac{1}{q_{s}}\frac{q_{s}E_{z}\hat{z}\times B_{\theta}\hat{\theta}}{B^{2}}
\end{equation}
Again, the azimuthal magnetic field has a lurking negative sign in the definition so we arrive at,
\begin{equation}
    \vec{v}_{\eta} = \eta_{\perp}\frac{J_{z}}{B_{\theta}}\hat{r}
\end{equation}
The above expression admits the possibility for the existence of a vortical configuration which has a large axial current near the axis of the pinch, while still possessing a sufficiently small magnetic field thereby exciting large radial drifts in this region. One quantitative example of this is shown in the main article.  

\section*{Unsteady Bennett Vorticity}
\qquad From the perspective of mathematical physics, it is worth exploring possible extensions to the equilibrium which permit its study in an unsteady context, as well as a non-exhaustive list of systems for which the study would potentially be fruitful. Two systems which exemplify the properties that would potentially lead to fruitful study are the Korteweg-de-Vries\cite{KdV1895} (KdV),
\begin{equation}
    \partial_{t}u + \partial_{z}^{3}u - 6u\partial_{z} u= 0 
\end{equation}
and Kuramoto-Sivashinsky\cite{Kuramoto1976}\cite{Sivashinsky1977} (KS) equations, either in 1D,
\begin{equation}
    \partial_{t}u + \partial_{z}^{2}u + \partial_{z}^{4}u + \frac{1}{2}(\partial_{z}u)^{2} = 0
\end{equation}
or the fully biharmonic 3D system,
\begin{equation}
    \partial_{t}u + \nabla^{2}u + \nabla^{4}u + \frac{1}{2}|\nabla u|^{2} = 0 
\end{equation}
What makes these systems exemplary is their first-order, linear temporal structure while possessing spatial structures which admit highly nonlinear effects. The relaxability of the axial symmetry is a boon in this regard as modulations to one or more of the parameters necessary to describe the vortex can be used to connect this unsteady character to, e.g., the axial transport of the vortex, or even the full 3D flow field in the case of KS. This observed relaxibility is also the basis for the form of the spatial derivatives presented in the above 1D equations. In the KS case we can derive a consistent axisymmetric, cylindrical system from the fully 3D, biharmonic system. In the KdV case we can also treat the radial direction in 1D, but this must be done delicately due to the cylindrical geometry implied by the radius, and further work in this regard is outside the scope of this supplement.

\qquad The main question in regard to an unsteady vortex is how the time-dependence is modulated in the underlying flow field. There are of course other nonlinear systems beyond the aforementioned which exist, and some which possess higher-order time dependencies while still retaining the linear nature which is a boon to any analytical campaign attempting to study the consequences of the spatial nonlinearity over one involving nonlinear evolution terms. A treatment of any system in the manner described here is beyond the scope of this supplement, and this commentary is only meant to discuss possible means by which to explore the unsteady nature of Bennett vortices.

\qquad A postscript to note in this regard is the application of this functional form to the modelling of a time-only signal, e.g.,
\begin{equation}
    I(t) = I_0\frac{t^{2}}{(t + C_{B,I})^{2}}
\end{equation}
to provide a ramp-law. The ramp-time constant $C_{B,I} \neq C_{B,T}$ for reasons of dimensional mismatch so this will need to be determined from the boundary conditions of the system. Continuing in this postscriptoral vein, the time-signal can be also attached separately to the fundamental Bennett form,
\begin{equation}
    u_{z}(r, t) = A(t)u_{z,0}\frac{r^{2}}{(r + C_{B,T})^{2}}
\end{equation}
to describe a vortex-modulated amplitude excitation. This is perhaps one of the most tractable ways to study the unsteady regime as the separability lends the temporal character cleanly to any nonlinear system. This gives different ramp-forms that are naturally attached to the aforementioned nonlinear PDEs, 
\begin{align}
    &KdV: \ \dot{A}u_{B}(z) + (\partial_{zzz}u_{B})A - 6u_{B}u_{B}'A^{2} = 0 \\
    &KS: \ \dot{A}u_{B}(z) + A(u_{B}'' + u_{B}'''') + A^{2}\frac{1}{2}u_{B}'^{2} = 0 \\
    &KS(3D): \dot{A}u_{B}(r) + A\bigg(\frac{1}{r}\dv{(ru_{B}')}{r} + \frac{1}{r}\dv{}{r}r(\nabla^{2}u_{B})'\bigg) + A^{2}\frac{1}{2}u_{B}'u_{B}'^{\ast} = 0
\end{align}

\section*{Bennett-Grad-Shafranov Vorticity}
\qquad In toroidal coordinates the analogous axisymmetric ideal MHD system is given by the Grad-Shafranov equation, which has the same form as a Hick's equation for an axisymmetric inviscid fluid,
\begin{equation}
    \pdv[2]{\psi}{r} - \frac{1}{r}\pdv{\psi}{r} + \pdv[2]{\psi}{z} = -\mu_{0}r^{2}\dv{p}{\psi} - \frac{1}{2}\dv{F^{2}}{\psi}
\end{equation}
with the stream function, $\psi$, being introduced so that the incompressibility of the axisymmetric system,
\begin{equation}
    \nabla\cdot\vec{u} = 0 
\end{equation}
is built in to the formulation via,
\begin{equation}
    \vec{u} = \frac{1}{r}\nabla\psi\times\hat{\theta} 
\end{equation}
which gives,
\begin{align}
    u_{r} &= -\frac{1}{r}\pdv{\psi}{z} \\
    u_{z} &= \frac{1}{r}\pdv{\psi}{r}
\end{align}
In the case of the cubic, pure-flow vortices that we have been studying the stream function is supplied fundamentally by a toroidal flow that is locally axial,
\begin{equation}
    \psi(r,z) = \psi(0,z) + u_{z,0}\int_{0}^{r}\frac{r'^{3}}{(r' + C_{B,T})^{2}}dr'
\end{equation}
The accompanying radial flow is then given entirely by the free function along the pinch axis,
\begin{equation}
    u_{r} = -\frac{1}{r}\bigg(\pdv{\psi(0,z)}{z}\bigg)
\end{equation}
which introduces problems with regularity unless the stream function goes to zero sufficiently fast enough near the axis, or is treated as being separable into a form which annihilates the singularity, e.g.,
\begin{equation}
    \psi(r,z) \sim r^{2}F(z) + \psi_{0}(z)
\end{equation}
however, the above canonical form is still inconsistent with a Bennett vortex for the locally axial flow. An additional gauge must be added so that,
\begin{align}
    u_{z}(r) &= \frac{1}{r}\pdv{\psi}{r} \\
    &= \frac{1}{r}\bigg(2rF(z) + \pdv{\psi_{B}(r,z)}{r}\bigg) \\
    &= u_{B}(r)
\end{align}
which implies that,
\begin{equation}
    \psi_{B}(r,z) = \psi_{B}(0,z) - r^{2}F(z) + \int_{0}^{r}r'u_{B}(r')dr'
\end{equation}
so that the radial flow field is,
\begin{align}
    u_{r} &= -\frac{1}{r}\bigg(r^{2}F'(z) + \psi_{0}'(z) + \psi_{B}(r,z)_{,z}\bigg) \\
    &= -\frac{1}{r}\bigg(\psi_{0}'(z) + C'(z)\bigg) \\
    &= -\frac{1}{r}\tilde{\psi}_{0}'(z)
\end{align}
where,
\begin{align}
    \tilde{\psi}_{0}(z) &= \psi_{0}(z) + \psi_{B}(0,z) \\ 
    &= \psi_{0}(z) + C(z)
\end{align}
Interestingly, the structure of the breathing mode does not influence the radial flow field when the toroidal flow is locally describable by that of the given Bennett vortex. Rather, the radial flow field depends entirely on two free functions, and its regularity on their cancellation when an axial derivative is taken.

\section*{Magnetic Reconnection}
\qquad When a finite resistivity is incorporated into the plasma dynamics, more than just radial drifts develop. Specifically, the magnetic field can also assume an unsteady, and diffusive, character which arises when field line topologies are no longer "frozen-in" to the plasma, e.g., near regions of large magnetic shear. Bennett vortices naturally have such a region occur in the equilibrium as the magnetic shear remains small until the flow shear layer develops the plasma close to its edge state, and then the onset of relatively large amounts of enclosed current across a thin region results in the onset of relatively large amounts of magnetic field amplitude across a thin region, i.e., a current-sheet like structure.

\qquad In the case of a finite resistivity, governance of the magnetic field expands in scope as previously described, obtaining,
\begin{equation}
    \pdv{\vec{B}}{t} = \nabla\times(\vec{u}\times\vec{B}) + \eta\nabla^{2}\vec{B}
\end{equation}
The structure of the magnetic field's unsteady, and diffusive character depends on the underlying form of the vortex. We have studied three steady flow forms so far,
\begin{align}
    \vec{u} &= u_{z}(r)\hat{z} \\
    \vec{u} &= u_{r}(r,z)\hat{r} + u_{z}(r)\hat{z} \\
    \vec{u} &= u_{\rho}(\rho, \phi)\hat{\rho} + u_{\phi}(\rho, \phi)\hat{\phi} = u_{z}(r)\hat{z}
\end{align}
as well as commented on the possibility of attaching various temporal modulations to the flow field pattern by giving an unsteady character to some underlying plasma parameter that is encapsulated by the shear layer positioning,
\begin{equation}
    \vec{u}(r, t) = u_{z}(r,t)\hat{z} = u_{z,0}\frac{r^{2}}{(r + C_{B,T}(t))^{2}}\hat{z}
\end{equation}
or by an amplitude signal,
\begin{equation}
    u_{z}(r,t) = A(t)u_{B}(r)
\end{equation}
and whose consequences can be readily studied in the context of various nonlinear systems, e.g., KdV, or KS, as described previously. While these extensions provide natural routes to studying the evolution of a diffusive magnetic field, a quantitative treatment of reconnection requires resistive, and kinetic effects which are outside the scope of this investigation. This brief note is merely to highlight the numerous possibilities that the natural geometric structure of the theory furnishes for settings in which these processes may arise. 

\section*{Ball Lightning}
\qquad It is also ventured that perhaps ball lightning could be a Bennett Vortex. In fact, this is not just wild speculation as observations of the phenomenon are frequently characterized by reports of it moving slowly in a single direction\cite{SHMATOV2019105115}. This matches both with the purely axial flow of a Bennett Vortex, and also the requirement that a macroscopic, "squat" pinch
\begin{equation}
    r_{p} > L
\end{equation}
has for being small-$C_{B,T}$. Namely, that the flow speed must be small!
\begin{equation}
    C_{B,T} = C_{B,u}n_{0}u_{z,0}^{2}\frac{r_{p}^{3}}{T_{p}}
\end{equation}
For a terrestrial vortex, let us think of the ball lightning as being a magnetized glob of charge which possesses mass, momentum, energy, and thermal power as residues of a lightning strike. The environment will absorb some of this energy, and the rest will be put into sustaining a plasma for as long as the conditions afford. 

\qquad The sustainment of ball lightning for such long periods can be explained by the formation of a plasma with Bennett Vortices in it from the stray thermal and electrical power left over in the aftermath of a lightning strike. This is perhaps a creative tangent to attempt and make, but the security of a shear-flow stabilized Z-Pinch with the thermal power still on affords a means by which the diffusive magnetic equilibrium could sustain itself for macroscopic timescales if there was sufficient thermal energy to begin with to justify its existence locally. Ball lightning would also have some stray electrical power remnant leftover from the lightning strike. 

\qquad The confinement time, and properties of this plasma can be calculated. When making calculations, it is best to stick to a small-$C_{B,T}$ plasma as this will not require you to first solve for the $u_{z,0}$ roots. If we consider a plasma where,
\begin{align}
    n_{0} &= 10^{25} \ [m^{-3}] \\
    u_{z,0} &= 10^{-3} \ [m \ s^{-1}] \\
    r_{p} &= 1 \ [m] \\
    T_{p} &= 1000 \ [degK]
\end{align}
then the resulting confinement time is,
\begin{equation}
    \tau_{E} = 75.8 \ [s]
\end{equation}
a value which is in good accord with observations where confinement times range up to a minute. The parameter $C_{B,T} = 1.46*10^{-6}$ is small, and the total power of $P = 164 \ [MW]$ is a very sensible number for the residual power of a lightning strike which could involve gigawatts of electrical power being discharged through the air. 

\qquad To connect this more strongly to the formation process for ball lightning in the aftermath of a lightning strike, then viscous, and thermal plasma chemistry effects would also need to be incorporated as the air becomes superheated and begins to drag on the plasma, with the energy leading to excitation of the plasma species, and possible chemical reactions. The lightning could potentially form in an evacuated pocket left behind when the current is done flowing into the ground. If not all of this current was able to be absorbed by the locality of the strike, then some of it would be ejected into this pocket with the leftover energy of the strike, alongside material from the ground which will advect with the plasma if it becomes ionic, thereby naturally being confined by the strong edge magnetic field.  

\section*{Bennett-Tipler Vortices}
\qquad The circumstances under which the investigation of the Shumlak criterion proceeded in order to yield the structure of these vortices are somewhat analogous to that of the emergence of Closed Timelike Curves (CTCs) in a Tipler cylinder where sufficiently strong rotation in an infinitely long cylinder causes light cones to tilt back over on themselves and close. While the present work is entirely within the context of non-relativistic, classical, ideal magnetohydrodynamics the analogy between strong vorticity and frame-dragging effects suggests a broader perspective by which strong rotational flow fields organize structure.  

\qquad In light of the Casimir Effect, which shows that a negative energy density appears between two conducting plates that are sufficiently close together due to the reduction of certain electromagnetic modes in the interstitial quantum vacuum, it bears investigating the relativistic consequences that two infinitely long rotating concentric cylinders of plasma have for this interstitial vacuum. It is important to emphasize that these negative energy densities arise from quantum boundary conditions on the fluctuation fields, and not classical plasma dynamics. Regardless, the study of highly sheared, rotating plasma configurations in this state provides a conceptual bridge towards understanding how energy distributions are influenced by rotating structure in more general field-theoretic settings.

\qquad The author's primary motivation for making this suggestion, beyond to expand the frontiers of gravitational and plasma physics, is to consider whether such configurations could, in principle, inform discussions of exotic energy conditions in relativistic spacetimes. Any potential relevance to constructs such as traversable wormholes would require a fully quantum field-theoretic treatment unified, and consistently coupled with the plasma dynamics presented in this article, and lies outside the scope of the present supplementary material. 

\section*{Relativistic Bennett Vorticity}
\qquad A fully relativistic theory of Bennett vorticity involves two levels of treatment. In the simplest flat spacetime treatment we cannot study the effect of mass on the local curvature of a configuration, and so to study the dynamics of vortices within a mass-dominated context we would need to extend our treatment to the second level, namely, that of the domain of general relativity. This would need to happen before we could study the boundary between filamentary plasmas and the gravitating, vortical systems which are observed together at the nodes of this filamentary network in the context of the cosmic web. This is beyond the scope of the current work, and neither shall we start from a first principles framework for the relativistic conservation of mass, momentum and energy, based on energy-momentum, electromagnetic stress, and material flux throughout a metered spacetime.

\qquad Instead, we can turn our attention to the electron inertia, which in the idealized MHD system we have studied heretofore is considered trivial. As the electron speed increases to relativistic scales we must treat the momentum as,
\begin{equation}
    p_{e} = \gamma_{z} m_{e}u_{z}
\end{equation}
where,
\begin{equation}
    \gamma_{z}(r) = \frac{1}{\sqrt{1 - \frac{u_{z}^{2}(r)}{c^{2}}}}
\end{equation}
If we take $n_{0}$ to be the proper density, i.e., the density measured in the electron frame then the relativistic contraction of space requires the plasma current density to be corrected,
\begin{equation}
    J_{z} = q\gamma_{z}(r)n_{0}u_{z}(r)
\end{equation}
and the kinetic energy density,
\begin{equation}
    \varepsilon_{K,e} = (\gamma_{z} - 1)\gamma_{z}n_{0}m_{e}c^{2}
\end{equation}
In regimes where $\gamma >> 1$ the relativistic inertia of the electrons then require modifications to the pressure balance, magnetic field, and plasma current density. 

\section*{Kadomtsev Stability}
\qquad A Z-pinch can be made stable to the $m = 0$ mode by tailoring the pressure profile in a certain way as B.B. Kadomtsev discovered\cite{kadomtsev1966hydromagnetic}
\begin{equation}
    \label{eqn:kadomtsev_stability}
    -\dv{\ln(p)}{\ln(r)} \geq \frac{4\Gamma}{2 + \Gamma\beta}
\end{equation}
which can be manipulated into the form,
\begin{equation}
    B^{2} \leq -\frac{\mu_{0}\Gamma p}{2\Gamma + \dv{\ln(p)}{\ln(r)}}\dv{\ln(p)}{\ln(r)}
\end{equation}
for $\beta = \frac{2\mu_{0}p}{B^{2}}$. For $T = C_{T}r^{3}$ the above can be written as,
\begin{equation}
    B^{2} + \frac{3\mu_{0}\Gamma(n_{0}k_{B}C_{T})^{2}}{2\Gamma + 3 n_{0}k_{B}C_{T}}r^{3} \leq 0
\end{equation}
evidently leaving only one possible path forward in pursuing the above line of attack, since both $B^{2}$, and the cubic term are positive-definite for $C_{T} > 0$. Namely, finding roots to the equation,
\begin{equation}
    B^{2}(r) + \frac{3\mu_{0}\Gamma(n_{0}k_{B}C_{T})^{2}}{2\Gamma + 3 n_{0}k_{B}C_{T}}r^{3} = 0
\end{equation}
which for the specific case of $T = C_{T}r^{3}$ can be represented by,
\begin{equation}
    \label{eqn:kadomtsev_stable_bennett_vortex}
    \frac{\mu_{0}^{2}e^{2}n^{2}_{0}u^{2}_{z,0}}{2r^{2}(r + C_{B,T})^{2}}f^{2} + \frac{3\mu_{0}\Gamma(n_{0}k_{B}C_{T})^{2}}{(2\Gamma + 3n_{0}k_{B}C_{T})}r^{3} = 0
\end{equation}
The highest-order in $f^{2}$ is sixth, however, the denominator attached to $f^{2}$ has fourth-order terms lurking which must be multiplied through, leading to,
\begin{align*}
    r^{7} &+ \mu_{0}e^{2}u_{z,0}^{2}(\frac{\Gamma + 3n_{0}C_{T}k_{B}}{6\Gamma k_{B}^{2}C_{T}^{2}})f^{2}(r, C_{B,T}) \\
    &\qquad + 2 C_{B,T}r^{6} + C_{B,T}^{2}r^{5} = 0
\end{align*}
the roots of this equation would give relationships amongst the parametric values of the model that represent states where the Kadomtsev condition is exactly met. No analytic formula is, or can be, known for root-finding with arbitrary coefficients for polynomials of degree greater than or equal to 5\cite{Ruffini1813}. Therefore, roots to Equation (\ref{eqn:kadomtsev_stable_bennett_vortex}) would need to be found numerically. This computation is of interest as these roots describe Bennett Vortices which are also at the very limit of what Kadomtsev stability to the $m = 0$ mode would accept.     

\section*{Sawteeth}
\qquad Sawteeth naturally emerge in chains of these vortices. A single tooth looks like, 
% FIGURE OUT WHAT THIS LOOKS LIKE
\begin{equation}
    \triangle(r) = \begin{cases}
        \frac{u_{0}}{r_{p}}r & \text{if } 0 < r \leq r_{p} \\
        -\frac{u_{0}}{r_{p}}r + 2u_{0} & \text{if } r_{p} < r < 2r_{p} \\
        0 & \text{else}
    \end{cases}
\end{equation}
Writing this as a chain of core pure-flow cubic vortex, and bulk, negative, pure-flow cubic vortex we have,
\begin{equation}
    u(r) = \begin{cases}
        u_{z,0}\frac{r^{2}}{(r + C_{B,T})^{2}} & \text{if } 0 < r \leq r_{p} \\
        u_{0} - u_{z,0}\frac{r^{2}}{(r + C_{B,T})^{2}} & \text{if } r_{p} < r \leq 2r_{p} \\
        0 & \text{else}
    \end{cases}    
\end{equation}
which is equivalent to,
\begin{equation}
    u(r) = u_{z}^{(2)}(r)\square(r) + u_{z}^{(2,-)}(r-r_{p})\square(r - r_{p})
\end{equation}
where,
\begin{equation}
    \square(r) = \begin{cases}
        1 & \text{if } 0 \leq r \leq r_{p} \\
        0 & \text{else}
    \end{cases}
\end{equation}
At the pinch radius we have,
\begin{align}
    u_{z}^{(2,-)}(r_{p}-r_{p}) = u_{z}^{(2,-)}(0) = u_{0} \\
    u_{z}^{(2)}(r_{p}) = u_{z,0}\frac{r_{p}^{2}}{(r_{p} + C_{B,T})^{2}}
\end{align}
Equating these two we have,
\begin{equation}
    u_{0} = u_{z,0}\frac{r_{p}^{2}}{(r_{p} + C_{B,T})^{2}}
\end{equation}
where the ratio of the core speed to the flow modulus of the second vortex naturally emerges as a key parameter,
\begin{equation}
    Ur = \frac{u_{0}}{u_{z,0}}
\end{equation}
However, we can study the structure of the normalized system instead of having to pin down a value for the above, 
\begin{equation}
    \tilde{u}_{z}^{(2,-)}(0) = \frac{u_{z}^{(2,-)}(0)}{u_{0}} = 1
\end{equation}
and,
\begin{equation}
    \tilde{u}_{z}^{2}(\phi_{p}) = \frac{u_{z}(r_{p})}{u_{z,0}} = \frac{\phi_{p}^{2}}{(\phi_{p} + 1)^{2}} 
\end{equation}
with,
\begin{equation}
    \phi_{p} = \frac{r_{p}}{C_{B,T}}
\end{equation}
Independent of the flow ratio number we then have the requirement that,
\begin{align}
    \phi_{p}^{2} = (\phi_{p} + 1)^{2} \\
    \implies 2\phi_{p} = -1 \\
    \therefore \phi_{p} = -\frac{1}{2}
\end{align}
which is impossible to obtain in the positive half-chord. This implies that an arbitrary 2-chain of vortices will inevitably construct a non-uniform sawtooth as the lack of a physically-realizable match at the boundary of the vortices means that either one, or both, of the vortices will under-shoot their target depending on the exact structure of the system. Naturally, this introduces a jump discontinuity into the system at the interface that suggests a mechanism by which shocks can naturally form.

\section*{L-H Mode Transition}
\qquad The existence of a high confinement "H" mode in tokamaks is coincident with the development of a transport barrier at the edge of the plasma that appears when the plasma is sufficiently heated. This transport barrier is further accompanied by the observation of enhanced thermodynamic gradients in the plasma. The current physical explanation for this process is that sheared $E\times B$, and zonal flows, alongside large edge plasma pressure gradients leads to suppression of the turbulent transport, but the exact self-organization of this system is unexplained. Large shears in the plasma current density near the pedestal have also been observed to occur over millimeter scales\cite{Groebner_2023} just before ELMs suggesting that this shear-dominated regime can be naturally connected to radial fluxes of particles, momentum, energy, and heat.

\qquad This remains an active area of investigation, and the structure of the Bennett vortex family is highly suggestive in this context. As such, it could serve as a useful model which provides a local analytic form for describing the shear-layer structure that appears in this transition. In particular, the coexistence of these structures with large thermal edge gradients suggests a possible route by which sufficiently strong heating can drive the plasma into a shear-dominated, transport-suppressed regime. 

\qquad A quantitative investigation of the relevant growth rates, balanced against the turbulent decorrelation of the sheared plasma dynamics, is beyond the scope of this present work, and left for future study. However, to make notes about this future study observe that,
\begin{align}
    \frac{r}{r + C_{B,T}} &= \frac{\psi_{N}}{\psi_{N} + \lambda} \\
    \psi_{N} &= \frac{r}{r_{p}} \\
    \lambda &= C_{B,N} = \frac{C_{B,T}}{r_{p}}
\end{align}
means the flux-coordinate, and pinch-normalized forms here are exactly the same as the cylindrical axisymmetric Z-pinch forms under the given transformation. The plasma pressure gradient in this representation becomes,
\begin{align}
    \dv{p}{\psi_{N}} &= -\frac{e^{2}n_{0}^{2}u_{z,0}^{2}\mu_{0}}{2}\frac{\psi_{N}}{(\psi_{N}+\lambda)^{3}}f(r, C_{B,T}) \\
    &= -\frac{e^{2}n_{0}^{2}u_{z,0}^{2}r_{p}^{3}\mu_{0}}{2}\frac{\psi_{N}}{(\psi_{N}+\lambda)^{3}}f(\psi_{N}, \lambda) \\
    &= -8n_{0}k_{B}C_{B,T}T_{p}\frac{\psi_{N}}{(\psi_{N}+\lambda)^{3}}f(\psi_{N}, \lambda)
\end{align}
where we have used the relation,
\begin{equation}
    \mu_{0}e^{2}n_{0}u_{z,0}^{2}r_{p}^{3} = 16k_{B}C_{B,T}T_{p}
\end{equation}

\section*{Thermal Lifetime \& Vortex Plume Structure}
\qquad The implication that each vortex has a finite thermal lifetime,
\begin{equation}
    \tau_{E} = \frac{3}{2}\frac{p_{0}{V_{p}}}{\int_{A_{L}}\vec{q}\cdot d\vec{A}}
\end{equation}
and its properties as a shear-flow stabilized Z-pinch means that the process of an $m = 0$ mode quenching the pinch at the end of its lifetime can interpreted kinematically in reverse as watching plume expansion from a compact core. This observation is useful in the context of computational applications, e.g., fusion propulsion, waterfall reactors, cosmology, etc. for obtaining an insight into the state of the magnetohydrodynamic fluid, especially at the surface of the compressive necking process, as an adjoint construction to it. In the context of engineering studies, this would give numerical insight into the aspect ratio of the simulated process, defined by the pinch radius near the collapse point over the equilibrium pinch radius,
\begin{equation}
    R^{\ast} = \frac{r_{neck}}{r_{p}} 
\end{equation}
the specific way to obtain $r_{core}$ computationally is outside the scope of this article, and left for a future work.

\qquad The structure of the thermal lifetime is of the utmost importance for any application of this profile, and so is the usage of this to extend the thermal lifetime of reactors, as well as to better understand the physics behind the process of an ideal, shear-flow stabilized thermal lifetime captured by a perpendicular thermal conductivity. The self-similarity of the equilibrium family solutions, meaning the class which retain the same locating of the shear layer, and same flow velocity, relative to each other, are identical in flow pattern, and can therefore have their plasma conditions varied, in particular plasma number density and temperature, and obtain the exact same flow pattern for a higher temperature and a larger density, as well as the converse. 

\qquad The presence of "inverse" parabolic temperature profiles in Zap DD shear-flow stabilized Z-pinch fusion plasma\cite{uri_2017} experiments, i.e., the "weak" form of the vortex, also motivates investigating the consequences of the parabolic temperature for the impact on thermal lifetime. Especially, to find ways to extend it naturally or to provide a setting for the state of the plasma during the collapse / expansion process to study the conditions of the plasma during an attempt to resuscitate the pinch, or just naturally as one part of the pulsed power cycle for a whole-device model where requirements for pinch current and repetition rate define the fusion gain, and adiabaticity of the fusion plasma process.  

\qquad The search for a maxima of pinch lifetime occurs in a 4D space of $(n_{0}, u_{z,0}, r_{p}, T_{p})$, and complex solutions to the root of a flow constant produce the same $C_{B,T}$ plasma relative to a real flow with the same magnitude, so their existence does not interfere with the definition of a thermal lifetime. One approach to resolving this space is to first fix the flow constant, which is a natural aid to an experimental investigation, and required for exactly self-similar solutions. However, the calculation for the pressure depends on logarithmic terms that introduce numerical instabilities into the solution, and there is not an immediately apparent fix as when this problem arose in making drift speed calculations.   

\qquad There are also numerous ways to treat the thermal lifetime that do not hinge on calculating the representative pressure in the above manner. There are also representative pressures for bulk vortices, and other orders of vortex that have been presented in this supplement, which establish how the norm of this could be taken to explore additional avenues of thermal lifetime study. Experimental values may also of course be taken as representative for the plasma pressure. 

\section*{Extended Vorticity}
\qquad The Shumlak criterion is what judges a plasma with axial flow to be shear-flow stabilized or not. In the case of the Z-pinch equilibrium the vorticity is,
\begin{equation}
    \vec{\omega} = \nabla\times\vec{u} = -\dv{u_{z}}{r}\hat{\theta}
\end{equation}
so that interestingly the Shumlak criterion can be written in this case as,
\begin{equation}
    |\vec{\omega}\cdot\hat{\theta}|=|\omega_{\theta}| > 0.1kV_{A}
\end{equation}
or more generally,
\begin{equation}
    |\vec{\omega}| > 0.1kV_{A}
\end{equation}
because the relevant shear naturally arises on the LHS, so each of these interpretations are equivalent in this specific case. This natural occurrence is highly suggestive of deeper mathematical structure, but care should be taken when interpreting the implications because in general the above is not true. It is only true when considering a pure Z-pinch.  

\qquad The most rigorous implication is that a wide variety of shear-flow stabilized Bennett-Shumlak vortices is possible, as any vorticity that possesses sufficient radial shear in the axial velocity can have cross-vorticities that exist alongside this shear-flow stabilized character which do not compromise it for the original forms presented in this work. For example, an axisymmetric system where $\pdv{}{\theta} \rightarrow 0$,
\begin{equation}
    \vec{\omega} = -\pdv{u_{\theta}}{z}\hat{r} + (\pdv{u_{r}}{z} - \pdv{u_{z}}{r})\hat{\theta} + \frac{1}{r}\pdv{(ru_{\theta})}{r}\hat{z}
\end{equation}
that can also be taken as swirlless,
\begin{equation}
    \vec{\omega} = (\pdv{u_{r}}{z} - \pdv{u_{z}}{r})\hat{\theta}
\end{equation}
or relaxed entirely,
\begin{equation}
    \vec{\omega} = (\frac{1}{r}\pdv{u_{z}}{\theta}-\pdv{u_{\theta}}{z})\hat{r} + (\pdv{u_{r}}{z} - \pdv{u_{z}}{r})\hat{\theta} + (\frac{1}{r}\pdv{(ru_{\theta})}{r} - \frac{1}{r}\pdv{u_{r}}{\theta})\hat{z} 
\end{equation}
The growth of swirl from the fundamental physics of the cylindrical equilibrium has not been addressed yet so we can either focus on the swirlless, axisymmetric form in a cylindrical basis, or study the implications for the expression of this vector in a spherical one. One observation to note beforehand is that a broken azimuthal symmetry is the only direct avenue for the introduction of axial velocity gradients that do not contribute to the Shumlak criterion.  

\qquad Let us study these spherical implications then, where the axial plasma flow becomes,
\begin{equation}
    \vec{u} = u_{\rho}(\rho, \phi)\hat{\rho} + u_{\phi}(\rho,\phi)\hat{\phi}
\end{equation}
giving a vorticity,
\begin{equation}
    \vec{\omega} = (\frac{1}{\rho\sin(\theta)}\pdv{u_{\rho}}{\phi} - \frac{1}{\rho}{\pdv{(\rho u_{\phi})}{\rho}})\hat{\theta}
\end{equation}
demonstrating that the introduction of swirl does not arise from any consequence of the spherical description of a pure Z-pinch. However, it does suggest that the azimuthal vorticity can naturally be thought of as a reflection of two terms in this equilibrium. This is a meaningful observation to make as it suggests that in a quasi-equilibrium pinch axial gradients in the radial in/outflow will lead to an increase in azimuthal vorticity. In an equilibrium pinch of the kind we have studied, these radial flows will formally have no such gradients. The addition of an unsteady character to the system admits the possibility for the growth of strong electric fields in cross directions which could give rise to substantial swirling, or in/out-flowing components. 

\qquad Another observation to make is that the relaxation of axial symmetry in the cylindrical basis, e.g., 
\begin{equation}
    \vec{u}(r,z) = u_{r}(r,z)\hat{r} + u_{z}(r)\hat{z}
\end{equation}
does not introduce any fundamentally new structure to the spherical description,
\begin{align}
    \vec{u}(r,z) &= (u_{r}(\rho, \phi)\sin\phi + u_{z}(\rho,\phi)\cos\phi)\hat{\rho} + (u_{r}(\rho, \phi)\cos\phi - u_{z}(\rho, \phi)\sin\phi)\hat{\phi} \\
    &= \vec{u}(\rho, \phi)
\end{align}
although it does suggest further avenues to explore the possibility of flow stagnation on these directions. This is an important path to explore because an instant of stagnated flow in the polar direction will lead to the appearance of a purely radial flow,
\begin{equation}
    u_{\rho} = 2u_{r}\sin\phi
\end{equation}
defined by,
\begin{equation}
    u_{z} = u_{r}\tan\phi
\end{equation}
We can also study the conditions for (spherical) radial stagnation, which would leave us with a flow that appears to be completely polar,
\begin{align}
    &u_{\rho} = 0 \\ 
    \therefore &u_{r}\sin\phi = -u_{z}\cos\phi \\
    \implies &u_{z} = -u_{r}\tan\phi \\
    \therefore u_{\phi} &= u_{r}\cos\phi + u_{r}\sin\phi\tan\phi \\
    &= u_{r} \frac{\cos^{2}\phi + \sin^{2}\phi}{\cos\phi}\\
    &= \frac{u_{r}}{\cos\phi}
\end{align}
\qquad 
which is interesting because it suggests that in this regime certain polar angles will naturally lead to a singular explosion in the velocity which are not present in the converse situation.

\bibliography{sn-bibliography}% common bib file

\end{document}